%% file: viscoelastic-stability-mdpi.tex
\documentclass[preprints,article,accept,moreauthors,pdftex]{Definitions/mdpi} 

\firstpage{1} 
\pubvolume{xx}
\issuenum{1}
\articlenumber{5}
\pubyear{2019}
\copyrightyear{2019}
\history{Received: date; Accepted: date; Published: date}
\pdfoutput=1

\usepackage{amsmath,amssymb}
\usepackage{mathptmx}

\usepackage[english]{babel}
\usepackage{subfig}
\usepackage[active]{srcltx} % use TeX-souce-specials-mode

\usepackage{mathbbol}
\usepackage{bm} % sophisticated \boldsymbol
\usepackage{units}
\usepackage{tensor}
\usepackage{accents}

\input{vit-prusa-macros-experimental}

\input{viscoelastic-taylor-couette-stability-macros}

\usepackage{todonotes}

\let\cite\citet

\Title{Finite amplitude stability of internal steady flows of the Giesekus viscoelastic rate-type fluid}

\Author{Mark Dostal\'{\i}k$^{1}$,\orcidC{},
  V\'{\i}t Pr\r{u}\v{s}a$^{1,}$*\orcidA{},
  and Karel T\r{u}ma$^{1}$\orcidB{}
}

\AuthorNames{Mark Dostal\'{\i}k, V\'{\i}t Pr\r{u}\v{s}a and Karel T\r{u}ma}

\address[1]{%
$^{1}$ \quad Faculty of Mathematics and Physics, Charles University, Sokolovsk\'a 83,
Praha 8 -- Karl\'{\i}n, CZ 186\;75, Czech Republic;  prusv@karlin.mff.cuni.cz\\
}

\corres{Correspondence: prusv@karlin.mff.cuni.cz%; Tel.: (optional; include country code; if there are multiple corresponding authors, add author initials) +xx-xxxx-xxx-xxxx (F.L.)
}

\abstract{Using a Lyapunov type functional constructed on the basis of thermodynamical arguments we investigate the finite amplitude stability of internal steady flows of viscoelastic fluids described by the Giesekus model. Using the functional we derive bounds on the Reynolds and the Weissenberg number that guarantee the unconditional asymptotic stability of the corresponding steady internal flow, wherein the distance between the steady flow field and the perturbed flow field is measured with the help of the Bures--Wasserstein distance between positive definite matrices. The application of the theoretical results is documented in the finite amplitude stability analysis of Taylor--Couette flow.}


\begin{document}
%%%%%%%%%%%%%%%%%%%%%%%%%%%%%%%%%%%%%%%%%%

\input{viscoelastic-stability-body}

%\appendix

\externalbibliography{yes}
\bibliography{vit-prusa,karel-tuma,mark-dostalik}

%%%%%%%%%%%%%%%%%%%%%%%%%%%%%%%%%%%%%%%%%%
\authorcontributions{V. P. devised the research objective, the main conceptual ideas and proof outline. K. T. performed the finite element simulations. All authors discussed the results and methods, and contributed equally to the final manuscript.}

%%%%%%%%%%%%%%%%%%%%%%%%%%%%%%%%%%%%%%%%%%
\funding{The authors acknowledge the support of the Czech Science Foundation project 18-12719S. Mark Dostal\'{\i}k and  Karel T\r{u}ma have been supported by Charles University Research program No.~UNCE/SCI/023.}

%%%%%%%%%%%%%%%%%%%%%%%%%%%%%%%%%%%%%%%%%%
\acknowledgments{The authors acknowledge the membership to the Nečas Center for Mathematical Modeling~(NCMM).}

%%%%%%%%%%%%%%%%%%%%%%%%%%%%%%%%%%%%%%%%%%
\conflictsofinterest{The authors declare no conflict of interest. The funders had no role in the design of the study; in the collection, analyses, or interpretation of data; in the writing of the manuscript, or in the decision to publish the results.}

\end{document}

%% file: vit-prusa-macros-experimental.tex
\DeclareMathOperator{\divergence}{div}

\DeclareMathOperator{\Tr}{Tr}

\DeclareMathOperator{\distanceop}{dist} % distance in a metric space

\newcommand{\charac}{\ensuremath{\mathrm{char}}} % characteristic quantity such as length scale, etc.
\newcommand{\reference}{\mathrm{ref}}

\newcommand{\bydefinition}{\mathrm{def}}
\newcommand{\traceless}[1]{{#1}_{\delta}}

\newcommand{\dimless}[1]{#1^\star}

\newcommand{\diff}{\mathrm{d}}
\newcommand{\Diff}[1][]{\mathrm{D}_{#1}} % For Frechet and Gateaux derivative
\renewcommand{\vec}[1]{\ensuremath{\mathbf{#1}}}

\makeatletter
\@ifpackageloaded{bm}% 
{\renewcommand{\vec}[1]{\ensuremath{\bm{#1}}}%
}{%
\relax% do nothing
}
\makeatother
\newcommand{\tensorq}[1]{{\ensuremath{\mathbb{#1}}}}      % tensorial quantity
\newcommand{\tensorc}[1]{\ensuremath{\mathrm{#1}}}      % tensorial quantity components  

\newcommand{\conjugate}[1]{#1^\star}
\newcommand{\transpose}[1]{#1^\top}

\newcommand{\inverse}[1]{#1^{-1}}

\newcommand{\identity}{\ensuremath{\tensorq{I}}}

\newcommand{\cstress}{\tensorq{T}}

\newcommand{\ecstress}{\tensorq{S}}

\newcommand{\fgrad}{\tensorq{F}}

\newcommand{\lcg}{\tensorq{B}}

\makeatletter
\@ifpackageloaded{bm}% 
{%
}{%

}

\@ifpackageloaded{bm}%
{%
 
}{%

}

\@ifpackageloaded{bm}%
{%
}{%

}
\makeatother

\newcommand{\generictensor}{{\tensorq{A}}}
\newcommand{\vecv}{\ensuremath{\vec{v}}}
\newcommand{\gradv}{\ensuremath{\nabla \vecv}}

\newcommand{\gradsym}{\ensuremath{\tensorq{D}}}

\newcommand{\gradvl}{\ensuremath{\tensorq{L}}}

\newcommand{\vhatp}[1][\vecvc]{{#1}^{\hat{\varphi}}}

\newcommand{\hatz}{\hat{z}}
\newcommand{\hatr}{\hat{r}}
\newcommand{\hatp}{\hat{\varphi}}

\newcommand{\cobvec}[1]{\vec{g}_{#1}} % covariant base vector
\newcommand{\R}{\ensuremath{{\mathbb R}}}
\newcommand{\Reynolds}{\mathrm{Re}}

\newcommand{\ienergy}{\ensuremath{e}} % specific internal energy (energy per unit mass)
\newcommand{\menergy}{\ensuremath{e}_{\mathrm{mech}}} % specific mechanical energy (energy per unit mass), kinetic energy plus internal energy minus thermal contribution
\newcommand{\fenergy}{\ensuremath{\psi}} % specific free energy
\newcommand{\entropy}{\ensuremath{\eta}} % specific entropy
\newcommand{\temp}{\ensuremath{\theta}} % temperature
\newcommand{\mns}{\ensuremath{m}} % mean normal stress
\newcommand{\nettenergy}{\ensuremath{E}_{\mathrm{tot}}} % net total energy
\newcommand{\netmenergy}{\ensuremath{E}_{\mathrm{mech}}} % net mechanical energy
\newcommand{\netentropy}{\ensuremath{S}} % net entropy
\newcommand{\cheatvol}{\ensuremath{c_{\mathrm{V}}}}
\newcommand{\hfluxc}{\vec{j}_{q}}     % heat flux, current configuration
\newcommand{\entfluxc}{\vec{j}_{\entropy}} % entropy flux, current configurtion 
\newcommand{\entprodc}{\xi} % entorpy production, current configuration
\newcommand{\entprodctemp}{\zeta} % entorpy production times temperature, current configuration

\newcommand{\pd}[2]{\ensuremath{\frac{\partial {#1}}{\partial {#2}}}}

\newcommand{\dd}[2]{\ensuremath{\frac{\diff {#1}}{\diff {#2}}}}

\newcommand{\fid}[1]{\ensuremath{\accentset{\triangledown}{#1}}}

\makeatletter
\@ifpackageloaded{tensor}% tensor is a package for a better typesetting of tensors
{

}{%

}
\makeatother

\makeatletter
\@ifpackageloaded{tensor}% tensor is a package for a better typesetting of tensors
{

}{%

}
\makeatother

\newcommand{\norm}[2][]{\ensuremath{\left\|#2\right\|_{#1}}}
\newcommand{\absnorm}[1]{\ensuremath{\left|#1\right|}}

\newcommand{\distance}[3][]{\distanceop_{#1}\left(#2, #3\right)} % distance in a metric space

\makeatletter
\@ifundefined{volume}{%
}%
{%
}
\makeatother

\newcommand{\cvolumee}{\diff \mathrm{v}}

\newcommand{\csurfacees}{\diff \mathrm{s}}

\newcommand{\intcvolume}[2][\volume]{\int_{#1} #2\; \cvolumee} % volume integral, current configuration
\newcommand{\tensortensor}[2]{\ensuremath{#1 \otimes #2}}
\makeatletter
\@ifpackageloaded{MnSymbol} % : as binary operator needs MnSymbol package
{
\newcommand{\tensordot}[2]{\ensuremath{#1 \vdotdot #2}} 
}{%
\newcommand{\tensordot}[2]{\ensuremath{#1 : #2}} 
}
\makeatother
\newcommand{\vectordot}[2]{\ensuremath{#1 \bullet #2}}

\newcommand{\sleb}[2]{\ensuremath{L}^{#1} \left(#2 \right)}             % Lebesgue space

%% file: viscoelastic-taylor-couette-stability-macros.tex
\newcommand{\nplacer}{\kappa_{\mathnormal{p}(\mathnormal{t})}}  % placer, natural configuration

\newcommand{\lcgnc}{\ensuremath{\lcg_{\nplacer}}} % left Cauchy--Green tensor, natural -> current
\newcommand{\lcgncc}{\ensuremath{\lcgc_{\nplacer}}} % left Cauchy--Green tensor, natural -> current, components

\newcommand{\lcgncs}{\ensuremath{\tensorq{Z}_{\nplacer}}} % shifted left Cauchy--Green tensor, natural -> current
\newcommand{\fgradnc}{\ensuremath{\fgrad_{\nplacer}}} % deformation gradient, natural -> current
\renewcommand{\lcgncc}{\tensorc{B}}

\newcommand{\Brrhat}[1][\lcgncc]{\tensor{{#1}}{^\hatr_\hatr}}
\newcommand{\Brphat}[1][\lcgncc]{\tensor{{#1}}{^\hatr_\hatp}}

\newcommand{\Bprhat}[1][\lcgncc]{\tensor{{#1}}{^\hatp_\hatr}}
\newcommand{\Bpphat}[1][\lcgncc]{\tensor{{#1}}{^\hatp_\hatp}}

\newcommand{\Bzzhat}[1][\lcgncc]{\tensor{{#1}}{^\hatz_\hatz}}

\newcommand{\Weissenberg}{\mathrm{Wi}}

%% file: viscoelastic-stability-body.tex
\section{Introduction}
\label{sec:introduction}

Flows of viscoelastic fluids exhibit the phenomenon dubbed ``elastic turbulence'' or ``inertia-less turbulence''. The flows of viscoelastic fluids can become---unlike the flows of the standard viscous fluids---unstable or ``turbulent'' at very low values of the Reynolds number. This behaviour indicates that the instability or transition to ``turbulence'' is driven by a nonstandard mechanism. Namely, it is not driven by the nonlinearity due to the \emph{inertial term} in balance of linear momentum, but it must be attributed to the \emph{nonlinearity in the governing equation for the ``elastic'' part of the Cauchy stress tensor}. The key challenge is to identify the parameter values that prohibit the onset of ``elastic'' instability or that trigger the ``elastic'' instability, and to describe the transition scenarios leading from the laminar to the ``turbulent'' flow. This task requires one to perform some sort of nonlinear stability analysis, since a nonlinear interaction between the finite amplitude perturbations might be decisive.

The phenomenon of ``elastic turbulence'' has been thoroughly investigated both from the experimental as well as theoretical point of view, see reviews by \cite{petrie.cjs.denn.mm:instabilities}, \cite{larson.rg:instabilities}, \cite{shaqfeh.esg:purely}, \cite{morozov.an.saarloos.w:introductory} or \cite{li.x.li.f.ea:very-low-re}. In particular, the experimental results reported by~\cite{groisman.a.steinberg.v:elastic*1} has stimulated enormous research activity regarding the elastic turbulence. On the other hand, theoretical results mainly follow from direct numerical simulations based on various viscoelastic rate-type models, see~\cite{biancofiore.l.brandt.l.ea:streak}, \cite{valente.pc.silva.cb.ea:energy}, \cite{lee.sj.zaki.ta:simulations} and~\cite{plan.elcvm.musacchio.s.ea:emergence} for some recent contributions. The need to resort to sophisticated numerical simulations in order to get qualitative insight into the flow dynamics is not surprising.

The reason is that the instabilities in viscoelastic fluids are very likely of \emph{subcritical} nature, see~\cite{meulenbroek.b.storm.c.ea:weakly}. The \emph{subcritical} nature of the instability implies, as remarked by~\cite{morozov.an.saarloos.w:introductory}, that
\begin{quote}
  [Linear stability] (if it exists) is not very relevant for the existence of dynamics of the patterns that typically arise before the instability of the base state occurs.
\end{quote}
(Note that a similar issue arises even for the standard Navier--Stokes fluid, see for example~\cite{baggett.js.trefethen.ln:low-dimensional}.) This means that the linear stability analysis, that is stability analysis with respect to \emph{infinitesimal} perturbations, is of limited applicability in the investigation of the transition scenarios, albeit it can still provide important insight into the problem. (See for example~\cite{beris.an.avgousti.m.ea:spectral}, \cite{blonce.l:linear}, \cite{grillet.am.bogaerds.acb.ea:stability} and \cite{pourjafar.m.sadeghy.k:taylor-couette}, \cite{pourjafar.m.sadeghy.k:dean} for linear stability analysis of flows of viscoelastic fluids described by the Giesekus viscoelastic rate-type model.) Moreover, quoting again~\cite{morozov.an.saarloos.w:introductory} 
\begin{quote}
  [Subcritical instability] is governed by all kinds of nonlinear self-enhancing interactions and so there is almost never a simple approximation scheme that allows one to explore the infinite dimensional space of interactions in all details, and determine which direction corresponds to the smallest threshold [for instability]. Thus, in practice, one can explore such situations, in theoretical studies as well as in experiments, only for a given class of perturbations.
  % page 128
\end{quote}
On top of that, even if the technique such as weakly nonlinear analysis is apparently successful, then, as \cite{meulenbroek.b.storm.c.ea:weakly} put it,
\begin{quote}
  One should also keep in mind that our expansion is only carried out to lowest order in the nonlinearity, so one may wonder about the robustness of these results as long as higher order terms in the expansion are unknown.  
\end{quote}

In what follows we want to address the lack of analytical results for the stability problem of flows of viscoelastic fluids subject to \emph{finite amplitude perturbations}. In particular, using a Lyapunov type technique, we investigate the stability of internal \emph{steady} flows of viscoelastic fluids described by the Giesekus model, and \emph{we derive bounds on the values of the Reynolds number and the Weissenberg number that guarantee the flow stability subject to any (finite) perturbation}. The result provides a \emph{sufficient condition} for stability, hence it can be seen as complementary result to the search for the smallest threshold for instability via approximation methods. The derived bounds are interesting not only on their own. What is perhaps equally interesting is the way the bounds are derived. The \emph{derivation heavily relies on the underlying thermodynamical arguments} and the notion of energy, which is an approach that seems to be discouraged in the nonlinear stability analysis of viscoelastic fluids, see~\cite{doering.cr.eckhardt.b.ea:failure}. %The remedy for the apparent inapplicability thermodynamics in the stability analysis is based on a trivial observation regarding the thermodynamical nature of the system of interest. The flowing viscoelastic fluid is from a thermodynamical point of view a \emph{thermodynamically open system in a spatially inhomogeneous non-equilibrium steady state}. Consequently, thermodynamical methods developed mainly for the stability analysis of \emph{spatially homogeneous equilibrium steady states}, see~\cite{coleman.bd:on} and~\cite{gurtin.me:thermodynamics*1,gurtin.me:thermodynamics} are of no use. Clearly, a method suitable for analysis of thermodynamically open systems must be used. In this regard we follow the method proposed by~\cite{bulcek.m.malek.j.ea:thermodynamics}.

The paper is organised as follows. In Section~\ref{sec:giesekus-model} we describe the Giesekus model, and we briefly comment on its thermodynamical underpinnings. In particular, we identify the \emph{energy storage} mechanisms and the \emph{entropy production} mechanisms that are implied by the evolution equations for the Giesekus model. Once the thermodynamical background is summarised, we formulate the governing equations for an internal steady flow, see Section~\ref{sec:equilibrium-steady-state}, and we proceed with the stability analysis of this non-equilibrium steady state. The stability is analysed using a Lyapunov type functional $\mathcal{V}_{\mathrm{neq}}$ constructed by the thermodynamically based method proposed by~\cite{bulcek.m.malek.j.ea:thermodynamics}. The functional used in the stability analysis of a steady flow $\widehat{\vec{v}}$ in a domain $\Omega$ is constructed in Section~\ref{sec:construction-lyapunov-functional}, and it is given by the formula
\begin{equation}
  \label{eq:140}
  \mathcal{V}_{\mathrm{neq}}
  \left(
    \left.
      \widetilde{\vec{W}}
    \right\|
    \widehat{\vec{W}}
  \right)
  =
  \int_{\Omega}
  \frac{1}{2}
  \absnorm{\widetilde{\vec{v}}}^2
  \,
  \cvolumee
  +
  \int_{\Omega}
  \frac{\Xi}{2}
  \left[
    -
    \ln \det \left( \identity + \inverse{\widehat{\lcgnc}}\widetilde{\lcgnc} \right)
    +
    \Tr 
    \left( 
      \inverse{\widehat{\lcgnc}} 
      \widetilde{\lcgnc}
    \right)
  \right]
  \,
  \cvolumee
  ,
\end{equation}
where $\widetilde{\vec{v}}$ denotes the perturbation of the velocity field, $\widetilde{\lcgnc}$ is related to the perturbation of the stress field, and~$\widehat{\lcgnc}$ is related to the stress field in the steady flow. (%Quantity $\lcgnc$ is the left Cauchy–Green tensor associated to the elastic part of the fluid response, and it is a \emph{symmetric positive definite tensor}. One might also interpret it as the conformation tensor.
See the corresponding sections for the notation.) The fact that~\eqref{eq:140} can serve as a Lyapunov type functional is closely related to the proper choice of the \emph{distance function} that characterises the proximity of the perturbation $\vec{W}$ and the corresponding steady state~$\widehat{\vec{W}}$. In our case the distance function is introduced using the Bures--Wasserstein distance
\begin{equation}
  \label{eq:47}
  \distance[\tensorq{P}(d),\, \mathrm{BW}]{\tensorq{A}}{\tensorq{B}}
  =_{\bydefinition}
  \left\{
    \Tr \tensorq{A}
    +
    \Tr \tensorq{B}
    -
    2 \Tr
    \left[
      \left(
        \tensorq{A}^{\frac{1}{2}}
        \tensorq{B}
        \tensorq{A}^{\frac{1}{2}}
      \right)^{\frac{1}{2}}
    \right]
  \right\}^{\frac{1}{2}}
  ,
\end{equation}
see~\cite{bhatia.r.jain.t.ea:on}, which measures the distance between the symmetric positive semidefinite matrices $\tensorq{A}$ and~$\tensorq{B}$. Once we generalise~\eqref{eq:47} to the setting of spatially distributed fields of symmetric positive semidefinite matrices, we exploit the concept of Lyapunov functional, and we derive bounds on the Reynolds number and the Weissenberg number that guarantee the flow stability with respect to \emph{any} perturbation, see Theorem~\ref{thm:1}. These bounds are universal for any flow geometry.%, and it turns out that once the distance function is properly identified, then the stability analysis is relatively simple.

The bounds on the Reynolds number and the Weissenberg number are then explicitly evaluated in Section~\ref{sec:taylor-couette-flow} in the case of Taylor--Couette type flow. Further, we also perform direct numerical simulations that allow us to quantitatively document some features of the perturbation dynamics. The results are commented in Section~\ref{sec:conclusion}.

\section{Giesekus model}
\label{sec:giesekus-model}

\subsection{Governing equations}
\label{sec:govern-equat-mech}
The governing equations for the Giesekus fluid, see~\cite{giesekus.h:simple}, in the absence of external force read
\begin{subequations}
  \label{eq:viscoelastic-model-governing-equations-intro}
  \begin{align}
    \label{eq:continuity-equation-intro}
    \divergence \vec{v} &= 0, \\
    \label{eq:momentum-equation-intro}
    \rho \dd{\vec{v}}{t}
                        &=
                          \divergence \cstress, \\
    \label{eq:lcgnc-equation-intro}
    \nu_1 \fid{\overline{\lcgnc}} 
                        &=
                          -
                          \mu 
                          \left[ 
                          \alpha \lcgnc^2 + (1 - 2 \alpha) \lcgnc - (1 - \alpha) \identity
                          \right] 
                          ,
  \end{align}
  where $\vec{v}$ denotes the velocity, $\rho$ denotes the density, and $\lcgnc$ is an extra tensorial quantity whose physical meaning will be given later. Finally, the symbol $\cstress$ denotes the Cauchy stress tensor that is given by the formulae
  \begin{equation}
    \label{eq:cstress-formula-intro}
    \cstress 
    = 
    \mns \identity + \traceless{\cstress}, 
    \qquad
    \traceless{\cstress} 
    = 
    2 \nu \gradsym + \mu \traceless{\left( \lcgnc \right)},
  \end{equation}
\end{subequations}
where $\mns$ denotes the mean normal stress (pressure) and $\gradsym =_{\bydefinition} \frac{1}{2} ( \gradv + \transpose{\left(\gradv\right)})$ denotes the symmetric part of the velocity gradient. Symbols $\nu$, $\nu_1$, $\mu$ and $\alpha$, $\alpha \in (0,1)$, denote material parameters. Note that if $\alpha=0$, then one recovers the standard Maxwell/Oldroyd-B models. \emph{The value $\alpha = 0$ is however not covered in the presented stability analysis}.

The remaining notation is the standard one, $\dd{}{t}=_{\bydefinition} \pd{}{t} + \vectordot{\vec{v}}{\nabla}$ denotes the material time derivative, and 
\begin{equation}
  \label{eq:upper-convected-derivative}
  \fid{\generictensor}=_{\bydefinition} \dd{\generictensor}{t} - \gradvl \generictensor - \generictensor \transpose{\gradvl},
\end{equation}
denotes the upper convected derivative, where $\gradvl=_{\bydefinition} \nabla \vec{v}$, and the symbol
$
\traceless{\generictensor} =_{\bydefinition} \generictensor - \frac{1}{3} \left( \Tr \generictensor \right) \identity
$
denotes the traceless part of the corresponding tensor. In virtue of the incompressibility constraint~\eqref{eq:continuity-equation-intro} we have~$\traceless{\gradsym} = \gradsym$. Note that if one uses a simple substitution $\ecstress =_{\bydefinition} \mu (\lcgnc - \identity)$, and if one redefines the pressure, $p =_{\bydefinition} - \mns + \frac{1}{3} \left( \Tr \ecstress \right) \identity$, then~\eqref{eq:lcgnc-equation-intro} and \eqref{eq:cstress-formula-intro} transform to
$
\lambda \fid{\overline{\ecstress}}
+
\ecstress
+
\frac{\alpha \lambda}{\nu_1} \ecstress^2
=
2 \nu_1 \gradsym
$
and
$
\cstress 
= 
- p \identity + 2 \nu \gradsym + \ecstress
$,
where $\lambda = _{\bydefinition} \frac{\nu_1}{\mu}$. This is another frequently used form of the governing equations for the Giesekus fluid.

\subsection{Thermodynamic basis}
\label{sec:deriv-gies-model}
The Giesekus model has been originally derived without any reference to thermodynamics. However, we want to design a Lyapunov type functional using concepts from non-equilibrium thermodynamics, hence we need to explore thermodynamical underpinnings of the model. The issue of finding a thermodynamic basis for viscoelastic-rate type models is claimed to be resolved by a plethora of theories for thermodynamics of complex fluids, see for example~\cite{leonov.ai:nonequilibrium}, \cite{mattos.hsc:thermodynamically}, \cite{wapperom.p.hulsen.ma:thermodynamics}, \cite{dressler.m.edwards.bj.ea:macroscopic} or \cite{ellero.m.espanol.p.ea:thermodynamically}. (Notably the treatise by \cite{dressler.m.edwards.bj.ea:macroscopic} contains a rich bibliography, and describes the issue from the viewpoint of the GENERIC formalism, see~\cite{grmela.m.ottinger.hc:dynamics}, \cite{ottinger.hc.grmela.m:dynamics} and~\cite{pavelka.m.klika.v.ea:multiscale}.) In the present analysis we exploit the approach proposed by~\cite{rajagopal.kr.srinivasa.ar:thermodynamic} that is relatively simple and that provides one a purely phenomenologically based concept of \emph{visco}-\emph{elastic} response.

The fact that the approach by~\cite{rajagopal.kr.srinivasa.ar:thermodynamic} closely follows the phenomenological concept of \emph{visco}-\emph{elastic} material is best seen in the interpretation of the quantity~$\lcgnc$ that appears in the formula for the Cauchy stress tensor. This quantity can be interpreted as the left Cauchy--Green tensor associated with the \emph{elastic} part of the fluid response. Using the approach by~\cite{rajagopal.kr.srinivasa.ar:thermodynamic}, the derivation of Maxwell/Oldroyd-B type models has been discussed by \cite{malek.j.rajagopal.kr.ea:on}, see also~\cite{hron.j.milos.v.ea:on}. More complex viscoelastic rate-type models that document the applicability of the approach in more involved settings are discussed in~\cite{malek.j.prusa.v.ea:thermodynamics,malek.j.rajagopal.kr.ea:derivation} and in~\cite{dostalk.m.prusa.v.ea:on}. Following~\cite{dostalk.m.prusa.v.ea:on}, we know that the Giesekus fluid is a fluid with the specific Helmholtz free energy $\fenergy$ in the form
\begin{equation}
  \label{eq:free-energy-constant-specific-heat}
  \fenergy 
  =_{\bydefinition} 
  - 
  \cheatvol^{\mathrm{iNSE}} \temp 
  \left[ 
    \ln \left( \frac{\temp}{\temp_{\reference}} \right) - 1 
  \right]
  + 
  \frac{\mu}{2\rho}
  \left(
    \Tr \lcgnc
    -
    3
    -
    \ln \det \lcgnc
  \right)
  ,
\end{equation}
where $\temp$ denotes the absolute temperature, $\temp_{\reference}$ denotes a constant reference temperature,  $\cheatvol^{\mathrm{iNSE}}$ is a positive material parameter (specific heat capacity at constant volume) and $\mu$ is another positive material parameter. The specific Helmholtz free energy describes the \emph{energy storage ability} of the fluid, and the chosen \emph{ansatz} is the same as for the standard Maxwell/Oldroyd-B fluid. This implies that the Giesekus fluid and Maxwell/Oldroyd-B fluids differ, from the perspective of the current approach, only in their \emph{entropy production} mechanisms, see below.

Specifying the Helmholtz free energy as a function of $\temp$ and $\lcgnc$, one can use the standard thermodynamical identities $\entropy = -\pd{\fenergy}{\temp}$ and $\ienergy = \fenergy + \temp \entropy$, and obtain explicit formulae for the specific entropy $\entropy$, and the specific internal energy $\ienergy$
\begin{subequations}
  \label{eq:energy-and-entropy}
  \begin{equation}
    \entropy = 
    \cheatvol^{\mathrm{iNSE}} \ln \left( \frac{\temp}{\temp_{\reference}} \right) 
    ,
    \qquad
    \ienergy = 
    \cheatvol^{\mathrm{iNSE}} \temp
    +
    \frac{\mu}{2\rho}
    \left(
      \Tr \lcgnc
      -
      3
      -
      \ln \det \lcgnc
    \right)
    .
  \end{equation}
  Note that adding the kinetic energy to the mechanical part of the internal energy~$\ienergy$, that is to the term $\frac{\mu}{2\rho} \left(\Tr \lcgnc - 3 - \ln \det \lcgnc \right)$, we can define the specific \emph{mechanical energy} via
  \begin{equation}
    \label{eq:30}
    \menergy =_{\bydefinition}
    \frac{1}{2} \absnorm{\vec{v}}^2
    +
    \frac{\mu}{2 \rho}
    \left(
      \Tr \lcgnc
      -
      3
      -
      \ln \det \lcgnc
    \right)
    .
  \end{equation}
\end{subequations}

Once the Helmholtz free energy and consequently also the internal energy is specified, one can derive the evolution equation for the entropy that has the structure
%\begin{equation}
%  \label{eq:21}
$
\rho \dd{\entropy}{t} + \divergence \entfluxc = \entprodc,
$
% \end{equation}
where $\entfluxc$ denotes the entropy flux and $\entprodc$ stands for the entropy production. In the case of Giesekus fluid the entropy production is given by the formula
 $\entprodc = \frac{\entprodctemp}{\temp}$, where
\begin{equation}
  \label{eq:entropy-production}
  \entprodctemp
  =_{\bydefinition}
  2 \nu \tensordot{\gradsym}{\gradsym}
  +
  \frac{\mu^2}{2 \nu_1}
  \Tr 
  \left[
    \alpha \lcgnc^2 + (1 - 3 \alpha) \lcgnc + (1 - \alpha) \inverse{\lcgnc} + (3 \alpha - 2) \identity
  \right]
  +
  \kappa
  \frac{\absnorm{\nabla \temp}^2}{\temp}  
  .
\end{equation}
(We use the notation $\tensordot{\generictensor}{\generictensor} =_{\bydefinition} \Tr \left( \generictensor \transpose{\generictensor} \right)$ for the scalar product on the space of matrices, and $\absnorm{\generictensor}$ for the corresponding Frobenius norm.) Since $\lcgnc$ is a symmetric positive definite matrix it is easy to check that the entropy production~\eqref{eq:entropy-production} is a nonegative quantity hence the second law of thermodynamics is satisfied.

The fact that $\lcgnc$ is a symmetric positive definite matrix follows directly from the governing equations~\eqref{eq:viscoelastic-model-governing-equations-intro} via an argument similar to that of~\cite{boyaval.s.lelievre.t.ea:free-energy-dissipative}. It is also a~consequence of the fact that~$\lcgnc$ is in the approach by~\cite{rajagopal.kr.srinivasa.ar:thermodynamic} constructed as the left Cauchy--Green tensor, which means that~$\lcgnc$ can be decomposed as $\lcgnc = \fgradnc \transpose{\fgradnc}$. We exploit the positivity of~$\lcgnc$ quite frequently in our analysis.

Finally, we introduce three more important quantities that play crucial role in the construction of Lyapunov type functional via the method proposed by~\cite{bulcek.m.malek.j.ea:thermodynamics}. Namely we introduce the \emph{net total energy} $\nettenergy$, the \emph{net mechanical energy} $\netmenergy$ and the \emph{net entropy} $\netentropy$ of the fluid occupying the domain $\Omega$,
\begin{equation}
  \label{eq:net-total-energy-and-entropy}
  \nettenergy
  =_{\bydefinition}
  \int_{\Omega}
  \rho
  \left[
    \frac{1}{2} \absnorm{\vec{v}}^2
    +
    \ienergy
  \right]
  \,
  \cvolumee
  ,
  \qquad
  \netmenergy
  =_{\bydefinition}
  \int_{\Omega}
  \rho
  \menergy
  \,
  \cvolumee
  ,
  \qquad
  \netentropy
  =_{\bydefinition}
  \int_{\Omega}
  \rho
  \entropy
  \,
  \cvolumee
  .
\end{equation}

\subsection{Scaling}
\label{sec:scaling}
Equations~\eqref{eq:viscoelastic-model-governing-equations-intro} governing the evolution of mechanical variables can be transformed to a dimensionless form by introducing the characteristic length $x_{\charac}$, characteristic time $t_{\charac}$. (Note that the tensor field $\lcgnc$ already is a dimensionless quantity.) Using the following relations between the original quantities and their dimensionless versions denoted by stars
$
  x = x_{\charac} \dimless{x}
$,
$  
t = t_{\charac} \dimless{t}
$,
$
  \vec{v} = \frac{x_{\charac}}{t_{\charac}} \dimless{\vec{v}}
$,
$
  \mns = \frac{\nu}{t_{\charac}} \dimless{\mns}
$,
we obtain
\begin{subequations}
  \label{eq:governing-equations-dimless}
  \begin{align}
    \label{eq:continuity-equation-dimless}
    \dimless{\divergence} \dimless{\vec{v}} &= 0, \\
    \label{eq:momentum-equation-dimless}
    \dd{\dimless{\vec{v}}}{\dimless{t}}
                                            &=
                                              \dimless{\divergence} \dimless{\cstress}
                                              , 
    \\
    \label{eq:lcgcn-equation-dimless}
    \dimless{\fid{\overline{\lcgnc}}} 
                                            &=
                                              -
                                              \frac{1}{\Weissenberg} 
                                              \left[
                                              \alpha \lcgnc^2 + (1 - 2 \alpha) \lcgnc - (1 - \alpha) \identity
                                              \right]
                                              ,
   \end{align}
where the dimensionless Cauchy stress tensor $\dimless{\cstress}$ is given by
\begin{equation}
  \label{eq:cstress-formula-dimless}
  \dimless{\cstress} 
  = 
  \frac{1}{\Reynolds} \dimless{\mns} \identity + \traceless{\left( \dimless{\cstress} \right)}, 
  \qquad
  \traceless{\left( \dimless{\cstress} \right)} 
  = 
  \frac{2}{\Reynolds} \traceless{\left( \dimless{\gradsym} \right)} 
  + 
  \Xi \, \traceless{\left( \lcgnc \right)}.
\end{equation}
\end{subequations}
In~\eqref{eq:governing-equations-dimless} we have introduced three dimensionless numbers---the \emph{Reynolds number} $\Reynolds$, the \emph{Weissenberg number} $\Weissenberg$ and dimensionless shear modulus $\Xi$---via the formulae 
$
    \Reynolds 
    =_{\bydefinition} 
      \frac{\rho x_{\charac}^2}{\nu t_{\charac}}
$,
$
    \Weissenberg
    =_{\bydefinition} 
      \frac{\nu_1}{\mu t_{\charac}}
$,
and
$
    \Xi
    =_{\bydefinition} 
      \frac{\mu t_{\charac}^2}{\rho x_{\charac}^2}
$. 
It remains to introduce a scaling factor for the net mechanical energy $\netmenergy$, which will be used for the construction of the Lyapunov type functional in Section~\ref{sec:construction-lyapunov-functional}. Using the scaling
$
  \netmenergy = \frac{\rho x_{\charac}^5}{t_{\charac}^2} \dimless{\netmenergy}
$,
we obtain
\begin{equation}
  \label{eq:netmenergy-dimless}
  \dimless{\netmenergy} \left( \dimless{\vec{W}} \right)
  =
  \int_{\dimless{\Omega}}
  \left[
    \frac{1}{2} \absnorm{\dimless{\vec{v}}}^2
    +
    \frac{\Xi}{2}
    \left(
      \Tr \lcgnc
      -
      3
      -
      \ln \det \lcgnc
    \right)
  \right]
  \,
  \dimless{\cvolumee}
  .
\end{equation}
Hereafter, we omit the star denoting dimensionless quantities unless otherwise specified.

The scaling is chosen in such a way that if $\Weissenberg \to 0+$, then $\lcgnc$ approaches the identity tensor. Indeed, if  $\Weissenberg \to 0+$ then~\eqref{eq:lcgcn-equation-dimless} implies that
\begin{equation}
  \label{eq:45}
  \tensorq{O} = \alpha \lcgnc^2 + (1 - 2 \alpha) \lcgnc - (1 - \alpha) \identity,
\end{equation}
and the solution of~\eqref{eq:45} is $\lcgnc = \identity$. Moreover, if $\lcgnc = \identity$, then the second term in~\eqref{eq:netmenergy-dimless}, that is the elastic contribution to the mechanical energy, vanishes, and the mechanical energy of the fluid reduces to the kinetic energy only. Finally, if $\lcgnc = \identity$, then the additional term in the Cauchy stress tensor~\eqref{eq:cstress-formula-dimless} vanishes. This means that for  $\Weissenberg \to 0+$ the governing equations~\eqref{eq:governing-equations-dimless} reduce to the standard incompressible Navier--Stokes equations.

\subsection{Boundary conditions}
\label{sec:boundary-conditions}
The governing equations~\eqref{eq:governing-equations-dimless} must be supplemented with boundary conditions for the velocity $\vec{v}$. We are interested in \emph{internal flow} problems, where one prescribes Dirichlet boundary conditions on a part of the flow domain  $\Omega \subset \R^3$, and periodic boundary conditions on another part of the domain. Such a domain is usually called the periodic cell. (For example, in the case of flow in between two infinite concentric rotating cylinders the Dirichlet boundary condition is prescribed on the surfaces of the cylinders, while the periodic boundary condition is prescribed in the direction of the axis of the cylinders.) On the parts of the boundary corresponding to the periodicity directions, say $\Gamma_1$, we therefore prescribe periodic boundary condition for~$\vec{v}$, while on the remaining part of the boundary, say~$\Gamma_2$, we prescribe the no-penetration and the no-slip boundary condition,
\begin{subequations}
  \label{eq:bc-no-slip}
  \begin{align}
    \label{eq:202}
    \left. \vectordot{\vec{v}}{\vec{n}} \right|_{\Gamma_2} &= 0, \\ 
    \label{eq:195}
    \left. (\identity - \tensortensor{\vec{n}}{\vec{n}}) \vec{v} \right|_{\Gamma_2} &= \vec{V}, 
  \end{align}
\end{subequations}
where $\vec{n}$ is the unit outward normal to the boundary of $\Omega$ and $\vec{V}$ is a given velocity in the tangential direction to the boundary. This means that the fluid adheres to the boundary, and, moreover, if $\vec{V} \not = \vec{0}$, then, in general, \emph{the energy is exchanged between the fluid and its surroundings}. Indeed, the balance of the net total energy reads
$
  \dd{\nettenergy}{t} = \int_{\partial \Omega} \vectordot{\left( \cstress \vec{v} \right)}{\vec{n}} \, \csurfacees - \int_{\partial \Omega} \vectordot{\hfluxc}{\vec{n}}  \, \csurfacees,
$
  where $\partial \Omega$ denotes the boundary of the domain~$\Omega$ and $\hfluxc$ denotes the heat flux. Consequently, if $\vec{v} \not = \vec{0}$ on the boundary, then the term $\int_{\partial \Omega} \vectordot{\left( \cstress \vec{v} \right)}{\vec{n}} \, \csurfacees$ does not, in general, vanish or is compensated by the second term on the right-hand side, and the \emph{net total energy} might even change in time.

Concerning the boundary conditions for the perturbation $\widetilde{\vec{v}}$ with respect to the reference state~$\widehat{\vec{v}}$, see below, we see that if $\widehat{\vec{v}}$ satisfies~\eqref{eq:bc-no-slip}, then the perturbed state $\vec{v} = \widehat{\vec{v}} + \widetilde{\vec{v}}$ also satisfies~\eqref{eq:bc-no-slip} provided that
\begin{equation}
  \label{eq:bc-perturbation-gamma2}
  \left. \widetilde{\vec{v}} \right|_{\Gamma_2} = \vec{0}.
\end{equation}
%(This means that unlike $\widehat{\vec{v}}$ the perturbation $\widetilde{\vec{v}}$ satisfies the homogeneous zero Dirichlet boundary condition on $\Gamma_2$.)
The periodic boundary condition on $\Gamma_1$ is preserved for the perturbation $\widetilde{\vec{v}}$.

In the following we frequently use the identity
\begin{equation}
  \label{eq:surface-integral-vanishes}
  \int_{\partial \Omega}
  \vectordot{\vec{f}}{\vec{n}}
  \,
  \csurfacees
  =
  0
  ,
\end{equation}
where $\vec{f}: \partial \Omega \to \R^3$ is a smooth function such that $\vec{f}$ fulfills the periodic boundary condition on~$\Gamma_1$ and $\vec{f} = \vec{0}$ on~$\Gamma_2$. Note that the identity holds even if one part of the boundary, no matter whether~$\Gamma_1$ or~$\Gamma_2$, is not present.

\section{Base flow -- non-equilibrium steady state}
\label{sec:equilibrium-steady-state}

\subsection{Notation for the stability analysis}
\label{sec:notation}
We are interested in the evolution of the triplet 
$
\vec{W}=_{\bydefinition}[\vec{v}, \mns, \lcgnc],
$
which solves the evolution equations~\eqref{eq:viscoelastic-model-governing-equations-intro}. We shall further use the notation
$
\widehat{\vec{W}} = [\widehat{\vec{v}}, \widehat{\mns}, \widehat{\lcgnc}]
$
for the triplet corresponding to a \emph{non-equilibrium steady state solution}, and
$
  \widetilde{\vec{W}} = [\widetilde{\vec{v}}, \widetilde{\mns}, \widetilde{\lcgnc}]
$
for the \emph{perturbation with respect to the non-equilibrium steady state}. This means that the triplet describing the complete perturbed state is given as a sum of the reference state $\widehat{\vec{W}}$ and the perturbation $\widetilde{\vec{W}}$ with respect to the reference state
\begin{subequations}
  \label{eq:perturbed-state}
  \begin{align}
    \label{eq:perturbed-state-1}
    \vec{W} &= \widehat{\vec{W}} + \widetilde{\vec{W}}, \\
    \label{eq:perturbed-state-2}
    [\vec{v}, m, \lcgnc] 
            &= 
              [\widehat{\vec{v}}, \widehat{\mns}, \widehat{\lcgnc}] 
              +
              [\widetilde{\vec{v}}, \widetilde{\mns}, \widetilde{\lcgnc}]
              .
  \end{align}
\end{subequations}
(Note that sometimes we will work only with the pair $\vec{W}=_{\bydefinition}[\vec{v}, \lcgnc]$, since the pressure is insubstantial in our analysis.) The term \emph{non-equilibrium steady state} is chosen in accordance with the practice in thermodynamics, and it means that the \emph{entropy is produced} at the steady state $\widehat{\vec{W}}$. In particular, the adjective non-equilibrium does not refer to the stability of the steady state.

\subsection{Governing equations in a steady state}
\label{sec:governing-equations}
\emph{The steady state $\widehat{\vec{W}} = [\widehat{\vec{v}}, \widehat{\mns}, \widehat{\lcgnc}]$ whose stability we want to investigate is a solution to the equations \eqref{eq:governing-equations-dimless} where the partial time derivatives are identically equal to zero.} In particular, we assume that the state described by the triplet $[\widehat{\vec{v}}, \widehat{\mns}, \widehat{\lcgnc}]$ solves the system
\begin{subequations}
  \label{eq:steady-state-equations}
  \begin{align}
    \label{eq:steady-state-equation-0}
    \divergence \widehat{\vec{v}}
    &= 0
      ,
    \\
    \label{eq:steady-state-equation-1}
    \left(
    \vectordot{\widehat{\vec{v}}}{\nabla}
    \right)    	
    \widehat{\vec{v}}
    & = 
      \divergence \cstress(\widehat{\vec{W}})
      ,
    \\
    \label{eq:steady-state-equation-2}
    \left( \vectordot{\widehat{\vec{v}}}{\nabla} \right) {\widehat{\lcgnc}}
    -
    \widehat{\gradvl}  \widehat{\lcgnc}
    - 
    \widehat{\lcgnc} \transpose{\widehat{\gradvl}}
    &=
      -
      \frac{1}{\Weissenberg}
      \left[ 
      \alpha \widehat{\lcgnc}^2 + (1 - 2 \alpha) \widehat{\lcgnc} - (1 - \alpha) \identity
      \right]
      .
  \end{align}
\end{subequations}
subject to boundary conditions~\eqref{eq:bc-no-slip} on $\Gamma_2$, that is
\begin{equation}
  \label{eq:bc-no-slip-steady}
  \left. \vectordot{\widehat{\vec{v}}}{\vec{n}} \right|_{\Gamma_2} = 0,  
  \qquad
  \left. (\identity - \tensortensor{\vec{n}}{\vec{n}}) \widehat{\vec{v}} \right|_{\Gamma_2} = \vec{V},
\end{equation}
and the periodic boundary conditions on $\Gamma_1$. Here the symbol $\cstress(\widehat{\vec{W}})$ denotes the Cauchy stress tensor induced by the triplet $[\widehat{\vec{v}}, \widehat{\mns}, \widehat{\lcgnc}]$ that is
\begin{equation}
  \label{eq:35}
  \cstress(\widehat{\vec{W}})
  =
  \frac{1}{\Reynolds} \widehat{\mns} \identity 
  +
  \frac{2}{\Reynolds} \widehat{\gradsym}
  + 
  \Xi \, \traceless{\left( \widehat{\lcgnc} \right)},
\end{equation}
where $\widehat{\gradsym} = \frac{1}{2} (\widehat{\gradvl} + \transpose{\widehat{\gradvl}})$, and $\widehat{\gradvl} = \nabla \widehat{\vec{v}}$. 

Note that if $\vec{V} = \vec{0}$, that is if no mechanical energy is supplied to the fluid, then the system would admit an \emph{equilibrium solution}
\begin{equation}
  \label{eq:36}
  [\widehat{\vec{v}}, \widehat{\mns}, \widehat{\lcgnc}] = [\vec{0}, c, \identity],
\end{equation}
where $c$ is an arbitrary number. (This is the standard ambiguity in the identification of the pressure well known from the case of Navier--Stokes fluid.) Here we use the adjective \emph{equilibrium} in order to emphasise that such a steady state would lead to \emph{zero entropy production}. Indeed, if $\lcgnc = \identity$ and $\vec{v} = \vec{0}$, then the (mechanical part) of the entropy production~\eqref{eq:entropy-production} vanishes. % As it is well known, see for example~\cite{coleman.bd:on}, the thermodynamically based stability analysis of such an spatially homogeneous equilibrium state is quite straightforward.

On the other hand, if $\vec{V} \not = \vec{0}$, then one must in general expect that the steady fields $\widehat{\vec{v}}$ and $\widehat{\lcgnc}$ are \emph{spatially inhomogeneous}, and consequently the entropy production~\eqref{eq:entropy-production} is positive. This means that the system \emph{produces the entropy}, hence it is, from the thermodynamical point of view, \emph{out of equilibrium}. Consequently, as discussed above, we use the adjective \emph{non-equilibrium} and we refer to the base flow as of non-equilibrium steady state. %In this case, the situation concerning the thermodynamically based stability analysis is much more complex, and it needs to be addressed by methods that go beyond the method introduced \hbox{by~\cite{coleman.bd:on}}.

\subsection{Concept of stability}
\label{sec:concept-stability}
Concerning the stability of the non-equilibrium steady state, we are interested in its \emph{asymptotic stability}. If we have a non-equilibrium steady state~$\widehat{\vec{W}}$ that solves~\eqref{eq:steady-state-equations}, then we want to know whether the perturbation~\hbox{$\vec{W} = \widehat{\vec{W}} + \widetilde{\vec{W}}$} of the non-equilibrium steady state~$\widehat{\vec{W}}$ tends back to the non-equilibrium steady state~$\widehat{\vec{W}}$ as time goes to infinity. In our case, the evolution of the perturbed state~$\vec{W}$ is governed by equations~\eqref{eq:governing-equations-dimless} that must be solved subject to the given boundary conditions~\eqref{eq:bc-no-slip} and subject to initial conditions
\begin{equation}
  \label{eq:64}
  \left. \vec{v} \right|_{t=0} = \widehat{\vec{v}} + \widetilde{\vec{v}}_0, \qquad
  \left. \lcgnc \right|_{t=0} = \widehat{\lcgnc} + \left(\widetilde{\lcgnc}\right)_0.
\end{equation}
The non-equilibrium steady state~$\widehat{\vec{W}}$ is said to be \emph{asymptotically stable} if the triplet $\vec{W}$ tends to $\widehat{\vec{W}}$ as time goes to infinity,
\begin{equation}
  \label{eq:67}
  \vec{W} \xrightarrow{t \to +\infty} \widehat{\vec{W}},
\end{equation}
for all initial data $\widetilde{\vec{v}}_0$ and $\left(\widetilde{\lcgnc}\right)_0$ chosen from a sufficiently small neighborhood of zero. %This means that the non-equilibrium steady state is recovered as time goes to infinity provided that the initial perturbation starts close to the non-equilibrium steady state.

Ideally, one would like to obtain stronger results. Namely one would like to have an \emph{unconditional} result that states that the non-equilibrium steady state is recovered as time goes to infinity \emph{regardless of the choice of initial perturbation}. This behaviour is expected if one deals with non-equilibrium steady states that are driven by a small energy inflow that is by a small boundary velocity~$\vec{V}$, or in other words if one deals with non-equilibrium steady states that are not far away from the equilibrium steady state. 

The key task in the stability analysis is the \emph{choice of a metric/norm on the state space} to give a meaning to the statement~\eqref{eq:67}. Namely, we need to answer the question as how to characterise the distance between~$\widehat{\vec{W}}$ and~$\vec{W}$, since~\eqref{eq:67} means
\begin{equation}
  \label{eq:46}
  \distance{\widehat{\vec{W}}}{\vec{W}} \xrightarrow{t \to +\infty} 0,
\end{equation}
where $\distance{\cdot}{\cdot}$ is a given metric that is not necessarily induced by a norm. Since~$\lcgnc$ is at a given spatial point~$\vec{x} \in \Omega$ a positive definite matrix, it seems reasonable to design the metric in such a way that it reflects this fact. This means that we have to rely on a \emph{metric on the set of positive definite matrices}. There are several possible definitions of the metric on these sets, see Appendix~\ref{sec:bures-wass-dist}. If we use the Bures--Wasserstein distance,
\begin{equation}
  \label{eq:8}
  \distance[\tensorq{P}(d),\, \mathrm{BW}]{\tensorq{A}}{\tensorq{B}}
  =_{\bydefinition}
  \left\{
    \Tr \tensorq{A}
    +
    \Tr \tensorq{B}
    -
    2 \Tr
    \left[
      \left(
        \tensorq{A}^{\frac{1}{2}}
        \tensorq{B}
        \tensorq{A}^{\frac{1}{2}}
      \right)^{\frac{1}{2}}
    \right]
  \right\}^{\frac{1}{2}}
  ,
\end{equation}
see~\eqref{eq:11} in Appendix~\ref{sec:bures-wass-dist}, and if we generalise this concept to the spatially distributed tensor fields, then we can define the distance between $\widehat{\vec{W}}$ and $\vec{W}$ as
\begin{equation}
  \label{eq:27}
  \distance{\widehat{\vec{W}}}{\vec{W}}
    =_{\bydefinition}
    \left(
      \norm[\sleb{2}{\Omega}]{\widehat{\vec{v}} - \vec{v}}^2
      +
      \left[
        \distance[\tensorq{P}_{\Omega}(d), \, \mathrm{BW}]{\widehat{\lcgnc}}{\lcgnc}
      \right]^2
    \right)^{\frac{1}{2}}
    ,
  \end{equation}
  see~\eqref{eq:19} and Appendix~\ref{sec:bures-wass-dist} for a discussion of the notation and correctness of this definition. It turns out that this concept of distance nicely fits to the dynamical system we are interested in.

The term ``stability'' is used in many other contexts, hence we will briefly comment on these other notions of stability. In particular, we would like to emphasise what is in the present work \emph{not} meant by the stability. First, we are not interested in the \emph{stability in the sense of continuous dependence on initial data}, which is the concept of stability investigated in~\cite{dafermos.cm:second} and various subsequent works especially in the theory of hyperbolic systems, see~\cite{dafermos.cm:hyperbolic}. The stability in the sense of continuous dependence on initial data means, see for example~\cite{schaeffer.dg.cain.jw:ordinary}, that
\begin{quotation}
  [\dots] if the initial data for an initial value problem are altered slightly, then the perturbed solution diverges from the original solution no faster than at a controlled exponential rate. % page 131
\end{quotation}
Apparently the \emph{asymptotic stability} we are interested in is a more ambitious concept, since we want the perturbed solution to \emph{converge} back to the original solution (non-equilibrium steady state). Second, we are not interested in the \emph{stability of the steady state subject to infinitesimal perturbations}, that is in the linearised stability. We are interested in the evolution of \emph{finite amplitude perturbations}. 

Finally, we emphasise that in our analysis \emph{we work with perturbations that are solution to the governing equations in the classical sense}. (All derivatives are understood as the classical derivatives, not as generalised derivatives such as distributional derivatives and so forth.) In particular, we \emph{do not} consider the perturbations that solve the governing equations only in a weak sense, although it is an important issue worth further investigation. The reader interested in the discussion of the state-of-the-art rigorous mathematical theory of equations governing the motion of viscoelastic fluids is kindly referred to~\cite{masmoudi.n:equations} or~\cite{barrett.jw.suli.e:existence*5}.

\section{Lyapunov functional}
\label{sec:construction-lyapunov-functional}

\subsection{Concept of Lyapunov functional}
\label{sec:conc-lyap-funct}
Let us briefly recall the concept of Lyapunov functional, see~\cite{henry.d:geometric}. We consider a system of governing equations in the form
\begin{equation}
  \label{eq:69}
  \dd{\vec{X}}{t} = \vec{F}(\vec{X}),
\end{equation}
where $\widehat{\vec{X}}$ is a steady state, that is $\vec{F}(\widehat{\vec{X}}) = \vec{0}$, and where~$\norm[\mathrm{st}]{\cdot}$ denotes a norm on the underlying state space. We say that the functional
$
\mathcal{V}(\widetilde{\vec{X}}\|\widehat{\vec{X}})
$
is a strict Lyapunov functional of the steady state $\widehat{\vec{X}}$ provided that:
\begin{subequations}
  \label{eq:lyapunov-functional-conditions}
  \begin{enumerate}
  \item There exists a neighborhood of $\widehat{\vec{X}}$ such that the functional is bounded from below by a function $f$ of the distance between the steady state $\widehat{\vec{X}}$ and the perturbation $\vec{X}$, that is
    \begin{equation}
      \label{eq:59}
      \mathcal{V}
      \left(
        \left.
          \widetilde{\vec{X}}
        \right\|
        \widehat{\vec{X}}
      \right) \geq 
      f\left(\norm[\mathrm{st}]{\widehat{\vec{X}} - \vec{X}}\right)
      ,
    \end{equation}
    where $f$ is a continuous strictly increasing function such that $f(0)=0$ and $f(r)>0$ whenever $r>0$. 
  \item The time derivative of $\mathcal{V}(\widetilde{\vec{X}}\|\widehat{\vec{X}})$ is negative and bounded from above by a function $g$ of the distance between the steady state $\widehat{\vec{X}}$ and the perturbation $\vec{X}$, that is 
    \begin{equation}
      \label{eq:68}
      \dd{}{t}
      \mathcal{V}
      \left(
        \left.
          \widetilde{\vec{X}}
        \right\|
        \widehat{\vec{X}}
      \right)
      \leq
      -
      g
      \left(
        \norm[\mathrm{st}]{\widehat{\vec{X}} - \vec{X}}
      \right)
      ,
    \end{equation}
    where $g$ is a continuous strictly increasing function such that $g(0)=0$ and $g(r)>0$ whenever $r>0$.
\end{enumerate}
\end{subequations}
If the given system of governing equations admits a strict Lyapunov functional near the state $\widehat{\vec{X}}$, then we know that the steady state $\widehat{\vec{X}}$ is asymptotically stable. This means that the solution $\vec{X} = \widetilde{\vec{X}} + \widehat{\vec{X}}$ that starts in the neighborhood of $\widehat{\vec{X}}$ satisfies
\begin{equation}
  \label{eq:71}
  \norm[\mathrm{st}]{\widehat{\vec{X}} - {\vec{X}}} \xrightarrow{t \to +\infty} 0.
\end{equation}

While the concept of Lyapunov type functional is very simple, it is difficult to apply in a particular setting. The main difficulty is to find a Lyapunov type functional. (Note that in the infinite dimensional setting one can not easily exploit LaSalle's invarinace principle since it requires precompactness of the trajectories, which is a qualitative property that goes beyond our assumption regarding the existence of the classical solution. Consequently, having a Lyapunov type functional with~\eqref{eq:68} replaced by the mere negativity everywhere except at the equilibrium,
$
\dd{}{t}
\mathcal{V}(\widetilde{\vec{X}}\|\widehat{\vec{X}})
<
0
$
for $\widetilde{\vec{X}} \not = \vec{0}$, is not a viable option.) Fortunately, since we are interested in equations describing a physical system, we can try to search for the functional using physical concepts.

If we were dealing with the stability of \emph{a homogeneous equilibrium steady state in a thermodynamically isolated system}, then a Lyapunov type functional could be constructed using the net entropy~$\netentropy$ and the net total energy~$\nettenergy$ functional. It can be shown that the appropriate Lyapunov type functional is in this setting reads
\begin{equation}
  \label{eq:73}
  \mathcal{V}_{\mathrm{eq}}
  \left(
    \left.
      \widetilde{\vec{W}}
    \right\|
    \widehat{\vec{W}}
  \right)
  =
  -
  \left[
    \netentropy(\widehat{\vec{W}} + \widetilde{\vec{W}})
    -
    \frac{1}{\widehat{\temp}}
    \left\{
      \nettenergy(\widehat{\vec{W}} + \widetilde{\vec{W}})
      -
      \nettenergy(\widehat{\vec{W}})
    \right\}
  \right]
  ,
\end{equation}
where $\widehat{\temp}$ is the \emph{spatially homogeneous temperature in the equilibrium steady state}, see~\cite{bulcek.m.malek.j.ea:thermodynamics} for details. The observation that~\eqref{eq:73} can be used as a suitable Lyapunov type functional for stability analysis of~\emph{homogeneous equilibrium steady states} in a thermodynamically \emph{isolated systems} or \emph{mechanically isolated systems immersed in a thermal bath} is well known, see for example \cite{silhavy.m:mechanics} or \cite{grmela.m.ottinger.hc:dynamics,ottinger.hc.grmela.m:dynamics}, and in the continuum thermodynamics setting it dates back to~\cite{coleman.bd:on} and \cite{gurtin.me:thermodynamics*1}, \cite{gurtin.me:thermodynamics}. (The core idea can be found in earlier works, see especially~\cite{duhem.p:traite}.) Unfortunately, the same functional cannot be used in stability analysis of \emph{non-equilibrium spatially inhomogeneous steady states} in thermodynamically \emph{open} systems. This fact is clear from the formula~\eqref{eq:73} itself. If one works with a \emph{spatially inhomogeneous} steady states, then $\widehat{\temp}$ in~\eqref{eq:73} is a \emph{function}, and~\eqref{eq:73} does not define a functional at all. %Further, the usability of the functional~\eqref{eq:73} as a Lyapunov type functional is based on the fact that in thermodynamically isolated systems one has $\dd{\netentropy}{t} \geq 0$ and $\dd{\nettenergy}{t} = 0$. \emph{This is not true in thermodynamically open systems}, that is in the systems that are allowed to exchange matter and energy with its surroundings. Still, the physically motivated Lyapunov type functional~\eqref{eq:73} can be used as a starting point for a construction of Lyapunov type functional that works also for non-equilibrium spatially inhomogeneous steady states in thermodynamically open systems.

\subsection{Construction of Lyapunov type functional  for stability analysis of a spatially inhomogeneous steady state}
\label{sec:constr-lyap-funct}
Recently, \cite{bulcek.m.malek.j.ea:thermodynamics} proposed a method for construction of Lyapunov type functionals for stability analysis of non-equilibrium spatially \emph{inhomogeneous steady states} in thermodynamically \emph{open} systems. In the ongoing analysis we use the same ideas as in~\cite{bulcek.m.malek.j.ea:thermodynamics}, but we restrict ourselves to the \emph{mechanical quantities only}. This is a matter of convenience, since we are interested in mechanical quantities only, and the temperature evolution has no feedback on the mechanical part of the system of governing equations. Consequently, we do not need to work with the Lyapunov type functional for the full thermomechanical problem, and we can construct a simpler Lyapunov type functional solely for the mechanical quantities.

Using the net mechanical energy functional $\netmenergy$ introduced in~\eqref{eq:netmenergy-dimless}, one can see that the net mechanical energy in a thermodynamically closed system must decay in time. Consequently, the functional
\begin{equation}
  \label{eq:60}
  \mathcal{V}_{\mathrm{eq}}
  \left(
    \left.
      \widetilde{\vec{W}}
    \right\|
    \widehat{\vec{W}}
  \right)
  =_{\bydefinition}
  \netmenergy 
  \left(
    \widehat{\vec{W}}
    +
    \widetilde{\vec{W}}
  \right)
  -
  \netmenergy 
  \left(
    \widehat{\vec{W}}
  \right)
\end{equation}
can serve as a Lyapunov type functional for stability analysis of equilibrium spatially homogeneous state~\eqref{eq:36}.

Following the methodology outlined in \cite{bulcek.m.malek.j.ea:thermodynamics} we use the Lyapunov type functional for the \emph{equilibrium} steady state~\eqref{eq:60}, and we define the candidate for the Lyapunov type functional for the \emph{non-equilibrium} steady state as
\begin{equation}
  \label{eq:lyapunov-functional-definition}
  \mathcal{V}_{\mathrm{neq}}
  \left(
    \left.
      \widetilde{\vec{W}}
    \right\|
    \widehat{\vec{W}}
  \right)
  =_{\bydefinition}
  \netmenergy 
  \left(
    \widehat{\vec{W}}
    +
    \widetilde{\vec{W}}
  \right)
  -
  \netmenergy 
  \left(
    \widehat{\vec{W}}
  \right)
  -
  \left.
    \Diff[\vec{W}] \netmenergy
    \left(
      \vec{W}
    \right)
  \right|_{\vec{W} = \widehat{\vec{W}}}
  \left[
    \widetilde{\vec{W}}
  \right]
  ,
\end{equation}
where
$
\left.
  \Diff[\vec{W}] \netmenergy
  \left(
    \vec{W}
  \right)
\right|_{\vec{W} = \widehat{\vec{W}}}
\left[
  \widetilde{\vec{W}}
\right]$
denotes the G\^ateaux derivative
% \footnote{%
% \label{fn:1}%
% Let us recall that the G\^ateaux derivative $\Diff \mathcal{M} (\vec{x})[\vec{y}]$ of a functional $\mathcal{M}$ at point $\vec{x}$ in the direction $\vec{y}$ is defined as
% $
%     \Diff \mathcal{M} (\vec{x})[\vec{y}]
%     =_{\bydefinition}
%     \lim_{s \to 0}
%     \frac{
%       \mathcal{M} (\vec{x} + s \vec{y}) - \mathcal{M} (\vec{x})
%     }
%     {
%       s
%     }
% $ 
% which is tantamount to 
% $
% \Diff \mathcal{M} (\vec{x})[\vec{y}]
% =_{\bydefinition}
% \left.
%   \dd{}{s}
%   \mathcal{M} (\vec{x} + s \vec{y})
% \right|_{s=0}
% $. If it is necessary to emphasize the variable against which we differentiate, we also write $\Diff[\vec{x}] \mathcal{M} (\vec{x})[\vec{y}]$ instead of $\Diff \mathcal{M} (\vec{x})[\vec{y}]$.
% } %footnote
at point $\widehat{\vec{W}}$ in the direction $\widetilde{\vec{W}}$. (This is essentially the affine correction trick introduced in a different context by~\cite{ericksen.jl:thermo-kinetic}, see also~\cite{dafermos.cm:second}.) The (dimensionless) explicit formulae for the individual terms in~\eqref{eq:lyapunov-functional-definition} read
\begin{subequations}
  \label{eq:lyapunov-functional-terms}
  \begin{align}
    \label{eq:lyapunov-functional-term-1}
    \netmenergy
    \left(
    \widehat{\vec{W}} + \widetilde{\vec{W}}
    \right)
    &=
      \int_{\Omega}
      \left[
      \frac{1}{2} 
      \absnorm{\widehat{\vec{v}} + \widetilde{\vec{v}}}^2
      +
      \frac{\Xi}{2}
      \left(
      \Tr \left( \widehat{\lcgnc} + \widetilde{\lcgnc} \right)
      -
      3
      -
      \ln \det \left( \widehat{\lcgnc} + \widetilde{\lcgnc} \right)
      \right)
      \right]
      \,
      \cvolumee
      ,\displaybreak[0]
    \\
    \label{eq:lyapunov-functional-term-2}
    \netmenergy
    \left(
    \widehat{\vec{W}}
    \right)
    &=
      \int_{\Omega}
      \left[
      \frac{1}{2} \absnorm{\widehat{\vec{v}}}^2
      +
      \frac{\Xi}{2}
      \left(
      \Tr \widehat{\lcgnc}
      -
      3
      -
      \ln \det \widehat{\lcgnc}
      \right)
      \right]
      \,
      \cvolumee
      ,
    \\
    \label{eq:lyapunov-functional-term-3}
    \left.
    \Diff[\vec{W}] \netmenergy
    \left(
      \vec{W}
    \right)
    \right|_{\vec{W} = \widehat{\vec{W}}}
    \left[
    \widetilde{\vec{W}}
    \right]
    &=
      \int_{\Omega}
      \left\{
      \vectordot{\widehat{\vec{v}}}{\widetilde{\vec{v}}}
      +
      \frac{\Xi}{2}
      \left[
      \Tr 
      \widetilde{\lcgnc}
      -
      \Tr 
      \left( 
      \inverse{\widehat{\lcgnc}} 
      \widetilde{\lcgnc}
      \right)
      \right]
      \right\}
      \,
      \cvolumee
      .
  \end{align}
\end{subequations}
Using~\eqref{eq:lyapunov-functional-terms} in~\eqref{eq:lyapunov-functional-definition} we get, after some algebraic manipulations, the explicit formula for the proposed Lyapunov type functional
\begin{equation}
  \label{eq:lyapunov-functional-explicit-formula}
  \mathcal{V}_{\mathrm{neq}}
  \left(
    \left.
      \widetilde{\vec{W}}
    \right\|
    \widehat{\vec{W}}
  \right)
  =
  \int_{\Omega}
  \frac{1}{2}
  \absnorm{\widetilde{\vec{v}}}^2
  \,
  \cvolumee
  +
  \int_{\Omega}
  \frac{\Xi}{2}
  \left[
    -
    \ln \det \left( \identity + \inverse{\widehat{\lcgnc}}\widetilde{\lcgnc} \right)
    +
    \Tr 
    \left( 
      \inverse{\widehat{\lcgnc}} 
      \widetilde{\lcgnc}
    \right)
  \right]
  \,
  \cvolumee
  .
\end{equation}
It remains to show that the functional~\eqref{eq:lyapunov-functional-explicit-formula} has the properties~\eqref{eq:lyapunov-functional-conditions} introduced in Section~\ref{sec:conc-lyap-funct}. First we show that the condition~\eqref{eq:59} holds for a neighborhood of $\widehat{\vec{W}}$, which means that we have to specify a norm on the state space.

The suitable norm is the norm introduced in Appendix~\ref{sec:bures-wass-dist} in Definition~\ref{dfn:3}. This norm is a ``standard'' Lebesgue type norm. However, it turns out that it is convenient to use this norm for a characterisation of the evolution of  a ``shifted'' state. The idea is the following. If we are given a \emph{constant-in-time} tensor field $\widehat{\lcgnc}$, which corresponds to the steady solution of~\eqref{eq:governing-equations-dimless}, and a state $\vec{W} = [\vec{v}, \lcgnc ]$, then we can introduce the shifted state~$\vec{Z}=[\vec{v}, \lcgncs]$ that is defined as
  \begin{equation}
    \label{eq:shifted-state}
    \vec{W}
    =
    \left[
      \vec{v}
      ,
      \lcgnc
    \right]
    \mapsto
    \vec{Z}
    =
    \left[
      \vec{v}
      ,
      \lcgncs
    \right]
    ,
    \qquad
    \lcgncs
    =_{\bydefinition}
    \left(
      \widehat{\lcgnc}^{-\frac{1}{2}} 
      \lcgnc
      \widehat{\lcgnc}^{-\frac{1}{2}}
    \right)^{\frac{1}{2}}
    .
  \end{equation}
  This shifted state seems to be ideal for the investigation of the perturbations to the steady state $\widehat{\lcgnc}$. Indeed, the steady state for~\eqref{eq:governing-equations-dimless} is $\widehat{\vec{W}} = [\widehat{\vec{v}}, \widehat{\lcgnc}]$, which is in virtue of~\eqref{eq:shifted-state} shifted to
  \begin{equation}
    \label{eq:26}
    \widehat{\vec{Z}} = [\widehat{\vec{v}}, \identity].
  \end{equation}
Now instead of investigating the behaviour of the perturbation $\lcgnc$ with respect to~$\widehat{\lcgnc}$ we can investigate the behaviour of the shifted perturbation~$\lcgncs$ with respect to the identity~$\identity$. 

\begin{Lemma}[Relation between the proposed Lyapunov functional and a norm]
  \label{lm:3}
  Let $\widehat{\vec{W}}$ and $\vec{W} = \widehat{\vec{W}} + \widetilde{\vec{W}}$ denote two states of the system governed by equations~\eqref{eq:governing-equations-dimless}, and let $\widehat{\vec{Z}}$ and $\vec{Z}$ denote the corresponding shifted states. Furthermore, let $\norm[\mathrm{st}]{\cdot}$ denote the norm introduced in Definition~\ref{dfn:3}. Then there exists a positive constant $D(\Xi)$ such that
  \begin{equation}
    \label{eq:52}
    \mathcal{V}_{\mathrm{neq}}
    \left(
      \left.
        \widetilde{\vec{W}}
      \right\|
      \widehat{\vec{W}}
    \right)
    \geq
    D
    \norm[\mathrm{st}]{\widehat{\vec{Z}}-\vec{Z}}^2.
  \end{equation}
\end{Lemma}

\begin{proof}
  We note that the formula~\eqref{eq:lyapunov-functional-explicit-formula} for the Lyapunov type functional can be rewritten as
  \begin{multline}
    \label{eq:28}
    \mathcal{V}_{\mathrm{neq}}
    \left(
      \left.
        \widetilde{\vec{W}}
      \right\|
      \widehat{\vec{W}}
    \right)
    =
    \frac{1}{2}
    \norm[\sleb{2}{\Omega}]{\widehat{\vec{v}} - \left(\widehat{\vec{v}} + \widetilde{\vec{v}}\right)}^2
    \\
    +
    \int_{\Omega}
    \frac{\Xi}{2}
    \left[
      \Tr 
      \left( 
        \widehat{\lcgnc}^{-\frac{1}{2}} 
        \left(\widehat{\lcgnc} + \widetilde{\lcgnc} \right)
        \widehat{\lcgnc}^{-\frac{1}{2}} 
      \right)
      -
      3
      -
      \ln \det \left( \widehat{\lcgnc}^{-\frac{1}{2}} \left( \widehat{\lcgnc} + \widetilde{\lcgnc} \right) \widehat{\lcgnc}^{-\frac{1}{2}} \right)
    \right]
    \,
    \cvolumee
    \\
    =
    \frac{1}{2}
    \norm[\sleb{2}{\Omega}]{\widehat{\vec{v}} - \left(\widehat{\vec{v}} + \widetilde{\vec{v}}\right)}^2
    +
    \int_{\Omega}
    \frac{\Xi}{2}
    \left[
      \Tr 
      \lcgncs^2
      -
      3
      -
      \ln \det \lcgncs^2
    \right]
    \,
    \cvolumee
    ,
  \end{multline}
  where we have used the cyclic property of the trace and the properties of the determinant. Making use of Lemma~\ref{lm:2}, inequality in the integrand of the last term in~\eqref{eq:28}, we see that
  \begin{equation}
    \label{eq:41}
    \mathcal{V}_{\mathrm{neq}}
    \left(
      \left.
        \widetilde{\vec{W}}
      \right\|
      \widehat{\vec{W}}
    \right)
    \geq
    \frac{1}{2}
    \norm[\sleb{2}{\Omega}]{\widehat{\vec{v}} - \left(\widehat{\vec{v}} + \widetilde{\vec{v}}\right)}^2
    +
    \frac{\Xi}{2}
    \int_{\Omega}
    \absnorm{\identity - \lcgncs}^2
    \,
    \cvolumee
    \geq
    \min
    \left\{
      \frac{1}{2}
      ,
      \frac{\Xi}{2}
    \right\}
    \norm[\mathrm{st}]{\widehat{\vec{Z}}- \vec{Z}}
    ^2
    .
  \end{equation}
\end{proof}

The less straightforward part of the analysis of properties of proposed Lyapunov type functional $\mathcal{V}_{\mathrm{neq}}$ is the evaluation of its time derivative $\dd{\mathcal{V}_{\mathrm{neq}}}{t}$. The formula for the time derivative is derived via a lengthy algebraic manipulation described in Appendix~\ref{sec:form-time-deriv}, and it is given below in Lemma~\eqref{lm:5}. Note that \emph{although we are working with a thermodynamically open system}, the formula for the time derivative~\emph{does not contain boundary terms}.

\begin{Lemma}[Explicit formula for the time derivative of the Lyapunov type functional]
  \label{lm:5}
  Let $\widehat{\vec{W}}$ and $\vec{W} = \widehat{\vec{W}} + \widetilde{\vec{W}}$ denote two states of the system governed by equations~\eqref{eq:governing-equations-dimless}, where the state $\widehat{\vec{W}}$ is the steady state, that is it solves~\eqref{eq:steady-state-equations}. The formula for the time derivative of the functional $\mathcal{V}_{\mathrm{neq}}(\widetilde{\vec{W}}\|\widehat{\vec{W}})$ introduced in~\eqref{eq:lyapunov-functional-explicit-formula} reads
  \begin{multline}
  \label{eq:9}
  \dd{\mathcal{V}_{\mathrm{neq}}}{t}
  \left(
    \left.
      \widetilde{\vec{W}}
    \right\|
    \widehat{\vec{W}}
  \right)
  =
  -
  \int_{\Omega}
  \frac{2}{\Reynolds} \tensordot{\widetilde{\gradsym}}{\widetilde{\gradsym}}
  \,
  \cvolumee
  -
  \int_{\Omega}
  \Xi \, \tensordot{\widetilde{\lcgnc}}{\widetilde{\gradsym}}
  \,
  \cvolumee
  -
  \int_{\Omega}
  \vectordot{\widehat{\gradsym} \widetilde{\vec{v}}}{\widetilde{\vec{v}}}
  \,
  \cvolumee
  \\
  -
  \int_{\Omega}
  \frac{\Xi}{2} \Tr
  \left[
    \inverse{\widehat{\lcgnc}} \widetilde{\lcgnc} \inverse{\widehat{\lcgnc}}
    \left(
      \vectordot{\widetilde{\vec{v}}}{\nabla}
    \right)
    \widehat{\lcgnc}
  \right]
  \,
  \cvolumee
  +
  \int_{\Omega}
  \frac{\Xi}{2}
  \tensordot{\inverse{\widehat{\lcgnc}}}{
    \left(
      \widetilde{\gradvl} \widetilde{\lcgnc} 
      + \widetilde{\lcgnc} \transpose{\widetilde{\gradvl}}
    \right)}
  \,
  \cvolumee
  \\
  -
  \int_{\Omega}
  (1 - \alpha) \frac{\Xi}{2 \Weissenberg}
  \Tr \left[ 
    \inverse{\left(\widehat{\lcgnc} + \widetilde{\lcgnc}\right)}
    \left(
      \widetilde{\lcgnc}
      \inverse{\widehat{\lcgnc}}
    \right)
    \transpose{
      \left(
        \widetilde{\lcgnc}
        \inverse{\widehat{\lcgnc}}
      \right)
    }
  \right]
  \,
  \cvolumee
  \\
  -
  \int_{\Omega}
  \alpha \frac{\Xi}{2 \Weissenberg} \Tr
  \left[
    \inverse{\widehat{\lcgnc}} \widetilde{\lcgnc}^2
  \right]
  \,
  \cvolumee
  .
\end{multline}
\end{Lemma}

\begin{proof}
  See Appendix~\ref{sec:form-time-deriv}.
\end{proof}

We note that the terms
\begin{subequations}
  \label{eq:87}
  \begin{gather}
    \label{eq:83}
    -
    \int_{\Omega}
    \frac{2}{\Reynolds} \tensordot{\widetilde{\gradsym}}{\widetilde{\gradsym}}
    \,
    \cvolumee
    ,
    \qquad
    -
    \int_{\Omega}
    \alpha \frac{\Xi}{2 \Weissenberg} \Tr
    \left[
    \inverse{\widehat{\lcgnc}} \widetilde{\lcgnc}^2
    \right]
    \,
    \cvolumee
    ,
    \\
    \label{eq:89}
    -
    \int_{\Omega}
    (1 - \alpha) \frac{\Xi}{2 \Weissenberg}
    \Tr \left[ 
    \inverse{\left(\widehat{\lcgnc} + \widetilde{\lcgnc}\right)}
    \left(
    \widetilde{\lcgnc}
    \inverse{\widehat{\lcgnc}}
    \right)
    \transpose{
    \left(
    \widetilde{\lcgnc}
    \inverse{\widehat{\lcgnc}}
    \right)
    }
    \right]
    \,
    \cvolumee
    ,
  \end{gather}
\end{subequations}
are negative provided that $\widetilde{\vec{v}} \not = \vec{0}$ and $\widetilde{\lcgnc} \not = \tensorq{O}$. If we were able to show that these damping terms are strong enough to balance all the remaining terms on the right-hand side of~\eqref{eq:9}, we would get the desired result~\eqref{eq:68} concerning the negativity of the time derivative. This should be possible at least for sufficiently small Reynolds number $\Reynolds$ and Weissenberg number $\Weissenberg$. The hypothesis follows from the observation that as~$\Reynolds$ and~$\Weissenberg$ tend to zero, then the magnitude of the damping terms increases, and it should outgrow the other terms in~\eqref{eq:9} that do not depend on $\Reynolds$ and $\Weissenberg$. This observation is consistent with the expectation that low Reynolds number and low Weissenberg number flows are stable.

Now the objective is to show that the hypothesis is true, and that the proposed functional indeed satisfies the condition~\eqref{eq:68}. In the quantification of the ``sufficient smallness'' of the Weissenberg number $\Weissenberg$ and the Reynolds number $\Reynolds$, we aim at a \emph{simple but a very rough estimate} based on the elementary use of Friedrichs--Poincar\'e, Cauchy--Schwarz, Young and Korn (in)equalities, see~\cite{necas.j.malek.j.ea:weak} or \cite{evans.lc:partial} or any other standard reference work on function spaces.
A precise characterisation of the Reynolds number and the Weissenberg number that guarantee the negativity of the time derivative, and hence the stability, could be obtained by a variational technique known from the standard energy method, see~\cite{joseph.dd:stability*1} or \cite{straughan.b:energy}. This is however beyond the scope of the current contribution.

\begin{Lemma}[Estimate on the time derivative]
  \label{lm:6}
  Let $\widehat{\vec{W}}$ and $\vec{W} = \widehat{\vec{W}} + \widetilde{\vec{W}}$ denote two states of the system governed by equations~\eqref{eq:governing-equations-dimless}, where the state $\widehat{\vec{W}}$ is the steady state, that is it solves~\eqref{eq:steady-state-equations}. Then there exist constants $C_1(\widehat{\vec{W}}, \Reynolds, \Weissenberg, \Xi, \Omega)$ and $C_2(\widehat{\vec{W}}, \Reynolds, \Weissenberg, \Xi)$ such that the time derivative of the functional $\mathcal{V}_{\mathrm{neq}}(\widetilde{\vec{W}}\| \widehat{\vec{W}})$ introduced in~\eqref{eq:lyapunov-functional-explicit-formula} can be estimated from above as
  \begin{equation}
    \label{eq:lyapunov-derivative-estimate}
    \dd{\mathcal{V}_{\mathrm{neq}}}{t}
    \left(
      \left.
        \widetilde{\vec{W}}
      \right\|
      \widehat{\vec{W}}
    \right)
    \leq
    C_1 {\norm{\nabla \widetilde{\vec{v}}}}_{\sleb{2}{\Omega}}^2
    +
    C_2 {\norm{\widetilde{\lcgnc}}}_{\sleb{2}{\Omega}}^2
    ,
  \end{equation}
  where we have denoted
  \begin{subequations}
    \label{eq:constants}
    \begin{equation}
      \label{eq:c1}
      C_1
      =_{\bydefinition} 
      -
      \frac{1}{\Reynolds}
      +
      C_P
      \sup_{\vec{x} \in \Omega} \absnorm{\lambda_{\min}( \widehat{\gradsym})}
      +
      \frac{\Xi}{2} \sup_{\vec{x} \in \Omega} \absnorm{\inverse{\widehat{\lcgnc}} - \identity}
      +
      C_P
      \frac{\Xi}{4}
      \sup_{\vec{x} \in \Omega}
      \absnorm{\inverse{\widehat{\lcgnc}}}^2
      \sup_{\vec{x} \in \Omega}
      \absnorm{\nabla \widehat{\lcgnc}}
      ,
    \end{equation}
    \begin{equation}
      \label{eq:c2}
      C_2
      =_{\bydefinition}
      -
      \alpha \frac{\Xi}{2 \Weissenberg}  
      \inf_{\vec{x} \in \Omega} \lambda_{\min} ( \inverse{\widehat{\lcgnc}} )
      +
      \frac{\Xi}{2} \sup_{\vec{x} \in \Omega} \absnorm{\inverse{\widehat{\lcgnc}} - \identity}
      +
      \frac{\Xi}{4}
      \sup_{\vec{x} \in \Omega}
      \absnorm{\inverse{\widehat{\lcgnc}}}^2
      \sup_{\vec{x} \in \Omega}
      \absnorm{\nabla \widehat{\lcgnc}}
      ,
    \end{equation}
  \end{subequations}
  and where $\lambda_{\min}(\cdot)$ denotes the minimal eigenvalue of the corresponding matrix and $C_P$ denotes the domain dependent constant from Friedrichs--Poincar\'e inequality.
\end{Lemma}

\begin{proof}
  See Appendix~\ref{sec:estim-time-deriv-1}.
\end{proof}

\begin{Lemma}[Estimate on the time derivative in terms of the norm on the shifted state space]
  \label{lm:7}
  Let $\widehat{\vec{W}}$ and $\vec{W} = \widehat{\vec{W}} + \widetilde{\vec{W}}$ denote two states of the system governed by equations~\eqref{eq:governing-equations-dimless}, where the state $\widehat{\vec{W}}$ is the steady state, that is it solves~\eqref{eq:steady-state-equations}, and let $\widehat{\vec{Z}}$ and $\vec{Z}$ denote the corresponding shifted states, see~\eqref{eq:shifted-state}. Let us further assume that the constants $C_1$ and $C_2$ in Lemma~\ref{lm:6} are negative. Then there exists a positive constant $C(C_1, C_2, \widehat{\lcgnc})$ such that
  \begin{equation}
    \label{eq:56}
    \dd{\mathcal{V}_{\mathrm{neq}}}{t}
    \left(
      \left.
        \widetilde{\vec{W}}
      \right\|
      \widehat{\vec{W}}
    \right)
    \leq
    -
    C
    \norm[\mathrm{st}]{\widehat{\vec{Z}}-\vec{Z}}^2,
  \end{equation}
  where $\mathcal{V}_{\mathrm{neq}}(\widetilde{\vec{W}}\| \widehat{\vec{W}})$ denotes the functional introduced in~\eqref{eq:lyapunov-functional-explicit-formula} and $\norm[\mathrm{st}]{\cdot}$ is the norm introduced in Definition~\ref{dfn:3}.
\end{Lemma}

\begin{proof}
  In virtue of Lemma~\ref{lm:6} we already know the estimate~\eqref{eq:lyapunov-derivative-estimate}. Making use of Friedrichs--Poincar\'e inequality 
$
{\norm{\vec{v}}}_{\sleb{2}{\Omega}}^2
  \leq
  C_P
  {\norm{\nabla \vec{v}}}_{\sleb{2}{\Omega}}^2,
$
where $C_P$ is the domain dependent constant, we see that if $C_1<0$ and $C_2<0$, then~\eqref{eq:lyapunov-derivative-estimate} implies
\begin{equation}
  \label{eq:57}
  \dd{\mathcal{V}_{\mathrm{neq}}}{t}
  \left(
    \left.
      \widetilde{\vec{W}}
    \right\|
    \widehat{\vec{W}}
  \right)
  \leq
  -
  \frac{\absnorm{C_1}}{C_P} {\norm{\widetilde{\vec{v}}}}_{\sleb{2}{\Omega}}^2
  -
  \absnorm{C_2} {\norm{\widetilde{\lcgnc}}}_{\sleb{2}{\Omega}}^2.
\end{equation}
Next we use Lemma~\ref{lm:8} which implies that
\begin{equation}
 \label{eq:32}
 \absnorm{\widetilde{\lcgnc}}
  =
  \absnorm{\widehat{\lcgnc} - \lcgnc}
  \geq
  \absnorm{\widehat{\lcgnc}^{-\frac{1}{2}}}^{-2}
  \absnorm{\identity - \left(\widehat{\lcgnc}^{-\frac{1}{2}}\lcgnc \widehat{\lcgnc}^{-\frac{1}{2}}\right)^{\frac{1}{2}}}
  =
  \absnorm{\widehat{\lcgnc}^{-\frac{1}{2}}}^{-2}
  \absnorm{\identity - \lcgncs}
,
\end{equation}
where $\lcgncs$ denotes the shifted state introduced in~\eqref{eq:shifted-state}. Consequently, we see that
\begin{equation}
  \label{eq:33}
   -
   \absnorm{C_2} {\norm{\widetilde{\lcgnc}}}_{\sleb{2}{\Omega}}^2
   =
   -
   \absnorm{C_2} 
   \int_{\Omega}
   \absnorm{\widetilde{\lcgnc}}^2
   \,
   \cvolumee
   \leq
   -
   \absnorm{C_2}
   \left(
     \sup_{\vec{x} \in \Omega}
     \absnorm{\widehat{\lcgnc}^{-\frac{1}{2}}}
   \right)^{-4}
   \int_{\Omega}
   \absnorm{\identity - \lcgncs}^2
   \,
   \cvolumee
   ,
 \end{equation}
 which means that~\eqref{eq:57} can be further manipulated to the form
 \begin{equation}
   \label{eq:44}
   \dd{\mathcal{V}_{\mathrm{neq}}}{t}
   \left(
     \left.
       \widetilde{\vec{W}}
     \right\|
     \widehat{\vec{W}}
   \right)
   \leq
   -
   \min
   \left\{
     \frac{\absnorm{C_1}}{C_P}
     ,
     \absnorm{C_2}
     \left(
       \sup_{\vec{x} \in \Omega}
       \absnorm{\widehat{\lcgnc}^{-\frac{1}{2}}}
     \right)^{-4}
   \right\}
   \left(
    {\norm{\widetilde{\vec{v}}}}_{\sleb{2}{\Omega}}^2
    +
    {\norm{\identity - \lcgncs}}_{\sleb{2}{\Omega}}^2
 \right)
 .
\end{equation}
Inequality~\eqref{eq:44} gives in virtue of the definition of the norm $\norm[\mathrm{st}]{\cdot}$, see Definition~\ref{dfn:3}, the proposition~\eqref{eq:56}. (Recall that the transformation~\eqref{eq:shifted-state} implies that $\widehat{\lcgncs} = \identity$.)
\end{proof}

\section{Main result}
\label{sec:main-result}
Using the estimate from Lemma~\ref{lm:7} and the relation between the metric and the functional $\mathcal{V}_{\mathrm{neq}}$, see Lemma~\ref{lm:3}, it is straightforward to prove the following theorem. 

\begin{Theorem}[Sufficient conditions for unconditional asymptotic stability]
  \label{thm:1}
  Let the pair $\widehat{\vec{W}}=[\widehat{\vec{v}}, \widehat{\lcgnc}]$ solve the governing equations for the steady state~\eqref{eq:steady-state-equations} with boundary conditions~\eqref{eq:bc-no-slip-steady}. If the Reynolds number $\Reynolds$, the Weissenberg number $\Weissenberg$ and the dimensionless shear modulus $\Xi$ are such that the constants $C_1$ and $C_2$ introduced in~\eqref{eq:constants} satisfy
  \begin{equation}
    \label{eq:114}
    C_1 < 0,
    \qquad
    C_2 < 0,
  \end{equation}
  then the spatially inhomogeneous non-equilibrium steady state $\widehat{\vec{W}}$ is unconditionally asymptotically stable, that is 
  \begin{equation}
    \label{eq:10}
     \distance{\widehat{\vec{W}}}{\vec{W}}
     \xrightarrow{t \to +\infty} 0,
   \end{equation}
holds for \emph{any} initial perturbation $\vec{W}$, where the metric $\distance{\cdot}{\cdot}$ is the metric introduced in~\eqref{eq:19}.
\end{Theorem}

\begin{proof}
  We first investigate the stability in the shifted state space, see~\eqref{eq:shifted-state}, that is we investigate perturbation~$\vec{Z}$ with respect to~$\widehat{\vec{Z}} = [\widehat{\vec{v}}, \identity]$. We introduce the functional $\mathcal{V}_{\mathrm{neq}}(\widetilde{\vec{W}}\| \widehat{\vec{W}})$, see~\eqref{eq:lyapunov-functional-explicit-formula} and the equivalent expression~\eqref{eq:28}. The functional satisfies condition~\eqref{eq:59}, see~Lemma~\ref{lm:3}. Furthermore, if $C_1$ and $C_2$ are negative, then Lemma~\ref{lm:7} implies that the functional~$\mathcal{V}_{\mathrm{neq}}(\widetilde{\vec{W}}\| \widehat{\vec{W}})$ decreases along the trajectories in a desired manner, that is it satisfies~\eqref{eq:68}. Consequently, the functional $\mathcal{V}_{\mathrm{neq}}(\widetilde{\vec{W}}\| \widehat{\vec{W}})$ is a genuine Lyapunov type functional for any neighborhood of the steady state $\widehat{\vec{Z}}$, hence $\widehat{\vec{Z}}$ is unconditionally asymptotically stable, $\norm[\mathrm{st}]{\widehat{\vec{Z}} - \vec{Z}} \xrightarrow{t \to +\infty} 0$.
  
  The convergence in the norm on the shifted space seems to be an obscure characterization of the approach to the equilibrium. However, if we exploit the definition of the shifted state, see~\eqref{eq:shifted-state}, and equality~\eqref{eq:49} proved in Lemma~\ref{lm:2}, we see that
  \begin{equation}
    \label{eq:62}
    \norm[\mathrm{st}]{\widehat{\vec{Z}} - \vec{Z}}
    =
    \left(
      \norm[\sleb{2}{\Omega}]{\widehat{\vec{v}} - \vec{v}}^2
      +
      \int_{\Omega}
      \left(
        \distance[\tensorq{P}(d),\, \mathrm{BW}]{\identity}{
          \widehat{\lcgnc}^{-\frac{1}{2}} 
          \lcgnc
          \widehat{\lcgnc}^{-\frac{1}{2}}
        }
      \right)^2
      \,
      \cvolumee
    \right)^{\frac{1}{2}}
    \geq
    E
    \distance{\widehat{\vec{W}}}{\vec{W}}
    ,
  \end{equation}
  where the last inequality follows from the estimate~\eqref{eq:38} in Lemma~\ref{lm:4}. Here $E$ is a positive constant that depends on $\widehat{\vec{W}}$, and $\distance{\cdot}{\cdot}$ is the metric introduced in Definition~\ref{dfn:2}, formula~\eqref{eq:19}. \emph{This metric is a natural one if we restrict ourselves to the set of positive definite tensor fields}. Inequality~\eqref{eq:62} then implies~\eqref{eq:10}.
\end{proof}

We note that if we want to investigate the spatially homogeneous steady state $\widehat{\vec{W}} = [\widehat{\vec{v}}, \widehat{\lcgnc}] =_{\bydefinition} [\vec{0}, \identity]$, that is if we set the boundary condition $\vec{V}=\vec{0}$, then
\begin{equation}
  \label{eq:12}
  C_1 = - \frac{1}{\Reynolds}, \qquad C_2 = - \alpha \frac{\Xi}{2 \Weissenberg},
\end{equation}
and the condition~\eqref{eq:114} is \emph{automatically satisfied without any restriction of the values of Reynolds number and the Weissenberg number}. If the steady state is non-trivial, that is if $\widehat{\vec{W}} = [\widehat{\vec{v}}, \widehat{\lcgnc}] \not = [\vec{0}, \identity]$, then the condition~\eqref{eq:114} must be evaluated. This is done in Section~\ref{sec:taylor-couette-flow} for the Taylor--Couette flow. (Note that $\widehat{\lcgnc}$ and $\widehat{\vec{v}}$ are solutions to~\eqref{eq:steady-state-equations} hence they depend on the Weissenberg and Reynolds number.) Naturally, one can expect that the condition will hold for a sufficiently small Reynolds number and Weissenberg number.

Having the Lyapunov type functional given by the formula~\eqref{eq:lyapunov-functional-explicit-formula} it is interesting to see how the functional works in the case of close to the equilibrium setting, that is for $\widehat{\lcgnc} \approx \identity$, and for small perturbations, that is for small $\widetilde{\lcgnc}$. We see that if $\widetilde{\lcgnc}$ is small and if $\widehat{\lcgnc}$ is close to the identity, then
\begin{equation}
  \label{eq:115}
  -
  \ln
  \det
  \left(
    \identity + \inverse{\widehat{\lcgnc}}\widetilde{\lcgnc}
  \right)
  +
  \Tr 
  \left( 
    \inverse{\widehat{\lcgnc}} 
    \widetilde{\lcgnc}
  \right)
  % \\
  % =
  % -
  % \ln
  % \left\{
  %   1
  %   +
  %   \Tr \left( \inverse{\widehat{\lcgnc}} \widetilde{\lcgnc} \right)
  %   +
  %   \frac{1}{2}
  %   \left[
  %     \left( \Tr \left( \inverse{\widehat{\lcgnc}} \widetilde{\lcgnc} \right) \right)^2
  %     -
  %     \Tr\left( \left(\inverse{\widehat{\lcgnc}} \widetilde{\lcgnc}\right)^2 \right)
  %   \right]
  %   +
  %   \cdots 
  % \right\}
  % +
  % \Tr 
  % \left( 
  %   \inverse{\widehat{\lcgnc}} 
  %   \widetilde{\lcgnc}
  % \right)
  % \\
  \approx
  \frac{1}{2}
  \Tr\left( \left(\inverse{\widehat{\lcgnc}} \widetilde{\lcgnc}\right)^2 \right)
  \approx
  \frac{1}{2}
  \absnorm{
    \widetilde{\lcgnc}
  }^2
  ,
\end{equation}
and the proposed Lyapunov type functional $\mathcal{V}_{\mathrm{neq}}$ can be approximated as 
$
\mathcal{V}_{\mathrm{neq}}
  \approx
  \mathcal{V}_{\mathrm{naive}}
$
where
\begin{equation}
  \label{eq:121}
  \mathcal{V}_{\mathrm{naive}}
  \left(
    \left.
      \widetilde{\vec{W}}
    \right\|
    \widehat{\vec{W}}
  \right)
  =_{\bydefinition}
  \int_{\Omega}
  \frac{1}{2}
  \absnorm{\widetilde{\vec{v}}}^2
  \,
  \cvolumee
  +
  \int_{\Omega}
  \frac{\Xi}{4}
  \absnorm{
    \widetilde{\lcgnc}
  }^2
  \,
  \cvolumee
  .
\end{equation}

The functional $\mathcal{V}_{\mathrm{naive}}$ might be the first candidate for the Lyapunov type functional if the stability is investigated using the popular ``energy method'', see for example~\cite{straughan.b:energy}. The functional is clearly nonnegative, and it vanishes if and only if the perturbation vanishes. Moreover, the candidate~$\mathcal{V}_{\mathrm{naive}}$ for the Lyapunov type functional is much simpler than $\mathcal{V}_{\mathrm{neq}}$. Indeed, the proximity of the perturbation to the non-equilibrium state $[\widehat{\vec{v}}, \widehat{\lcgnc}]$ is measured using the standard Lebesgue space norms, and $\mathcal{V}_{\mathrm{naive}}$ does not depend on the value of $\widehat{\lcgnc}$. Therefore it seems that $\mathcal{V}_{\mathrm{naive}}$ is a good candidate for the Lyapunov type functional for the analysis of arbitrary spatially inhomogeneous non-equilibrium steady state $[\widehat{\vec{v}}, \widehat{\lcgnc}]$.

The inappropriateness of  $\mathcal{V}_{\mathrm{naive}}$ for the stability analysis is in fact apparent even in a very trivial setting. Let us consider the \emph{spatially homogeneous equilibrium steady state} $\widehat{\lcgnc} = \identity$, $\widehat{\vec{v}} = \vec{0}$ in a mechanically isolated container, that is we set
$
  \vec{V} = \vec{0} 
$
in the boundary condition~\eqref{eq:bc-no-slip}. If we use the (exact) evolution equations for the perturbation velocity~\eqref{eq:evolution-equation-velocity-pertubation}, and if we evaluate the time derivative of $\mathcal{V}_{\mathrm{naive}}$, then we get
\begin{multline}
  \label{eq:124}
  \dd{\mathcal{V}_{\mathrm{naive}}}{t}
  \left(
    \left.
      \widetilde{\vec{W}}
    \right\|
    \widehat{\vec{W}}
  \right)
  =
  \int_{\Omega}
  \vectordot{\widetilde{\vec{v}}}{\pd{\widetilde{\vec{v}}}{t}}
  \,
  \cvolumee
  +
  \int_{\Omega}
  \frac{\Xi}{2}
  \Tr
  \left(
    \widetilde{\lcgnc}
    \pd{\widetilde{\lcgnc}}{t}
  \right)
  \,
  \cvolumee
  \\
  =
  -
  \int_{\Omega}
  \frac{2}{\Reynolds} \tensordot{\widetilde{\gradsym}}{\widetilde{\gradsym}}
  \,
  \cvolumee
  -
  \int_{\Omega}
  \Xi \, \tensordot{\widetilde{\lcgnc}}{\widetilde{\gradsym}}
  \,
  \cvolumee
  -
  \int_{\Omega}
  \vectordot{\widehat{\gradsym} \widetilde{\vec{v}}}{\widetilde{\vec{v}}}
  \,
  \cvolumee
  +
  \int_{\Omega}
  \frac{\Xi}{2}
  \Tr
  \left(
    \widetilde{\lcgnc}
    \pd{\widetilde{\lcgnc}}{t}
  \right)
  \,
  \cvolumee
  ,
\end{multline}
see also~\eqref{eq:lyapunov-time-derivative-first-term}. The last term on the right-hand side of~\eqref{eq:124} can be evaluated using the  (exact) evolution equation for~$\widetilde{\lcgnc}$, see~\eqref{eq:123}. Substituting~\eqref{eq:123} into~\eqref{eq:124} and using the fact that $\widehat{\lcgnc} = \identity$ and $\widehat{\vec{v}} = \vec{0}$ yields
\begin{multline}
  \label{eq:128}
  \dd{\mathcal{V}_{\mathrm{naive}}}{t}
  \left(
    \left.
      \widetilde{\vec{W}}
    \right\|
    \widehat{\vec{W}}
  \right)
  =
  -
  \int_{\Omega}
  \frac{2}{\Reynolds} \tensordot{\widetilde{\gradsym}}{\widetilde{\gradsym}}
  \,
  \cvolumee
  -
  \int_{\Omega}
  \frac{\Xi}{2}
  \Tr
  \left[
    \left(
      \left\{
        \vectordot{\widetilde{\vec{v}}}{\nabla}
      \right\}
      \widetilde{\lcgnc}
    \right)
    \widetilde{\lcgnc}
  \right]
  \,
  \cvolumee
  +
  \int_{\Omega}
  \Xi
  \Tr
  \left(
    \widetilde{\gradsym}
    \widetilde{\lcgnc}^2
  \right)
  \,
  \cvolumee
  \\
  -
  \int_{\Omega}
  \frac{\Xi}{2\Weissenberg} 
  \Tr
  \left[
    \alpha \widetilde{\lcgnc}^3
    +
    2\alpha \widetilde{\lcgnc}^2
    + 
    (1 - 2 \alpha) \widetilde{\lcgnc}^2
  \right]
  \,
  \cvolumee
  .
\end{multline}
Using the standard manipulation
\begin{equation}
  \label{eq:143}
  \int_{\Omega}
  \frac{\Xi}{2}
  \Tr
  \left[
    \left(
      \left\{
        \vectordot{\widetilde{\vec{v}}}{\nabla}
      \right\}
      \widetilde{\lcgnc}
    \right)
    \widetilde{\lcgnc}
  \right]
  \,
  \cvolumee
  % =
  % \int_{\Omega}
  % \frac{\Xi}{4}
  % \left\{
  %   \vectordot{\widetilde{\vec{v}}}{\nabla}
  % \right\}
  % \absnorm{\widetilde{\lcgnc}}^2
  % \,
  % \cvolumee
  % \\
  =
  \int_{\partial \Omega}
  \frac{\Xi}{4}
  \absnorm{\widetilde{\lcgnc}}^2
  \left(
    \vectordot{\widetilde{\vec{v}}}{\vec{n}}
  \right)
  \,
  \csurfacees
  -
  \int_{\Omega}
  \frac{\Xi}{4}
  \left(
    \divergence \widetilde{\vec{v}}
  \right)
  \absnorm{\widetilde{\lcgnc}}^2
  \,
  \cvolumee
  ,
\end{equation}
we see that the second term on the right-hand side of~\eqref{eq:128} vanishes in virtue of the incompressibility constraint~\eqref{eq:evolution-equation-velocity-pertubation-incompressibility} and the boundary condition for $\widetilde{\vec{v}}$. Consequently~\eqref{eq:128} reduces to
\begin{multline}
  \label{eq:144}
    \dd{\mathcal{V}_{\mathrm{naive}}}{t}
  \left(
    \left.
      \widetilde{\vec{W}}
    \right\|
    \widehat{\vec{W}}
  \right)
  =
  -
  \int_{\Omega}
  \frac{2}{\Reynolds} \tensordot{\widetilde{\gradsym}}{\widetilde{\gradsym}}
  \,
  \cvolumee
  +
  \int_{\Omega}
  \Xi
  \Tr
  \left(
    \widetilde{\gradsym}
    \widetilde{\lcgnc}^2
  \right)
  \,
  \cvolumee
  \\
  -
  \int_{\Omega}
  \frac{\Xi}{2\Weissenberg} 
  \Tr
  \left[
    \alpha \widetilde{\lcgnc}^3
    +
    2\alpha \widetilde{\lcgnc}^2
    + 
    (1 - 2 \alpha) \widetilde{\lcgnc}^2
  \right]
  \,
  \cvolumee
  .
\end{multline}

Let us now consider an initial perturbation that is chosen is such a way that
$
\int_{\Omega}
\Xi
\Tr
(
  \widetilde{\gradsym}
  \widetilde{\lcgnc}^2
)
\,
\cvolumee
>
0
$,
which can certainly be done.
% $\widetilde{\lcgnc}$ is nonzero in some part $\Omega^\prime$ of the domain $\Omega$, and that $\widetilde{\lcgnc}$ and $\widetilde{\gradsym}$ take in  $\Omega^\prime$ the values
% \begin{equation}
%   \label{eq:149}
%   \widetilde{\gradsym}
%   =
%   \begin{bmatrix}
%     \gamma & 0 & 0 \\
%     0 & -\gamma  & 0 \\
%     0 & 0 & 0
%   \end{bmatrix}
%   ,
%   \qquad
%   \widetilde{\lcgnc}
%   =
%   \begin{bmatrix}
%     \beta & 0 & 0 \\
%     0  & 0 & 0 \\
%     0 & 0 & 0
%   \end{bmatrix}
%   ,
% \end{equation}
% where $\gamma$ is a positive constant and $\beta > -1$. (Note that $\widetilde{\gradsym}$ is indeed a symmetric traceless matrix, and $\widehat{\lcgnc} + \widetilde{\lcgnc}$ is a symmetric positive definite matrix as required.) In this case, we get
% \begin{equation}
%   \label{eq:150}
%   \int_{\Omega}
%   \Xi
%   \Tr
%   \left(
%     \widetilde{\gradsym}
%     \widetilde{\lcgnc}^2
%   \right)
%   \,
%   \cvolumee
%   =
%   \Xi
%   \gamma \beta^2
%   \absnorm{\Omega^\prime}
%   ,
% \end{equation}
% which is a positive number.
This positive value will dominate the right-hand side of~\eqref{eq:144} provided that the Reynolds number and the Weissenberg number are large enough. Consequently, $\mathcal{V}_{\mathrm{naive}}$ will (initially) increase, and it would be useless as the Lyapunov type functional unless we \emph{a priori} limit ourselves to small perturbations.%

On the other hand, if we use the functional $\mathcal{V}_{\mathrm{neq}}$ in the case $\widehat{\lcgnc} = \identity$ and $\widehat{\vec{v}} = \vec{0}$, then we \emph{immediately} see that the constants $C_1$ and $C_2$ in~\eqref{eq:lyapunov-derivative-estimate} are negative, and that the equilibrium steady state is asymptotically stable with respect to \emph{any} perturbation and \emph{any} value of the Reynolds and the Weissenberg number! (Note that \cite{guillope.c.saut.jc:existence} obtained only a conditional stability result in a Sobolev space norm for the equilibrium rest state $\widehat{\lcgnc} = \identity$ and $\widehat{\vec{v}} = \vec{0}$, see their Corollary 3.5 and assumptions of Theorem 3.3.) Based on the analysis presented above, we can therefore claim that we have indeed benefited from a \emph{well constructed Lyapunov type functional~$\mathcal{V}_{\mathrm{neq}}$ and the choice of metric}. Unlike the naive Lyapunov type functional~$\mathcal{V}_{\mathrm{naive}}$, the proposed Lyapunov type functional~$\mathcal{V}_{\mathrm{neq}}$ seems to properly reflect the nonlinearity of the governing equations and the related energy storage mechanisms and the entropy production mechanisms.

\section{Taylor--Couette flow}
\label{sec:taylor-couette-flow}
\begin{figure}[b]
  \begin{center}
    \includegraphics[width=0.3\textwidth]{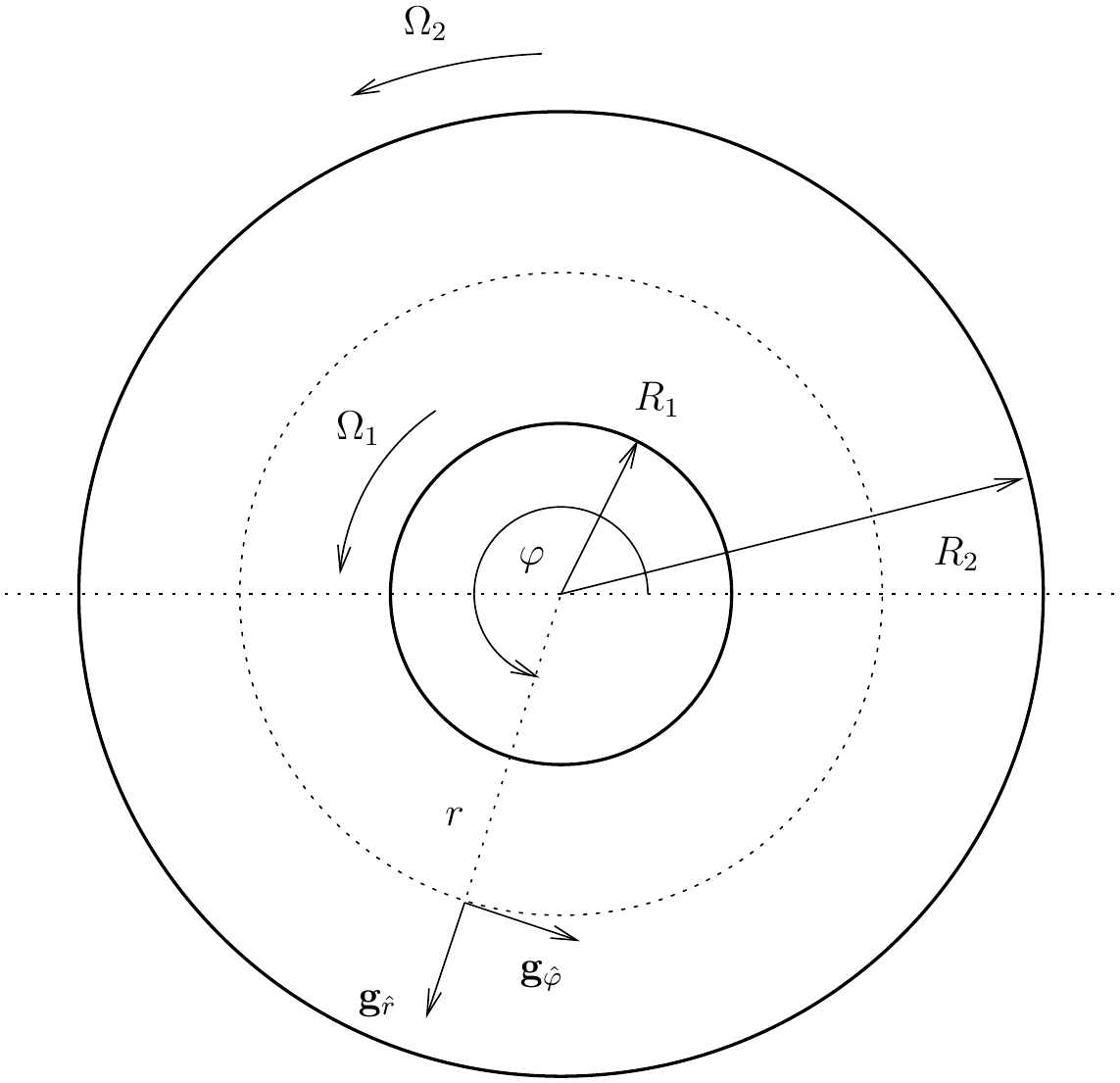}
  \end{center}
  \caption{Cylindrical Taylor--Couette flow.}
  \label{fig:couette-flow}
\end{figure}

Let us now consider a viscoelastic fluid described by the Giesekus model introduced in Section~\ref{sec:giesekus-model} with~$\alpha = \frac{1}{2}$, and let us investigate the stability of steady flow in the standard Taylor--Couette flow geometry, see Figure~\ref{fig:couette-flow}. The objective is to show as how the theory introduced above works in a specific setting. The choice $\alpha = \frac{1}{2}$ is motivated by the simplicity of the expressions for the corresponding steady state.

The fluid is placed in between two infinite concentric cylinders of radii $R_1$ and $R_2$, $R_1 < R_2$. The cylinders are rotating with the angular velocities $\Omega_1$ (inner cylinder) and $\Omega_2$ (outer cylinder) along the common axis. The geometry naturally leads to the use of cylindrical coordinates $(r, \varphi, z)$; the normed basis vectors are denoted as $\cobvec{\hatr}$, $\cobvec{\hatp}$ and $\cobvec{\hatz}$, see Figure \ref{fig:couette-flow}. Since the domain is unbounded in the $z$-direction we henceforth consider a periodic cell 
\begin{equation}
  \label{eq:periodic-cell}
  \Omega =_{\bydefinition} \{ (r, \varphi, z) \in \R^3 \, | \, R_1 < r < R_2, \, 0 \leq \varphi < 2 \pi, \, \absnorm{z} < h \}
\end{equation}
where $h > 0$ is arbitrary, and we use the notation 
$
\Gamma_1 =_{\bydefinition} \{ (r, \varphi, z) \in \R^3 \, | \, R_1 < r < R_2, \, 0 \leq \varphi < 2 \pi, \, \absnorm{z} = h \}
$
for the top and bottom base, and
$
    \Gamma_2 =_{\bydefinition} \{ (r, \varphi, z) \in \R^3 \, | \, r \in \{ R_1, R_2 \}, \, 0 \leq \varphi < 2 \pi, \, \absnorm{z} < h \}
$
for the cylindrical walls of the domain. The flow is driven by the rotation of the cylinders along the common axis.

\subsection{Base flow -- non-equilibrium steady state}
\label{sec:steady-solution-taylor-couette}
The first task in the stability analysis is to find the steady solution to the governing equations. This solution is the spatially inhomogeneous non-equilibrium steady state $\widehat{\vec{W}}$ as introduced in Section~\ref{sec:governing-equations}. The characteristic length and characteristic time have been chosen as
$
x_{\charac} =_{\bydefinition} R_1
$,
$
t_{\charac} =_{\bydefinition} \frac{1}{\Omega_1}
$.
We use the periodic boundary condition on $\Gamma_1$ and the no-slip boundary condition for velocity field $\vec{v}$ on~$\Gamma_2$, that is $\left. \vec{v} \right|_{r=R_1} = R_1 \Omega_1 \cobvec{\hatp}$, $\left. \vec{v} \right|_{r=R_2} = R_2 \Omega_2 \cobvec{\hatp}$. These boundary conditions are consistent with the requirements on boundary conditions specified in Section~\ref{sec:boundary-conditions}. In their dimensionless form the boundary conditions read
\begin{equation}
  \label{eq:boundary-conditions-dimless}
  \left. \dimless{\vec{v}} \right|_{\dimless{r}=1} = \cobvec{\hatp}, \qquad
  \left. \dimless{\vec{v}} \right|_{\dimless{r}=\frac{1}{\eta}} = \frac{\zeta}{\eta} \cobvec{\hatp},
\end{equation}
where we have introduced two dimensionless parameters $\eta =_{\bydefinition} \frac{R_1}{R_2}$ and $\zeta =_{\bydefinition} \frac{\Omega_2}{\Omega_1}$.
% \begin{equation}
%  \label{eq:dimless-params-eta-zeta}
%   \eta =_{\bydefinition} \frac{R_1}{R_2}, \qquad \zeta =_{\bydefinition} \frac{\Omega_2}{\Omega_1}.
% \end{equation}
Hereafter, we work exclusively with the dimensionless variables and thus, for the sake of simplicity, we omit the star denoting them.

Since the problem has the rotational symmetry, we search for the steady non-equilibrium state in a special form. Namely, the solution to~\eqref{eq:steady-state-equations} subject to boundary conditions~\eqref{eq:boundary-conditions-dimless} is sought in the form
  \begin{equation}
    \label{eq:solution-ansatz}
    \widehat{\vec{v}} = \vhatp(r) \cobvec{\hatp}, \qquad
    \widehat{\mns} = \widehat{\mns}(r), \qquad
    \widehat{\lcgnc}
    =
    \begin{bmatrix}
      \Brrhat(r) & \Brphat(r) & 0 \\
      \Bprhat(r) & \Bpphat(r) & 0 \\
      0 & 0 & \Bzzhat(r)
    \end{bmatrix}
  \end{equation}
Note that the chosen \emph{ansatz} for the velocity field automatically satisfies the incompressibility condition. The assumptions lead to the following expressions for the velocity gradient, the symmetric part of the velocity gradient, the convective term, the divergence of~$\widehat{\lcgnc}$, and the upper convected derivative of~$\widehat{\lcgnc}$,
\begin{multline}
  \label{eq:formulae-cylindrical}
    \nabla \widehat{\vec{v}}
    =
      \begin{bmatrix}
        0 & -\omega & 0 \\
        r \dd{\omega}{r}  + \omega & 0 & 0 \\
        0 & 0 & 0
      \end{bmatrix}
                ,
    \qquad
    \widehat{\gradsym}
    =
      \begin{bmatrix}
        0 & \frac{r}{2} \dd{\omega}{r} & 0 \\
        \frac{r}{2} \dd{\omega}{r} & 0 & 0 \\
        0 & 0 & 0
      \end{bmatrix}
                ,
    \qquad
    \dd{\widehat{\vec{v}}}{t}
    =
      \begin{bmatrix}
        -r \omega^2 \\
        0 \\
        0
      \end{bmatrix}
      ,
      \\
    \divergence \widehat{\lcgnc}
    =
      \begin{bmatrix}
        \frac{1}{r}
        \dd{}{r}
        \left(
          r \Brrhat
        \right)
        -
        \frac{\Bpphat}{r}
        \\
        \dd{\Bprhat}{r} + \frac{\Bprhat + \Brphat}{r} 
        \\
        0
      \end{bmatrix}
    ,
    \qquad
    \fid{\overline{\widehat{\lcgnc}}}
    =
      \begin{bmatrix}
        0 & -r \dd{\omega}{r} \Brrhat & 0 \\
        -r \dd{\omega}{r} \Brrhat & -2 r \dd{\omega}{r} \Bprhat & 0 \\
        0 & 0 & 0
      \end{bmatrix}
      ,
\end{multline}
where we have introduced the angular velocity $\omega(r)$, $\vhatp(r) =_{\bydefinition} \omega(r) r$. Using~\eqref{eq:formulae-cylindrical}, we see that the governing equations for the velocity field~\eqref{eq:steady-state-equation-1} reduce to
\begin{subequations}
  \label{eq:84}
  \begin{equation}
    \label{eq:85}
    \begin{bmatrix}
      - r \omega^2 \\
      0 \\
      0
    \end{bmatrix}
    =
    \begin{bmatrix}
      \dd{}{r}\left(\widehat{\mns}+\Xi \left(\Brrhat-\frac13(\Brrhat+\Bpphat+\Bzzhat) \right)\right)
      +
      \Xi
      \frac{\Brrhat - \Bpphat}{r}
      \\
      \frac{1}{r^2}
      \dd{}{r}\left( \frac{1}{\Reynolds} r^3 \dd{\omega}{r} + \Xi r^2 \Brphat \right)
      \\
      0
    \end{bmatrix}
    ,
  \end{equation}
  while the governing equations~\eqref{eq:steady-state-equation-2} for~$\widehat{\lcgnc}$ read
  \begin{multline}
    \scriptsize
    \label{eq:86}
    \begin{bmatrix}
      0 & -r \dd{\omega}{r} \Brrhat & 0 \\
      -r \dd{\omega}{r} \Brrhat & -2 r \dd{\omega}{r} \Bprhat & 0 \\
      0 & 0 & 0
    \end{bmatrix}
    =
    \\
    \scriptsize
    -\frac{1}{\Weissenberg}
    \setlength\arraycolsep{-6.5pt}
    \begin{bmatrix}
      \alpha \left((\Brrhat)^2+(\Brphat)^2\right) + (1-2\alpha)\Brrhat - (1-\alpha) & \alpha\Brphat(\Brrhat+\Bpphat) + (1-2\alpha)\Brphat & 0 \\
      \alpha\Brphat(\Brrhat+\Bpphat) + (1-2\alpha)\Brphat & \alpha\left((\Brphat)^2+(\Bpphat)^2\right) + (1-2\alpha)\Bpphat - (1-\alpha) & 0 \\
      0 & 0 & \alpha(\Bzzhat)^2 + (1-2\alpha)\Bzzhat - (1-\alpha)
    \end{bmatrix}
    .
  \end{multline}
\end{subequations}
Assuming that $\dd{\omega}{r} \neq 0$ in $(R_1, R_2)$, equation \eqref{eq:86} can be solved for $\Brrhat$, $\Brphat$, $\Bpphat$ and $\Bzzhat$. However, for general $\alpha \in (0, 1)$ the formulae for the aforementioned quantities are too complex to be written down here. Let us note however that for $\alpha = \frac{1}{2}$ the formulae simplify significantly; the solution to \eqref{eq:86} which satisfies the condition of~$\widehat{\lcgnc}$ being positive definite in this case reads
\begin{equation}
  \label{eq:96}
    \Bzzhat = 1, \qquad
    \Brphat = \frac{-1+\sqrt{1+c^2}}{c}, \qquad
    \Brrhat = \sqrt{\frac{2}{c} \Brphat}, \qquad
    \Bpphat = \sqrt{2 \left( c + \frac{1}{c} \right) \Brphat},
\end{equation}
where we have denoted $c =_{\bydefinition} 2 \Weissenberg \, r \dd{\omega}{r}$. Substituting \eqref{eq:96} into the second equation in \eqref{eq:85} then yields an ordinary differential equation for the angular velocity $\omega$
\begin{equation}
  \label{eq:omega-bvp}
  0 = \dd{}{r}\left( \frac{1}{\Reynolds} r^3 \dd{\omega}{r} + \Xi r^2 \frac{-1+\sqrt{1+4 \Weissenberg^2 \, r^2 \left( \dd{\omega}{r} \right)^2}}{2 \Weissenberg \, r \dd{\omega}{r}} \right),
\end{equation}
supplemented by the boundary conditions
$\left. \omega \right|_{r=1} = 1$, $\left. \omega \right|_{r=\frac{1}{\eta}} = \zeta$,
that follow from \eqref{eq:boundary-conditions-dimless} and from the fact that $\vhatp(r) = \omega(r) r$. Equation \eqref{eq:omega-bvp} together with the boundary conditions constitute a boundary value problem which needs to be solved numerically.

\subsection{Explicit criterion for the stability of spatially inhomogeneous non-equilibrium steady state}
\label{sec:taylor-couette-stability}
Here, we explicitly compute constants $C_1$, $C_2$ defined by \eqref{eq:c1} and \eqref{eq:c2} for the Taylor--Couette problem and for the specific values of the dimensionless numbers $\Xi$, $\Reynolds$ and $\Weissenberg$. Let us recall that for the sake of simplicity we have set $\alpha = \frac{1}{2}$, and we consider the steady tensor field~$\widehat{\lcgnc}$ given by \eqref{eq:96}. We fix the values for the geometric parameter $\eta$ and angular velocities ratio $\zeta$ as $\eta = \frac{1}{2}$ and $\zeta = 2$.
% \begin{equation}
%   \label{eq:geometry-parameters}
%   \eta = \frac{1}{2}, \qquad \zeta = 2.
% \end{equation}

\begin{figure}[h]
  \centering
  \subfloat[Tensor $\widehat{\lcgnc}$, component~$\Brrhat$.]{\includegraphics[scale=0.75]{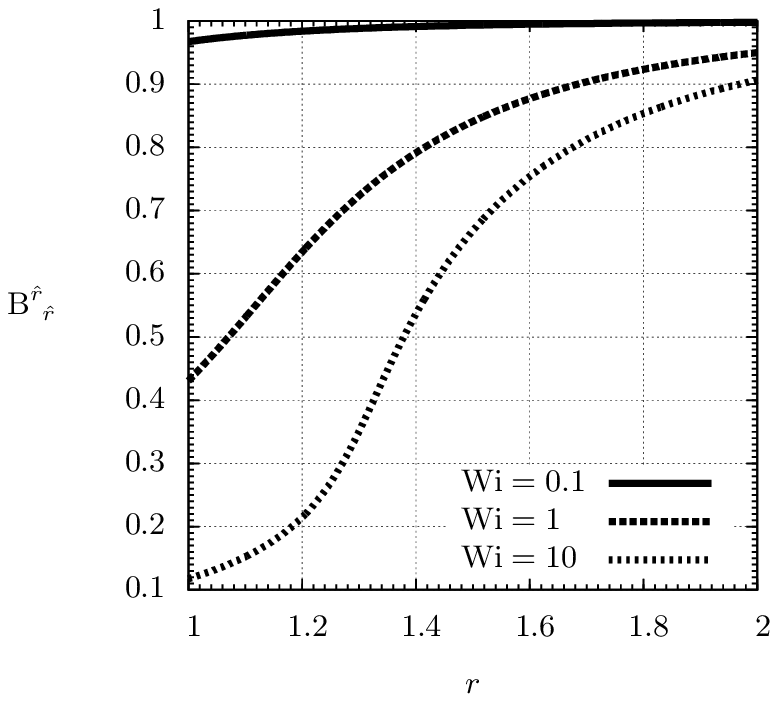}}
  \qquad
  \subfloat[Tensor $\widehat{\lcgnc}$, component~$\Bpphat$.]{\includegraphics[scale=0.75]{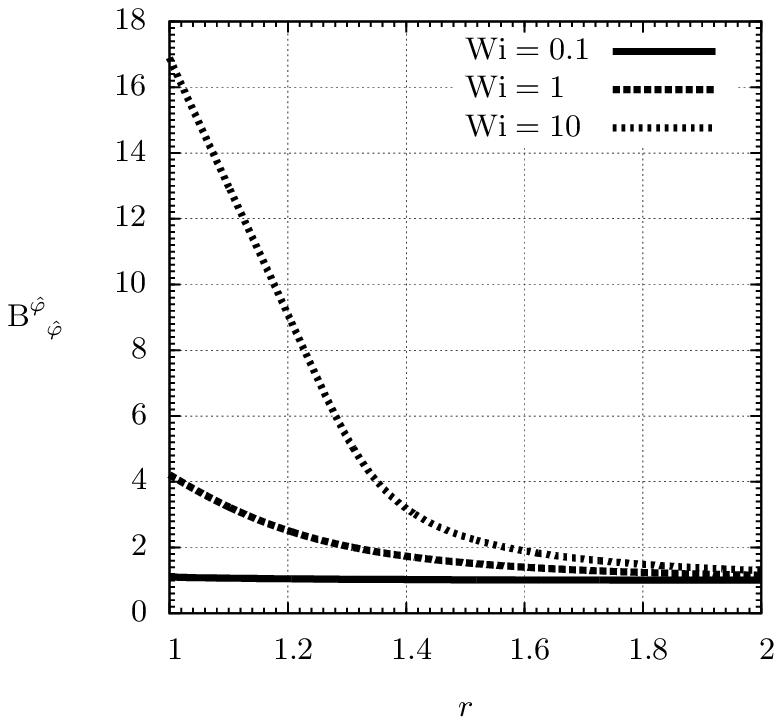}}
  \\
  \subfloat[Tensor $\widehat{\lcgnc}$, component~$\Brphat$.]{\includegraphics[scale=0.75]{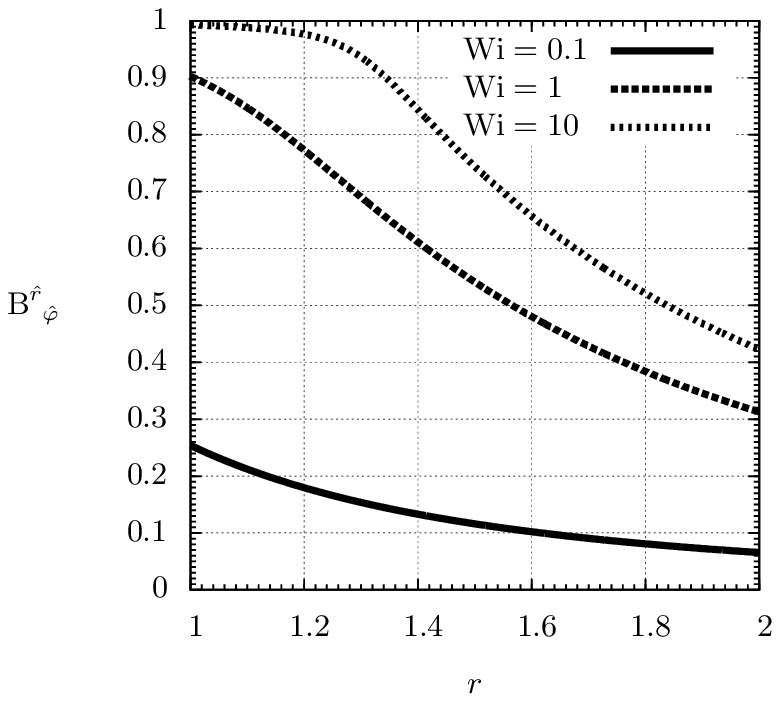}}
  \qquad
  \subfloat[Velocity $\widehat{\vec{v}}$, component~$\vhatp$.]{\includegraphics[scale=0.75]{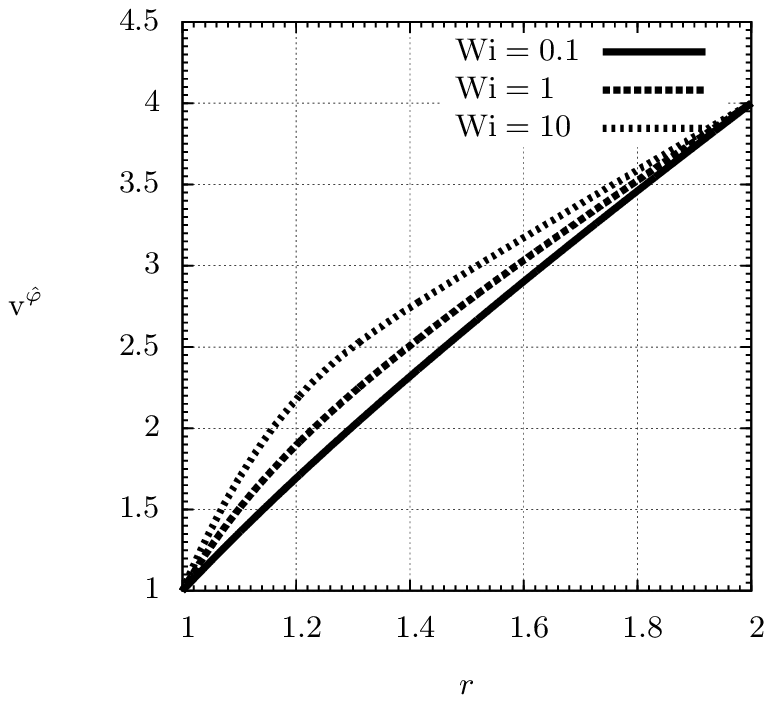}}
  \caption{Taylor--Couette flow, spatially inhomogeneous non-equilibrium steady state for various values of Weissenberg number $\Weissenberg$. Giesekus parameter $\alpha = \frac{1}{2}$, Reynolds number $\Reynolds = 100$, dimensionless shear modulus $\Xi = 0.1$, problem parameters $\eta = \frac{1}{2}$, $\zeta = 2$.}
  \label{fig:steady-solutions-plots}
\end{figure}

The angular velocity $\omega$ is obtained by solving \eqref{eq:omega-bvp} which is a boundary-value problem for a second order nonlinear differential equation. The problem is solved numerically using the function \texttt{solve\_bvp} from SciPy library version 1.0.0, which implements a fourth order collocation algorithm with the control of residuals as described in~\cite{kierzenka.j.shampine.lf:bvp}. With the angular velocity $\omega$ in hand, we immediately get the steady velocity field $\widehat{\vec{v}} = \omega(r) r \cobvec{\hatp}$, and the steady left Cauchy--Green tensor field $\widehat{\lcgnc}$ through formulae \eqref{eq:96}. The plots of the velocity field and the components of $\widehat{\lcgnc}$ are shown in Figure~\ref{fig:steady-solutions-plots}.

\begin{figure}[h]
  \centering
  \subfloat[\label{fig:stability-regions-a}Dimensionless shear modulus~$\Xi = 0.1$.]{\includegraphics[scale=0.82]{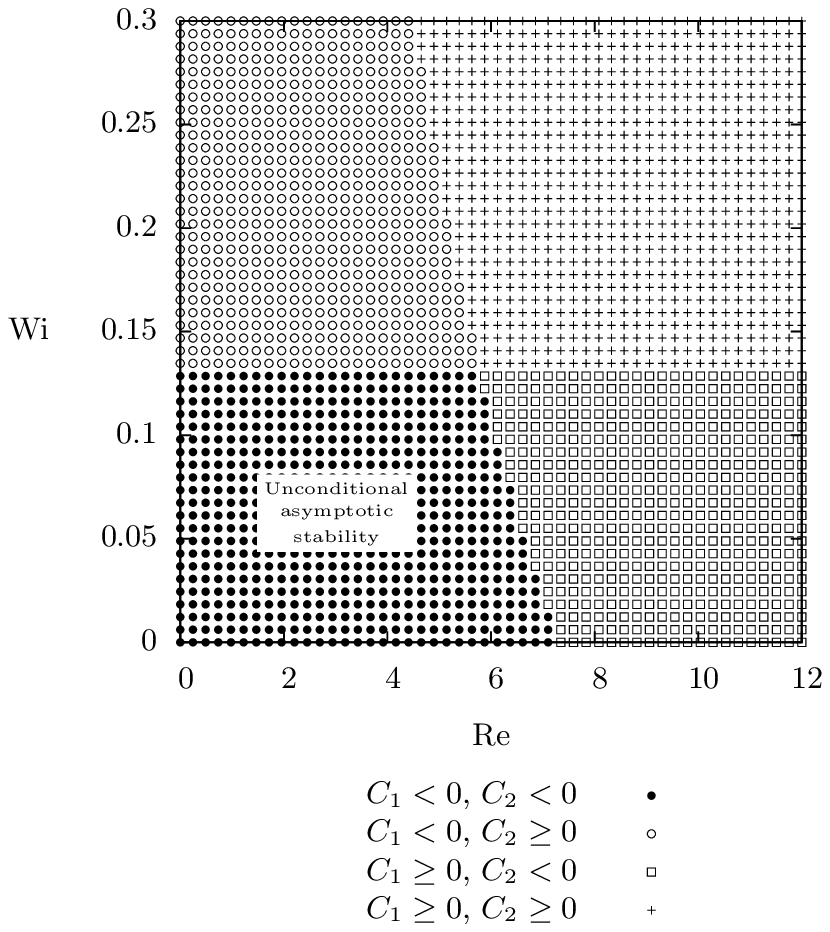}}
  \qquad
  \subfloat[\label{fig:stability-regions-b}Dimensionless shear modulus~$\Xi = 1$.]{\includegraphics[scale=0.82]{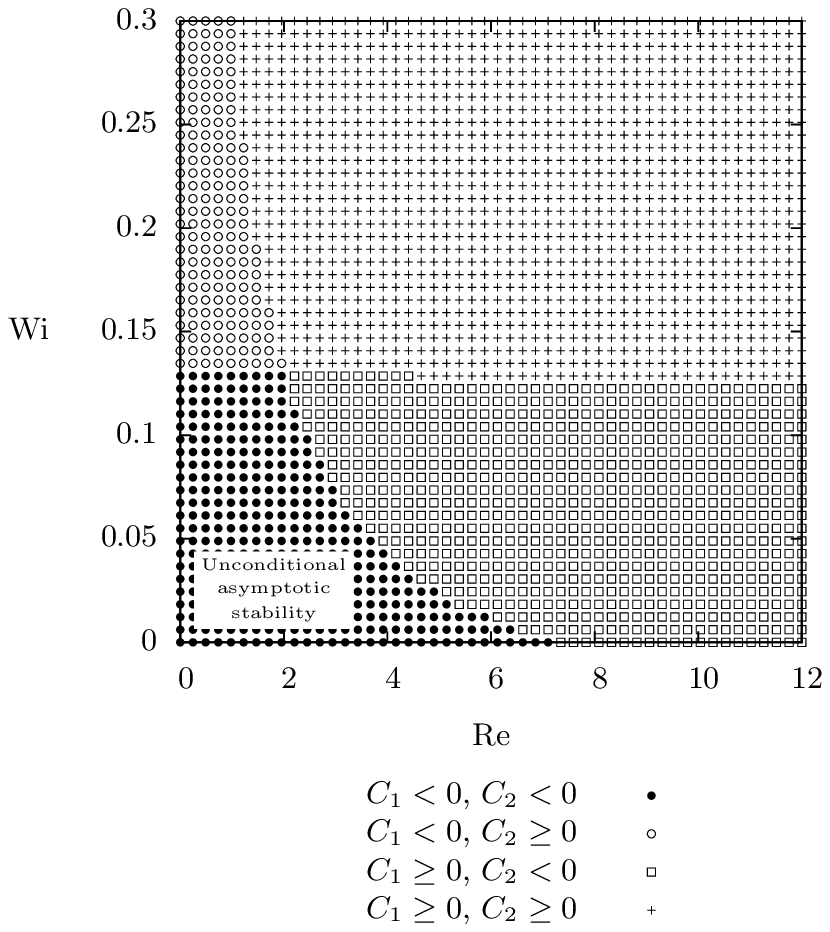}}
  \caption{Taylor--Couette flow, numerical values of constants $C_1$ and $C_2$ for various values of the Reynolds number $\Reynolds$, Weissenberg number $\Weissenberg$ and the dimensionless shear modulus $\Xi$. \emph{Unconditional asymptotic stability is granted provided that $C_1<0$ and $C_2<0$}, numerical values of constants $C_1$ and $C_2$ are evaluated using formulae~\eqref{eq:constants}. Giesekus parameter $\alpha = \frac{1}{2}$, problem parameters $\eta = \frac{1}{2}$, $\zeta = 2$.}
  \label{fig:stability-regions}
\end{figure}

Having computed the steady velocity field $\widehat{\vec{v}}$ and the corresponding steady field~$\widehat{\lcgnc}$, we can evaluate the constants $C_1$ and $C_2$ in the estimate~\eqref{eq:lyapunov-derivative-estimate}. The gradient of~$\widehat{\vec{v}}$ as well as the gradient of $\widehat{\lcgnc}$ are again computed numerically from the obtained numerical solution. Finally, the Poincar\'e constant for the cylindrical annulus is determined via an explicit solution of the corresponding eigenvalue problem $-\Delta u = \lambda u$ for the Laplace operator with Dirichlet boundary condition, which leads, for the geometrical parameter $\eta = \frac{1}{2}$, to the value
$
C_P \approx 0.1025.
$
The resulting stability regions in the $\Reynolds$--$\Weissenberg$ plane are shown in Figure~\ref{fig:stability-regions} for a fixed value of the dimensionless shear modulus $\Xi$. As one might expect the spatially inhomogeneous steady state is indeed unconditionally asymptotically stable if the Weissenberg number $\Weissenberg$ and the Reynolds number $\Reynolds$ are small enough.

\subsection{Numerical experiments -- evolution of various initial perturbations}
\label{sec:numer-simul}
Finally, we document the theoretically predicted behaviour by numerical experiments. The numerical experiments allow us to quantitatively track the evolution of key quantities such as the net kinetic energy, and also to quantitatively monitor the energy exchange between the fluid and its surroundings.  

The governing equations are numerically solved using standard techniques. The weak forms of the governing equations are discretised in the space using the finite element method, while the time derivatives are approximated with the backward Euler method. The two-dimensional domain~$\Omega$ is discretised by regular quadrilaterals. The mesh divides the annular region $\Omega$ into $80$ pieces in the radial direction, and in $720$ pieces in the azimuthal direction. The corresponding total number of degrees of freedom in all numerical experiments is over $1.3 \times 10^6$. The velocity field~$\vecv$ and the $\lcgnc$ field are approximated by biquadratic $Q2$ elements, and the pressure field $\mns$ is approximated by the piecewise linear discontinuous elements $P1^d$, see~\cite{korelc.j:manual} for details. The finite element pair that is used for the velocity/pressure fields satisfies the Babu\v{s}ka-Brezzi condition, the finite element for $\lcgnc$ field is chosen to be the same as for the velocity in order to provide rich enough finite element space for the solution. The same finite elements have been chosen for the two-dimensional simulation of other viscoelastic rate-type fluids (Oldroyd, Burgers and their various nonlinear versions) by \cite{hron.j.rajagopal.kr.ea:flow} and \cite{malek.j.rajagopal.kr.ea:thermodynamically*1}. %, while \cite{hron.j.milos.v.ea:on} discretised the domain with the triangles and used $P2$ elements for velocity and $\lcgnc$ fields and $P1$ for the pressure field.
In three-dimensional case low order elements can be used in order to decrease the overall cost of the calculation, see~\cite{tuma.k.stein.j.ea:motion}.

The numerical scheme is implemented in the AceGen/AceFEM system, see~\cite{korelc.j:Multi} and \cite{korelc.j:Automation}. The main advantage of the system is that it provides automatic differentiation used for the computation of the exact tangent matrix needed by the Newton solver that treats all nonlinearities. The final set of linear equations is solved by the direct solver Intel MKL Pardiso. The stopping criterion for the Newton solver is set to $10^{-9}$.

Using the numerical scheme we are ready to study the behaviour of various perturbations to the non-equilibrium steady state. In all scenarios described below we use the dimensionless parameters
\begin{equation}
  \label{eq:136}
  \Xi=0.1, \quad \Reynolds=50, \quad \Weissenberg=5, \quad \alpha = \frac{1}{2}
\end{equation}
and we fix the geometric parameter $\eta$ and angular velocities ratio $\zeta$ as in~Section~\ref{sec:taylor-couette-stability} that is $\eta = \frac{1}{2}$ and $\zeta = 2$. The chosen values of $\eta$, $\zeta$ and $\Xi$ correspond to the stability diagram shown in Figure~\ref{fig:stability-regions-a}. The values of Reynolds number and Weissenberg number are outside the region where we have \emph{proven} the decay of the proposed Lyapunov type functional. Nevertheless, as we shall see, the Lyapunov type functional is, in the cases being investigated below, still a decreasing function.

First, we start from the homogeneous steady state solution\footnote{%
\label{fn:2}%
More precisely, the initial condition is $\vec{v} = \vec{0}$ inside the domain $\Omega$, and~\eqref{eq:bc-no-slip-steady} on the boundary of~$\Omega$. After the first computational time step, which is chosen as $\Delta t = 0.05$, we get on the discrete level a divergence-free velocity field with the appropriate boundary condition. This discrete velocity field provides us a consistent initial condition for further computations. Therefore, we formally start the evolution not at $t=0$, but at $t = 0.05$.} % footnote
$[ \vecv, \lcgnc, \mns] = [ \vec{0}, \identity, 0]$, and we let the system to spontaneously evolve up to the time instant $t=1000$. At this time instant the system is almost relaxed and is close to the steady solution. The solution at $t=1000$ is used as a starting point for solving the steady governing equations (without the time derivatives) and the spatially inhomogeneous non-equilibrium \emph{steady} state is obtained just in two Newton iterations. (The finite element solution coincides with the semi-analytical steady solution obtained in Section~\ref{sec:steady-solution-taylor-couette}. This among others provides us a tool for the code verification.) Consequently, the finite element solution is in what follows used as the spatially inhomogeneous non-equilibrium steady state~$\widehat{\vec{W}}$.

Having obtained the numerical representation of the spatially inhomogeneous non-equilibrium steady state, we proceed with two scenarios concerning the specification of the initial perturbation.

\subsubsection{Scenario A -- localized perturbation of the left Cauchy--Green field} 
\begin{figure}[h]
  \centering
  \captionsetup[subfigure]{labelformat=empty}
  \subfloat[$t=0$]{\includegraphics[width=0.16\textwidth]{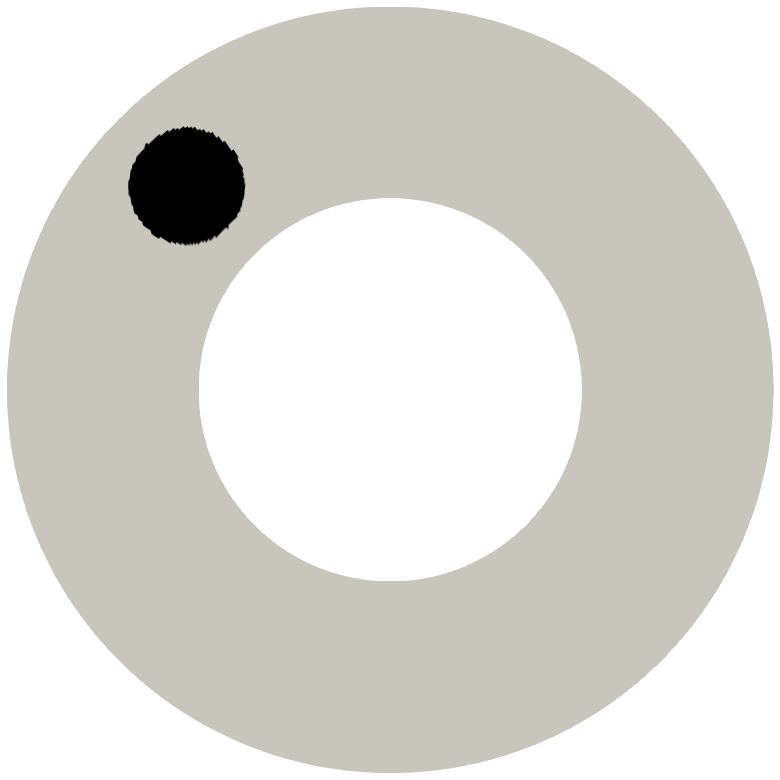}}
  \qquad
  \subfloat[$t=0.1$]{\includegraphics[width=0.16\textwidth]{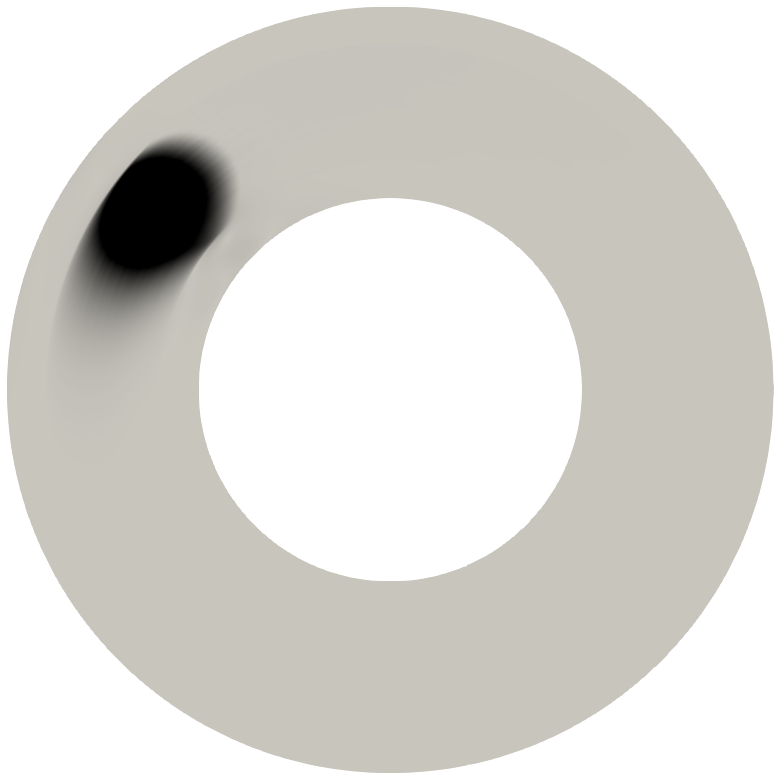}}
  \qquad
  \subfloat[$t=0.5$]{\includegraphics[width=0.16\textwidth]{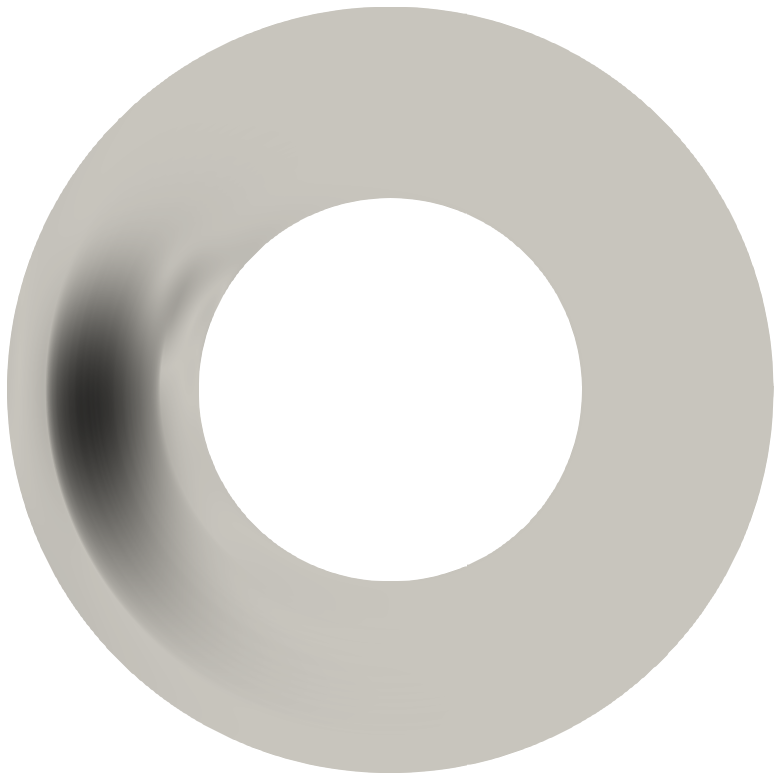}}
  \qquad
  \subfloat[$t=1$]{\includegraphics[width=0.16\textwidth]{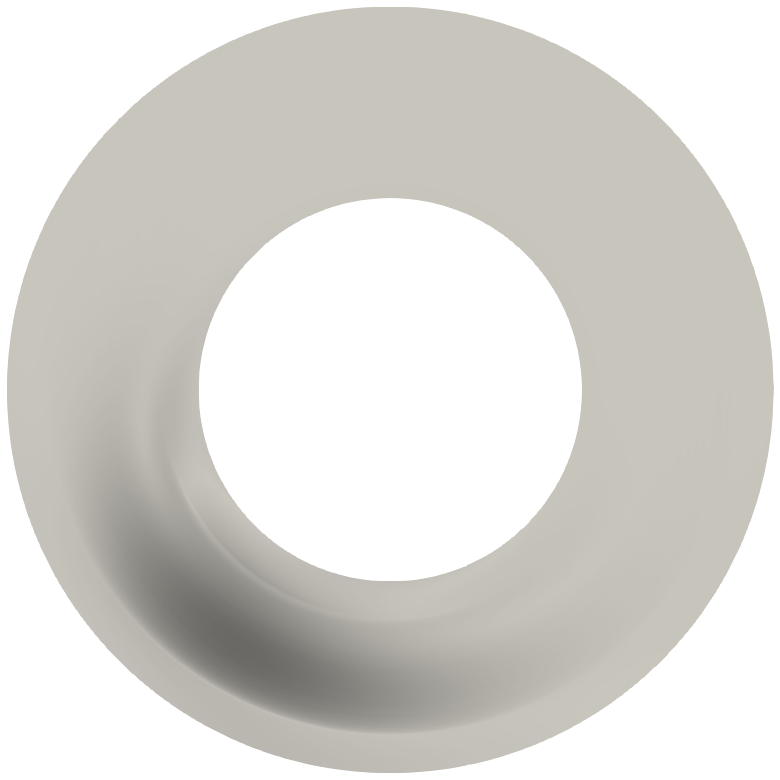}}
  \\
  \subfloat[$t=2$]{\includegraphics[width=0.16\textwidth]{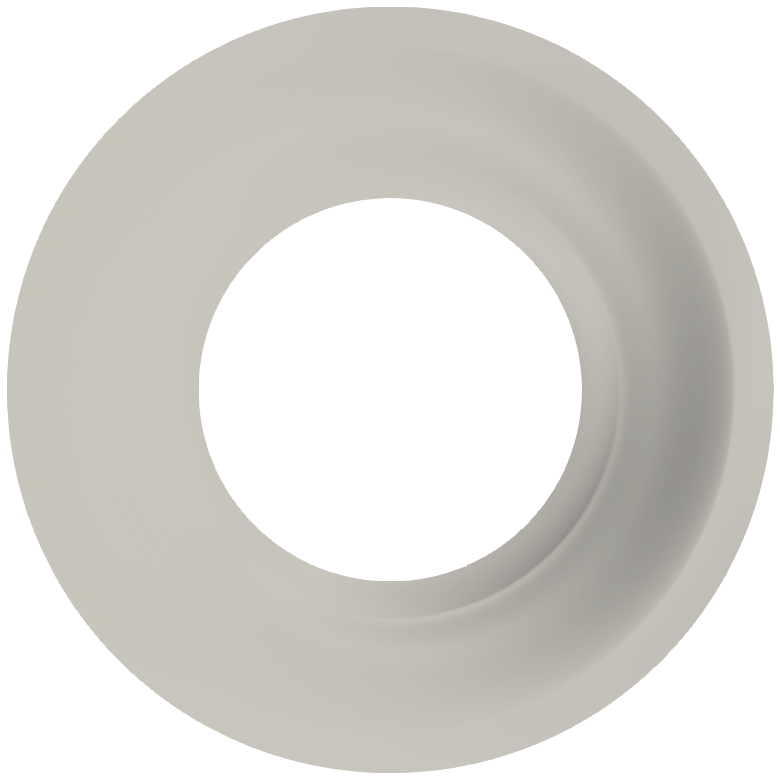}}
  \qquad
  \subfloat[$t=3$]{\includegraphics[width=0.16\textwidth]{snap_B_20-crop}}
  \qquad
  \subfloat[$t=5$]{\includegraphics[width=0.16\textwidth]{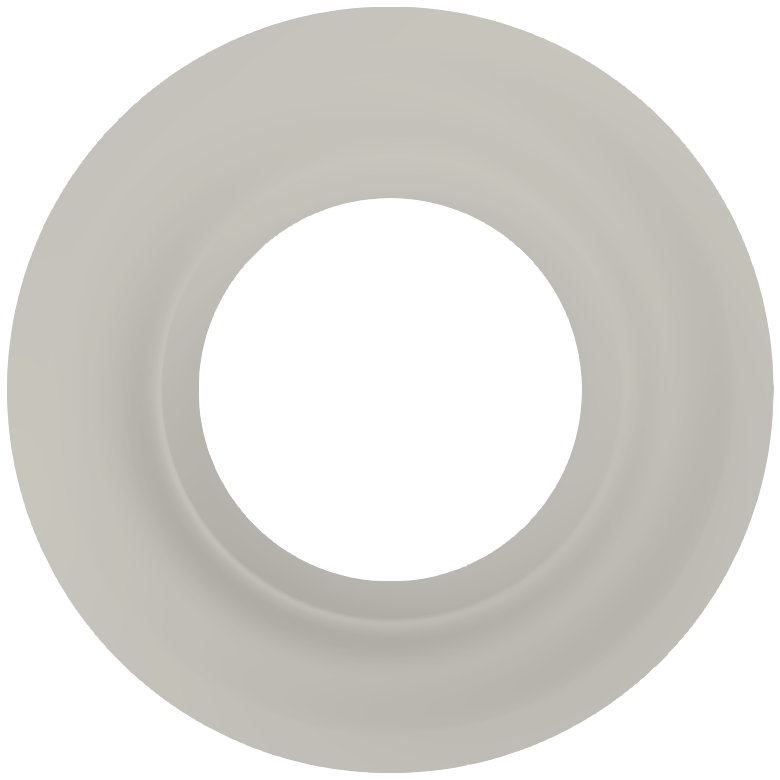}}
  \qquad
  \subfloat[$t=10$]{\includegraphics[width=0.16\textwidth]{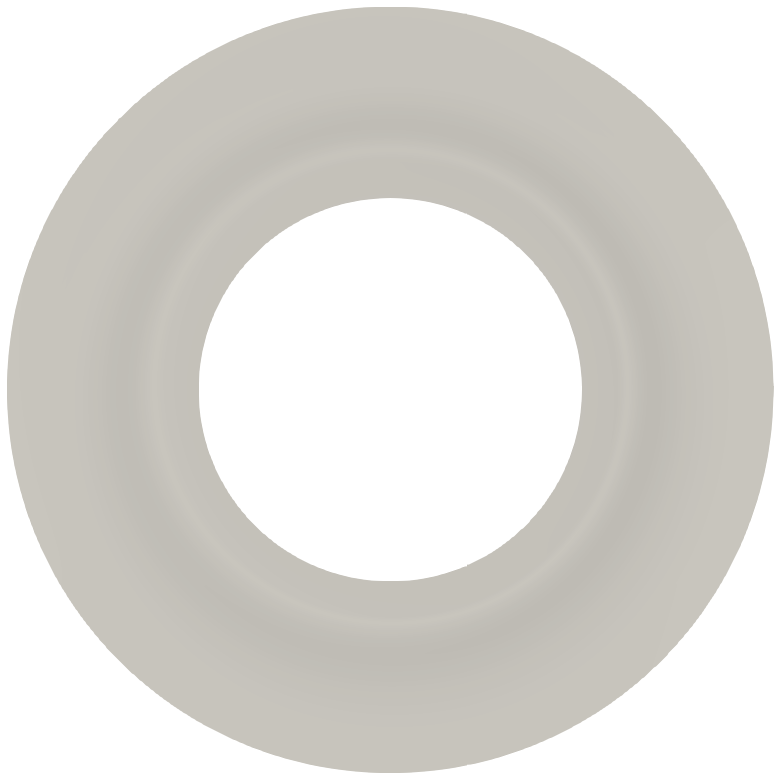}}
  \\
  \subfloat{\includegraphics[width=0.16\textwidth]{scale-lcgnc}}
  \caption{Scenario A, snapshots of $|\widetilde{\lcgnc}|$ at different time instants.}% (lightest gray 0, black 2)
\label{fig:snapshots1a}
\end{figure}

\begin{figure}[h]
  \centering
  \captionsetup[subfigure]{labelformat=empty}
  \subfloat[$t=0$]{\includegraphics[width=0.16\textwidth]{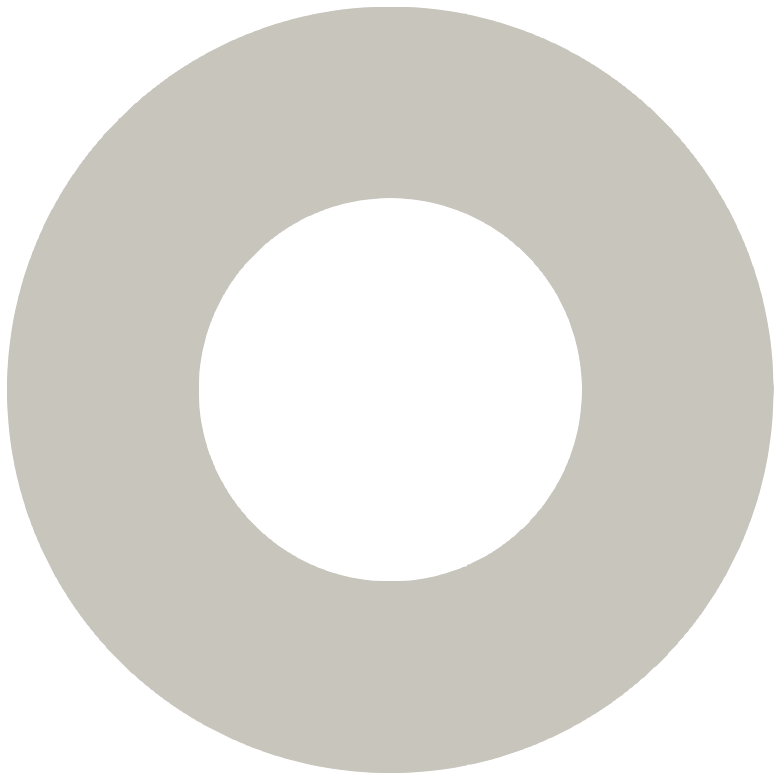}}
  \qquad
  \subfloat[$t=0.1$]{\includegraphics[width=0.16\textwidth]{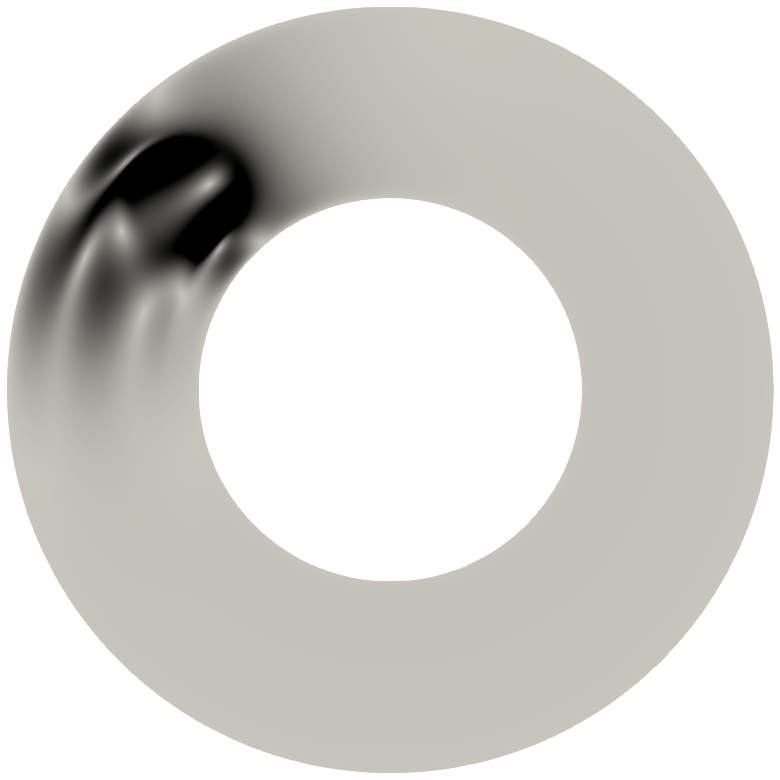}}
  \qquad
  \subfloat[$t=0.5$]{\includegraphics[width=0.16\textwidth]{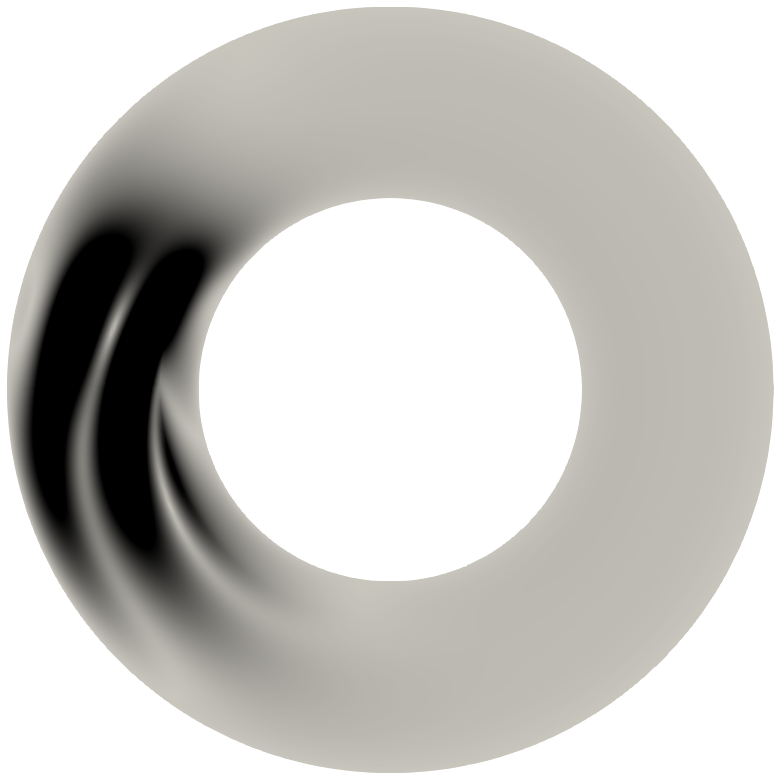}}
  \qquad
  \subfloat[$t=1$]{\includegraphics[width=0.16\textwidth]{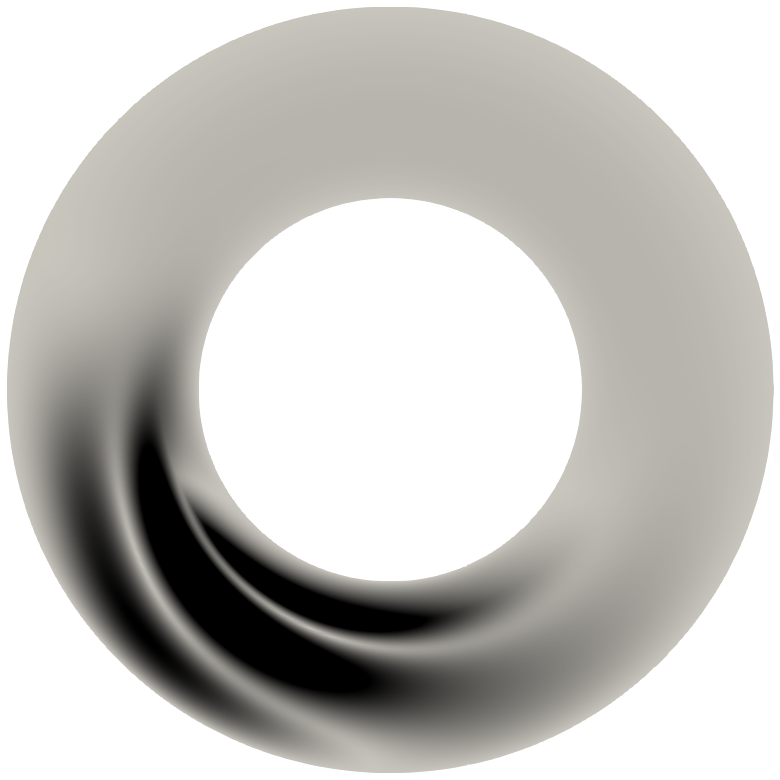}}
  \\
  \subfloat[$t=2$]{\includegraphics[width=0.16\textwidth]{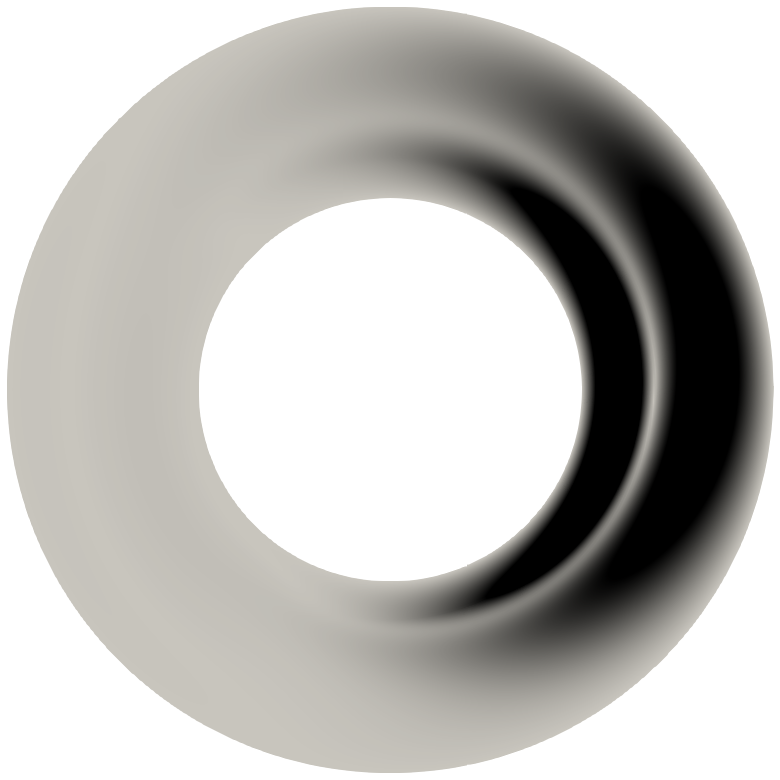}}
  \qquad
  \subfloat[$t=3$]{\includegraphics[width=0.16\textwidth]{snap_v_20-crop}}
  \qquad
  \subfloat[$t=5$]{\includegraphics[width=0.16\textwidth]{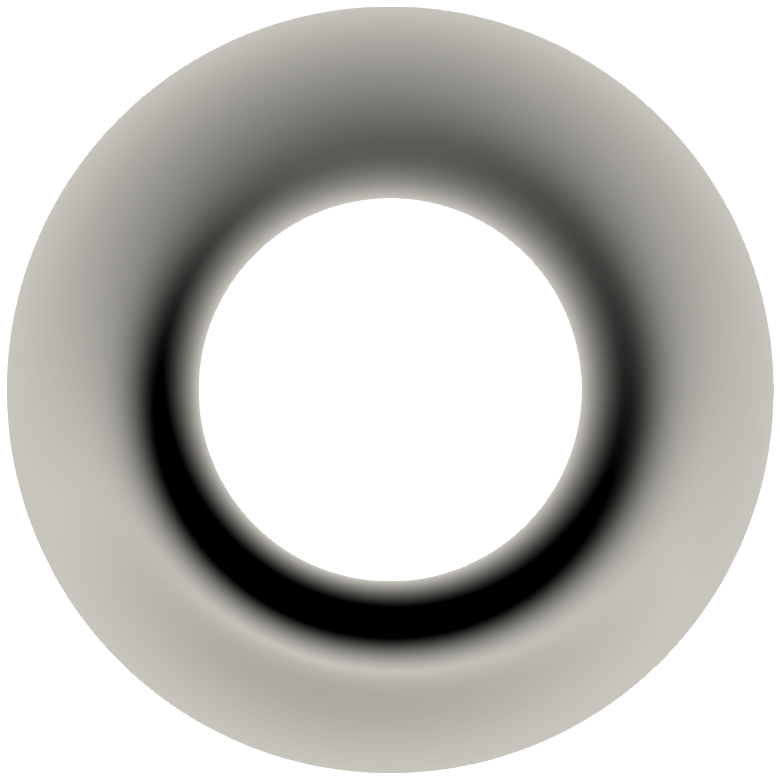}}
  \qquad
  \subfloat[$t=10$]{\includegraphics[width=0.16\textwidth]{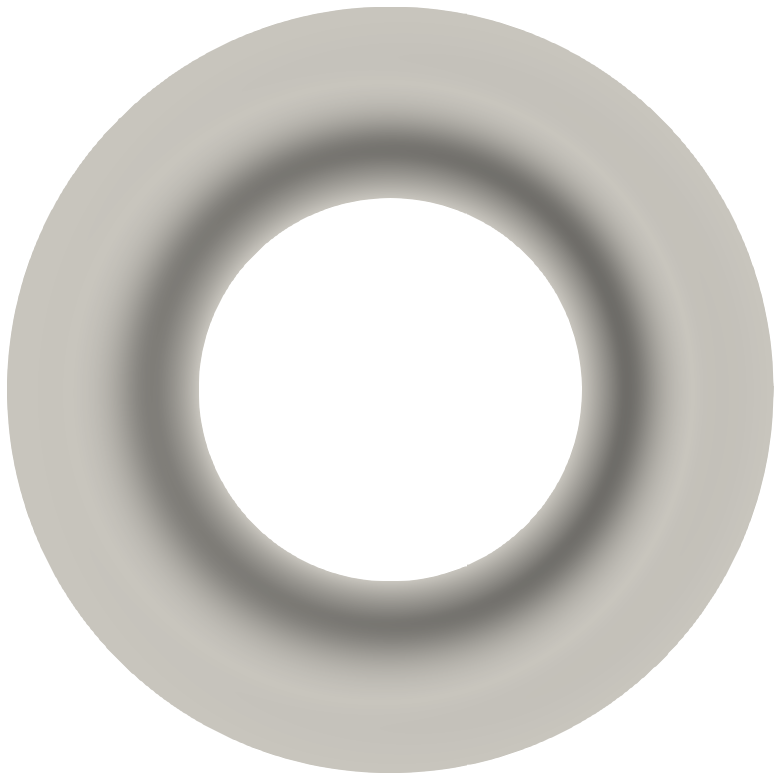}}
  \\
  \subfloat{\includegraphics[width=0.16\textwidth]{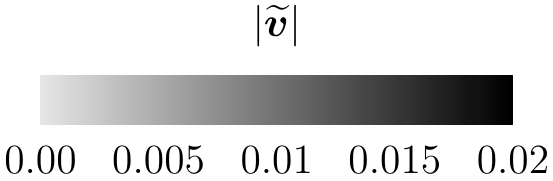}}
  \caption{Scenario A, snapshots of $|\widetilde{\vecv}|$ at different time instants.}% (lightest gray 0, black 2)
  \label{fig:snapshots1b}
\end{figure}

\begin{figure}[h]
  \centering
  \subfloat[Net ``elastic energy'' of the perturbation, $\int_{\Omega} \frac{\Xi}{4} \absnorm{\widetilde{\lcgnc}}^2\,\cvolumee$.]{\includegraphics[width=0.45\textwidth]{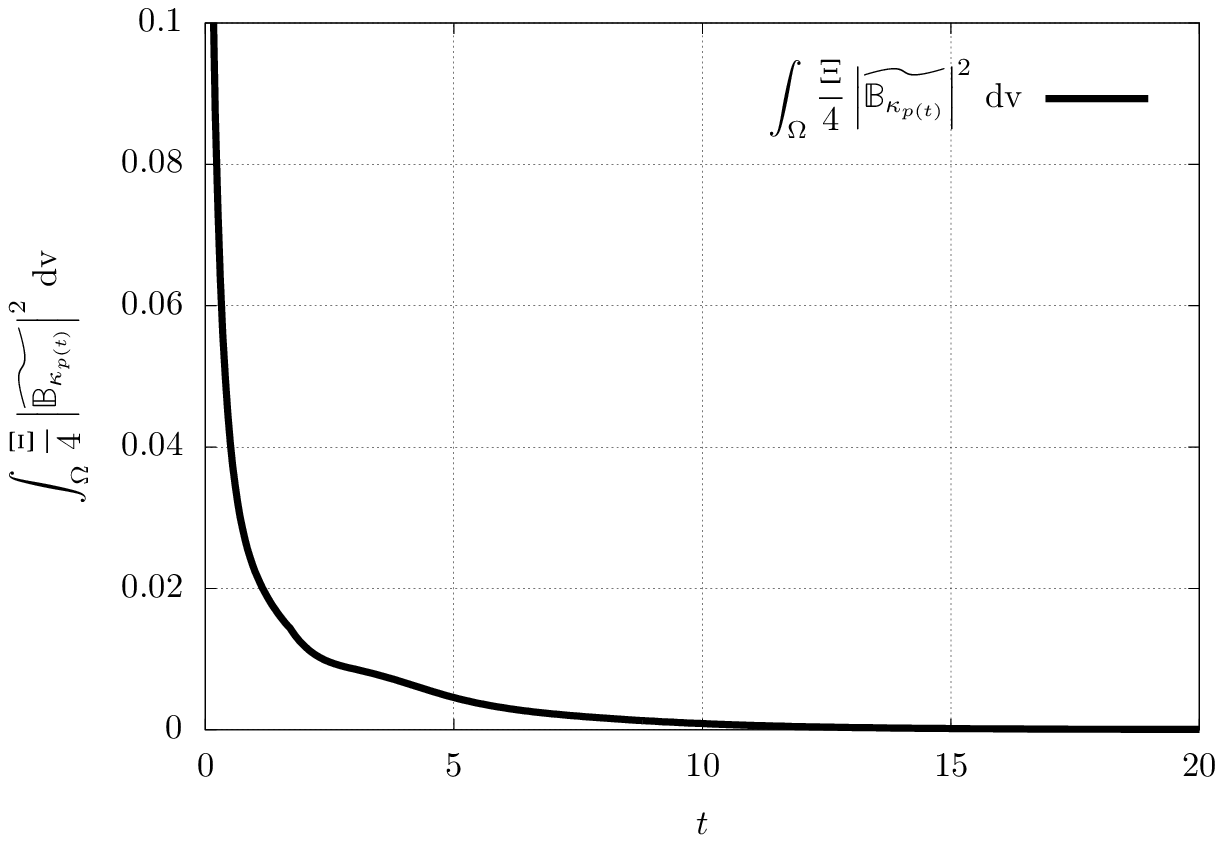}}
  \qquad
  \subfloat[\label{fig:graphs_perturb_stress-b}Net ``kinetic energy'' of the perturbation, $\int_{\Omega}\frac{1}{2}\absnorm{\widetilde{\vec{v}}}^2\,\cvolumee$.]{\includegraphics[width=0.45\textwidth]{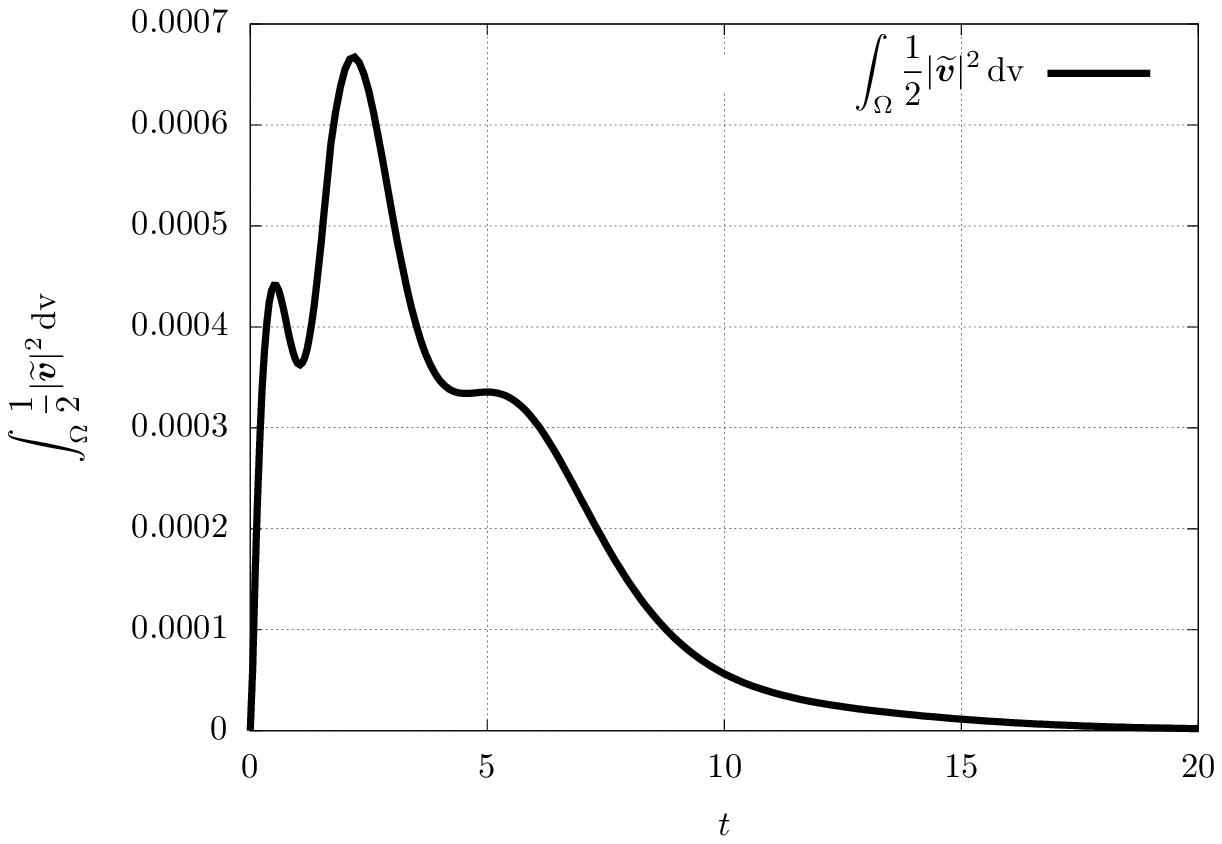}}
  \qquad
  \subfloat[Lyapunov type functional $\mathcal{V}_{\mathrm{neq}}$ and the functional $\mathcal{V}_{\mathrm{naive}}$.]{\includegraphics[width=0.45\textwidth]{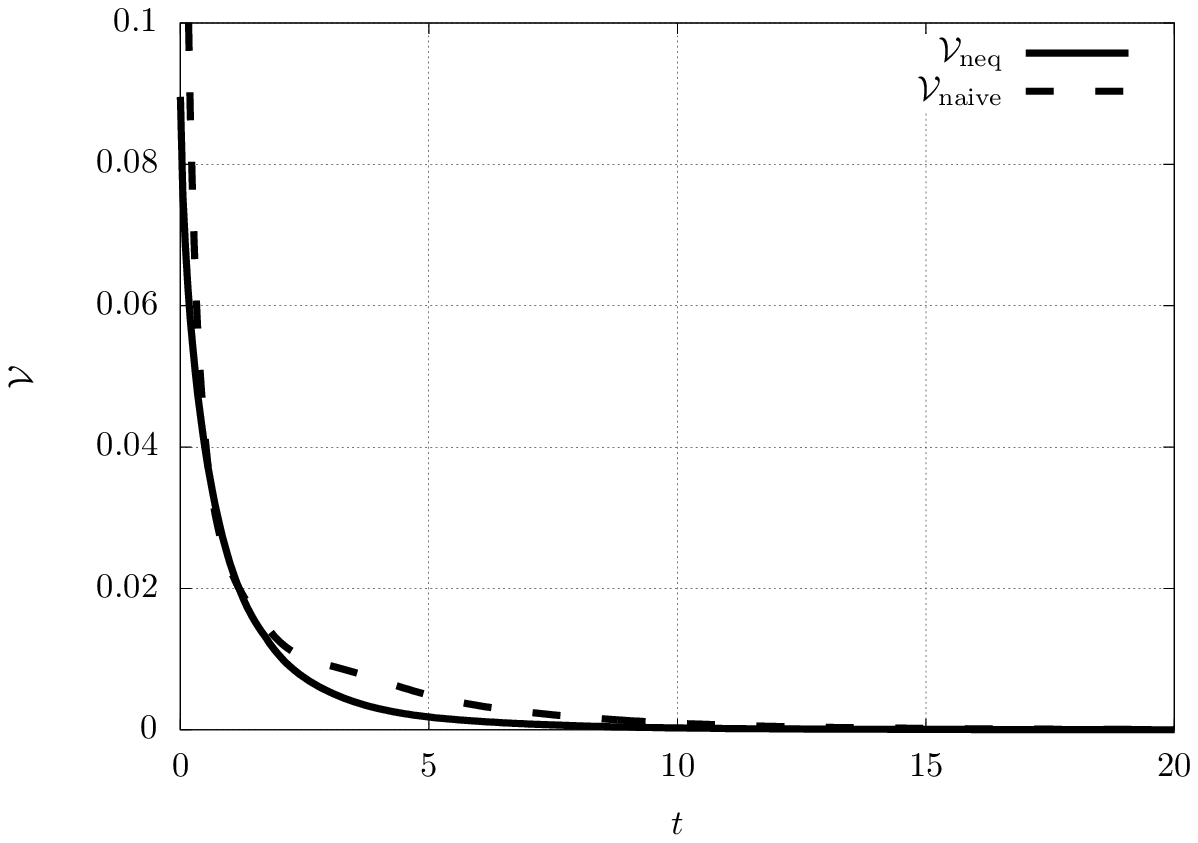}}
  \qquad
  \subfloat[\label{fig:graphs_perturb_stress-d}Net mechanical energy flux through the boundary, \hbox{$\int_{\partial \Omega} \vectordot{\cstress \vec{v}}{\vec{n}} \, \csurfacees$}.]{\includegraphics[width=0.45\textwidth]{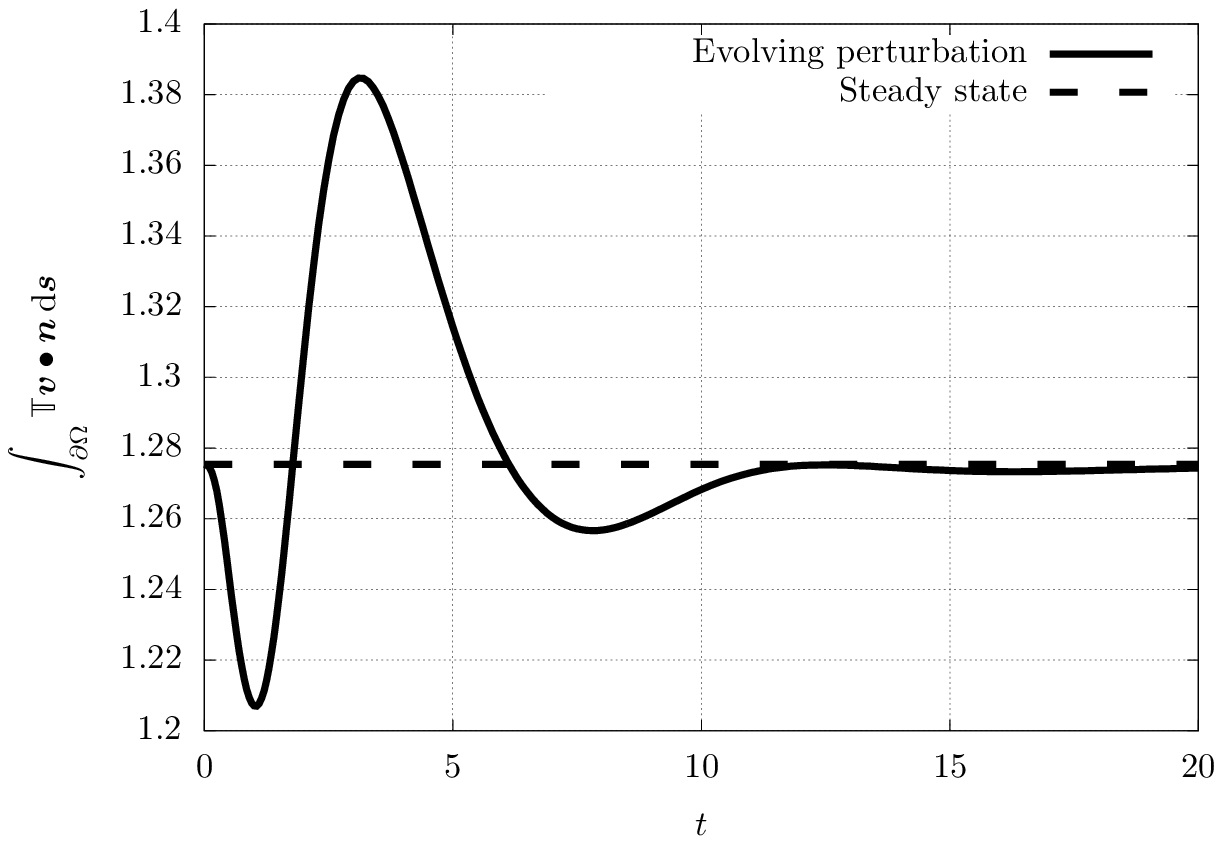}}
  \caption{Scenario A, time evolution of the net quantities.}
  \label{fig:graphs_perturb_stress}
\end{figure}

In the first scenario, we keep the initial velocity field perturbation equal to zero,
\begin{equation}
  \label{eq:135}
  \left. \widetilde{\vec{v}} \right|_{t=0} = \vec{0},
\end{equation}
while the initial perturbation in $\lcgnc$ is localised in space, see the first snapshot in Figure~\ref{fig:snapshots1a}. Since the system is fully coupled, the perturbation in the $\lcgnc$ field triggers for $t>0$ a nontrivial evolution of the velocity perturbation $\tilde{\vecv}$, see Figure~\ref{fig:snapshots1b}. This can be observed also in the plots showing the evolution of the net elastic stored energy and the net kinetic energy, see Figure~\ref{fig:graphs_perturb_stress}.

Finally, we also investigate the time evolution of the proposed Lyapunov type functional~$\mathcal{V}_{\mathrm{neq}}$ and the naive Lyapunov type functional~$\mathcal{V}_{\mathrm{naive}}$, and the net mechanical energy flux going through the boundary of $\Omega$, see Figure~\ref{fig:graphs_perturb_stress}. Although we work with the Reynolds number/Weissenberg number pair outside the guaranteed stability region, we see that the value of Lyapunov type functional~$\mathcal{V}_{\mathrm{neq}}$ still decreases in time, and that the perturbation vanishes for $t \to +\infty$. This indicates that the estimates on the time derivative of the proposed Lyapunov type functional are, at least for a class of perturbations, too strict and they might be improved. One should also note that the ``net kinetic energy'' of the perturbation, that is the functional $\int_{\Omega}\frac{1}{2}\absnorm{\widetilde{\vec{v}}}^2\,\cvolumee$, \emph{does not} decrease for all $t>0$, see Figure~\ref{fig:graphs_perturb_stress-b}. In fact, it experiences a transitional growth, and such a transient growth can be observed even for the Reynolds number/Weissenberg number values within the stability region. This is a natural observation. The elastic energy stored in the material can be released in the form of the kinetic energy. It is only the combination of the elastic energy and the kinetic energy that appears in the Lyapunov type functional that leads to a quantity that decays at any time.

Further, the net mechanical energy flux through the boundary fluctuates around the value that corresponds to the non-equilibrium steady state, and then it reaches the value that corresponds to the spatially inhomogeneous non-equilibrium steady state, see Figure~\ref{fig:graphs_perturb_stress-d}. This can again happen even if the Reynolds number/Weissenberg number take values within the stability region.

\subsubsection{Scenario B -- global perturbation of the velocity field}

\begin{figure}[h]
  \centering
  \captionsetup[subfigure]{labelformat=empty}
  \subfloat[$t=0$]{\includegraphics[width=0.16\textwidth]{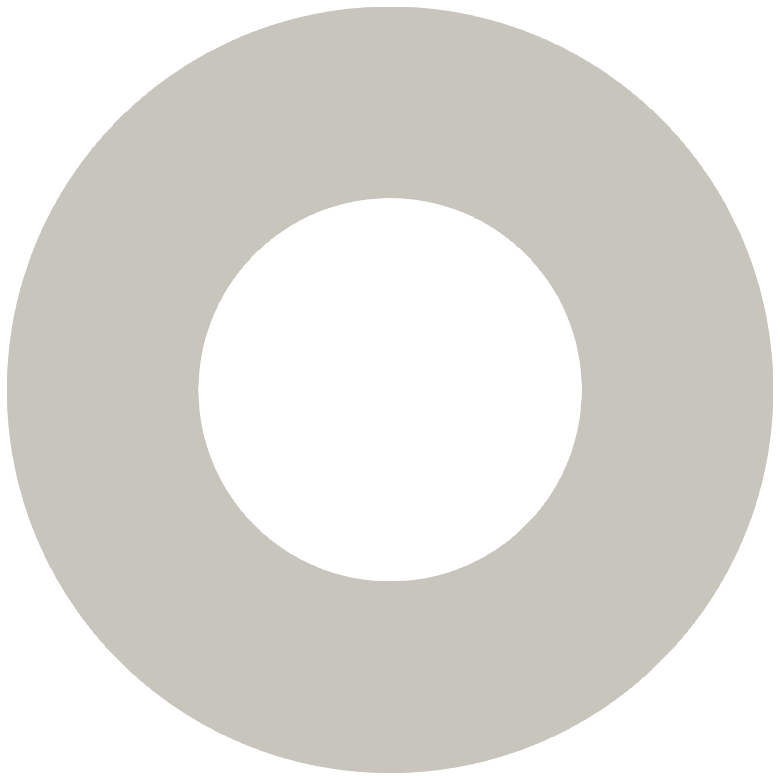}}
  \qquad
  \subfloat[$t=0.1$]{\includegraphics[width=0.16\textwidth]{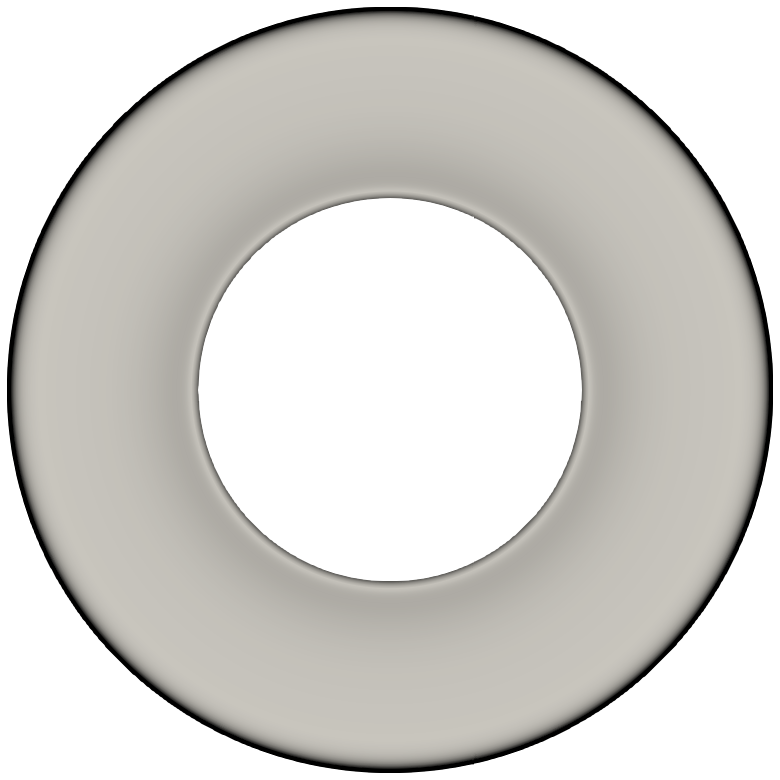}}
  \qquad
  \subfloat[$t=0.5$]{\includegraphics[width=0.16\textwidth]{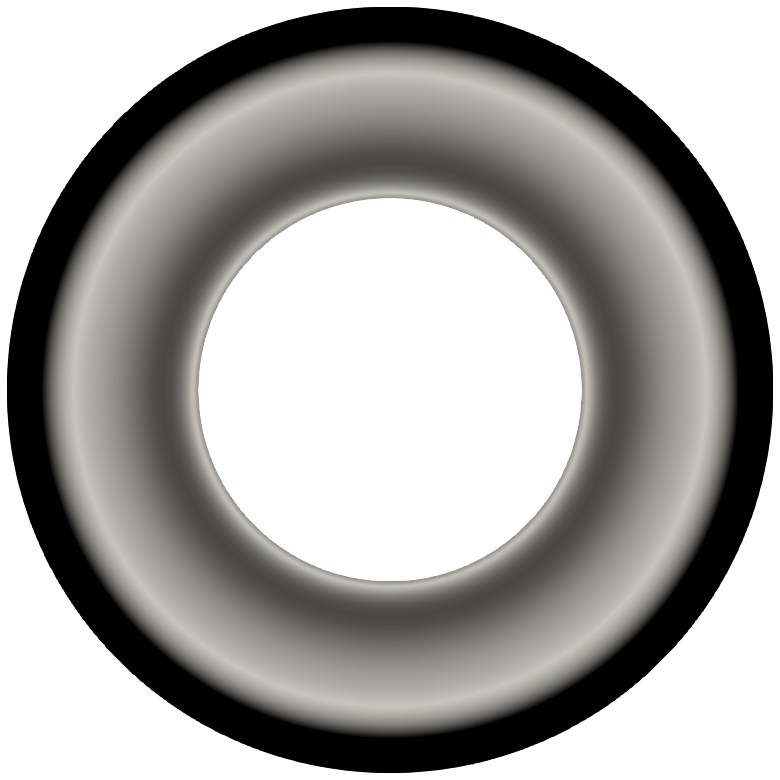}}
  \qquad
  \subfloat[$t=1$]{\includegraphics[width=0.16\textwidth]{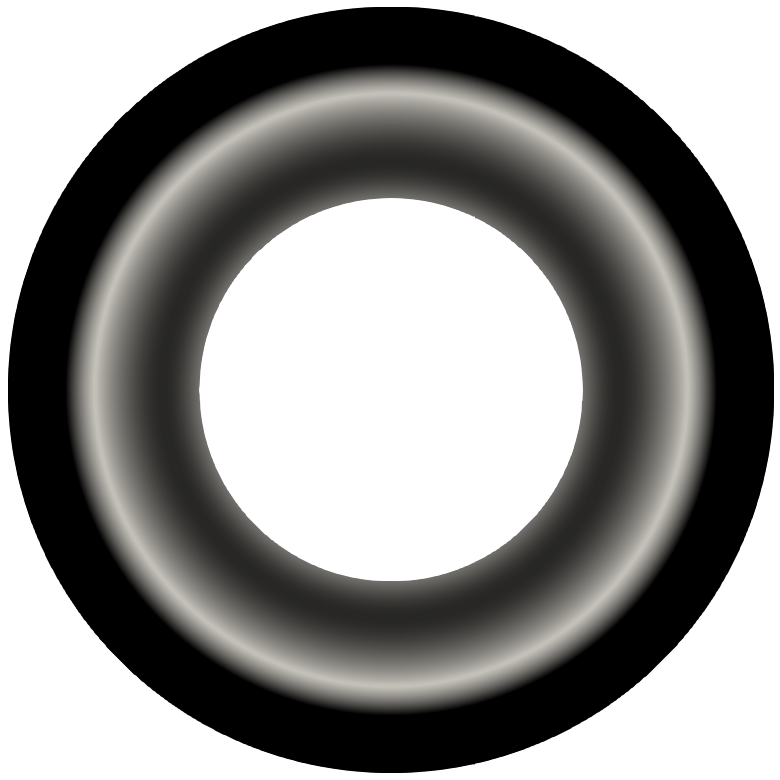}}
  \\
  \subfloat[$t=2$]{\includegraphics[width=0.16\textwidth]{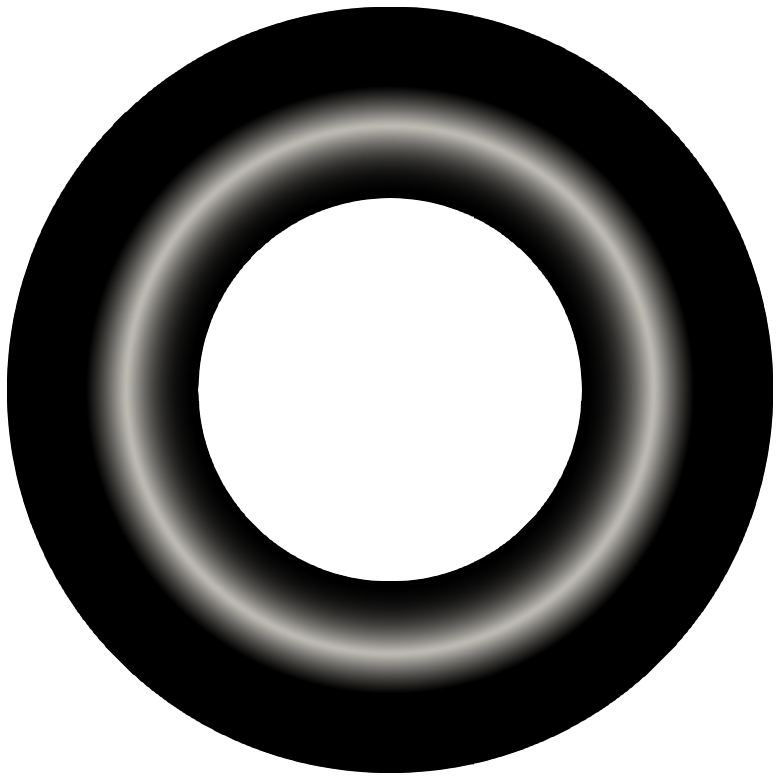}}
  \qquad
  \subfloat[$t=3$]{\includegraphics[width=0.16\textwidth]{snap_B_20_start-crop}}
  \qquad
  \subfloat[$t=5$]{\includegraphics[width=0.16\textwidth]{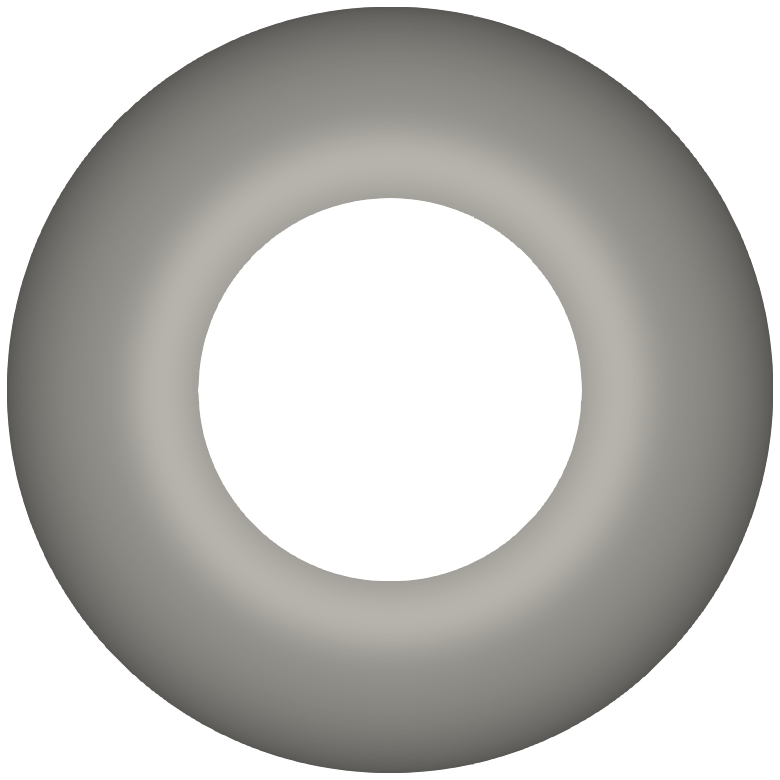}}
  \qquad
  \subfloat[$t=10$]{\includegraphics[width=0.16\textwidth]{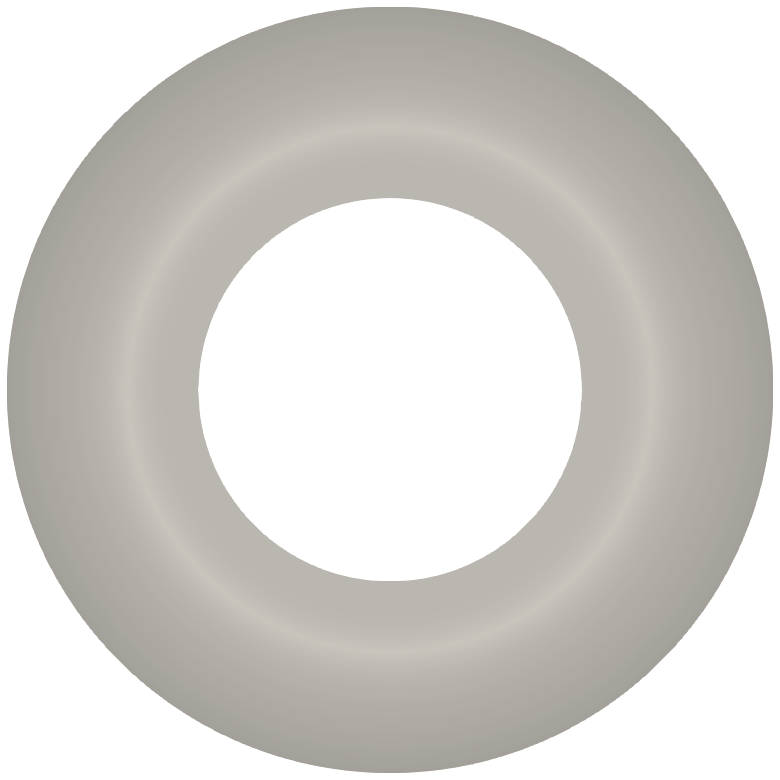}}
  \\
  \subfloat{\includegraphics[width=0.16\textwidth]{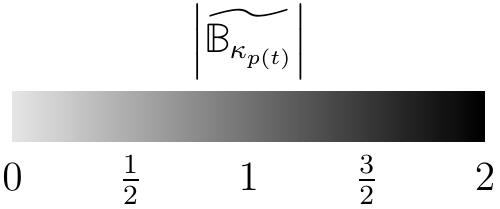}}
  \caption{Scenario B, snapshots of $|\widetilde{\lcgnc}|$ at different time instants.}% (lightest gray 0, black 2)
\label{fig:snapshots2a}
\end{figure}

\begin{figure}[h]
  \centering
  \captionsetup[subfigure]{labelformat=empty}
  \subfloat[$t=0$]{\includegraphics[width=0.16\textwidth]{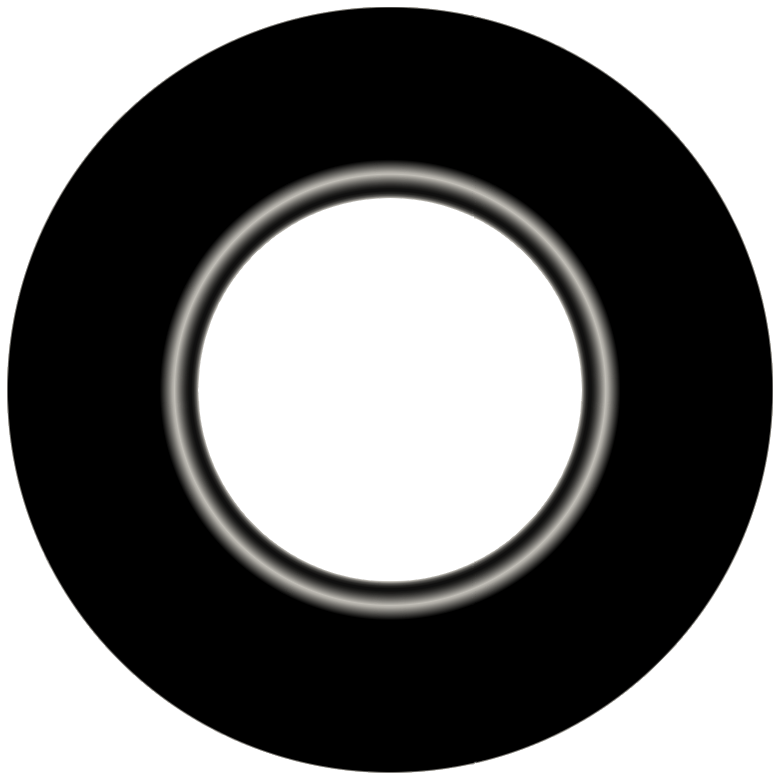}}
  \qquad
  \subfloat[$t=0.1$]{\includegraphics[width=0.16\textwidth]{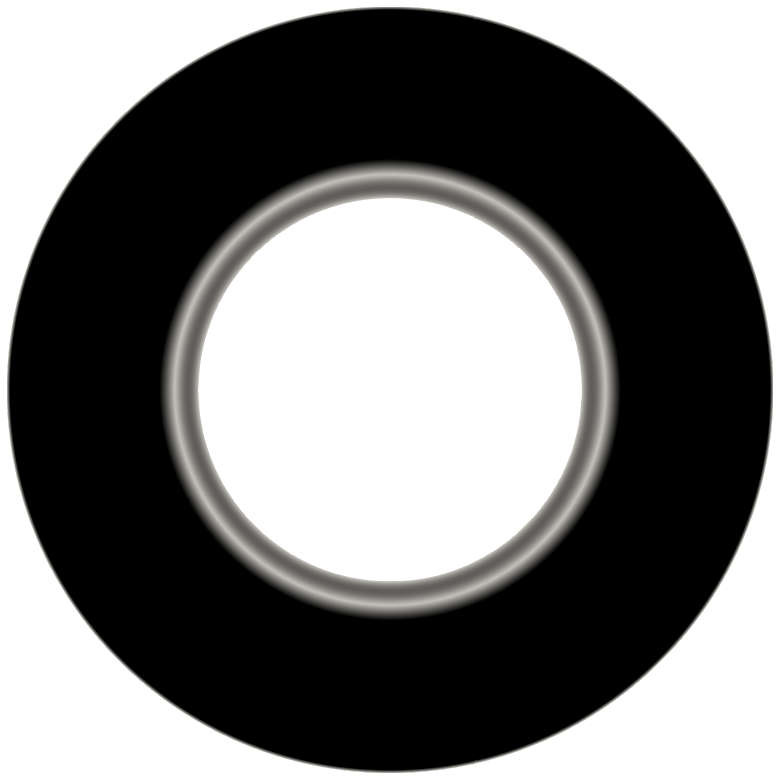}}
  \qquad
  \subfloat[$t=0.5$]{\includegraphics[width=0.16\textwidth]{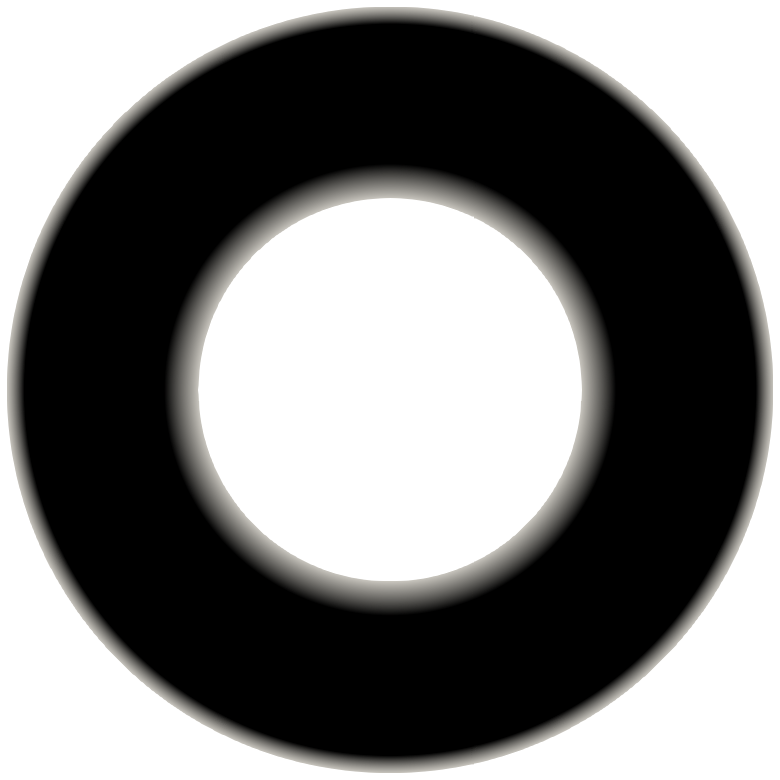}}
  \qquad
  \subfloat[$t=1$]{\includegraphics[width=0.16\textwidth]{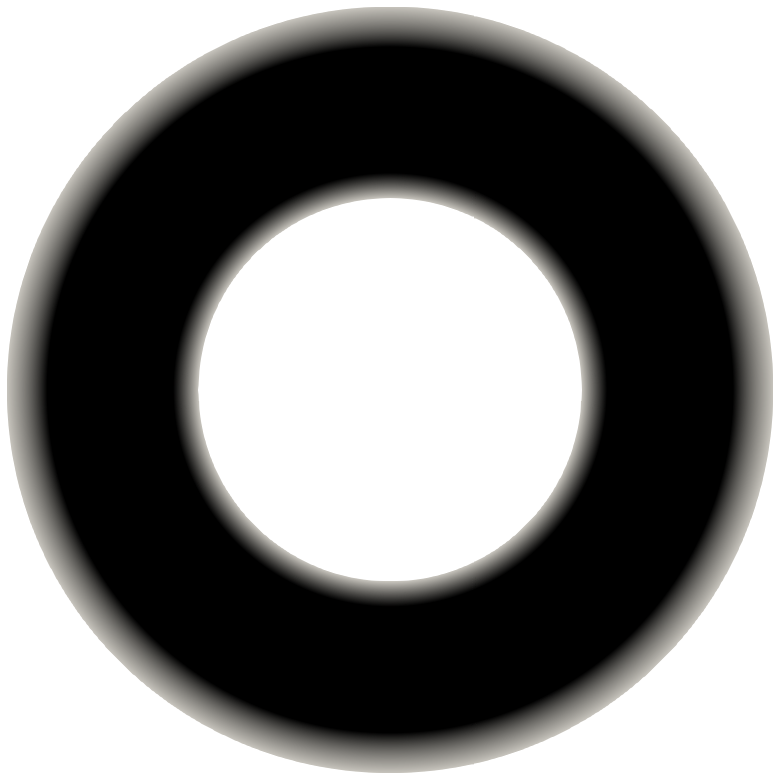}}
  \\
  \subfloat[$t=2$]{\includegraphics[width=0.16\textwidth]{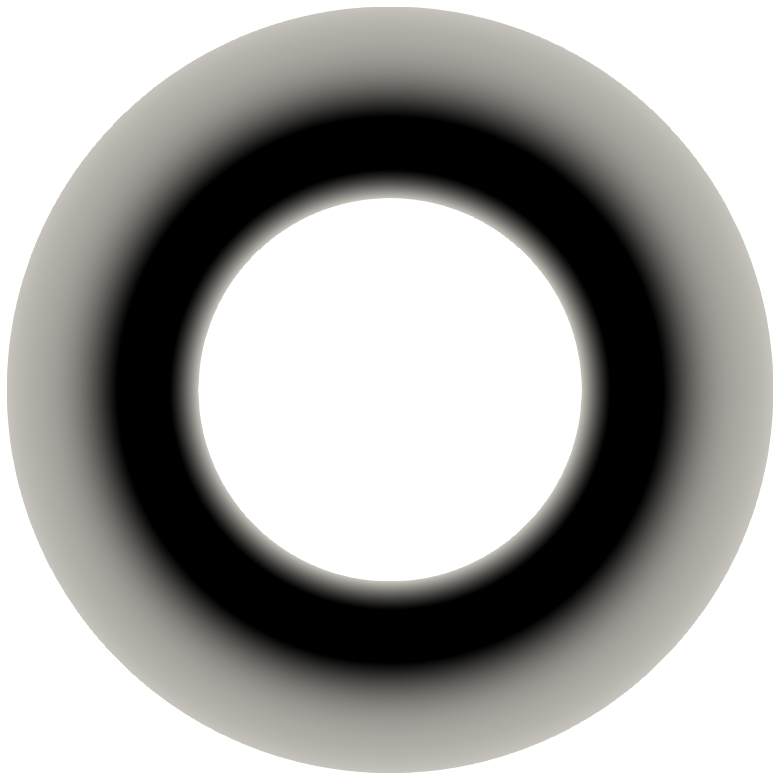}}
  \qquad
  \subfloat[$t=3$]{\includegraphics[width=0.16\textwidth]{snap_v_20_start-crop}}
  \qquad
  \subfloat[$t=5$]{\includegraphics[width=0.16\textwidth]{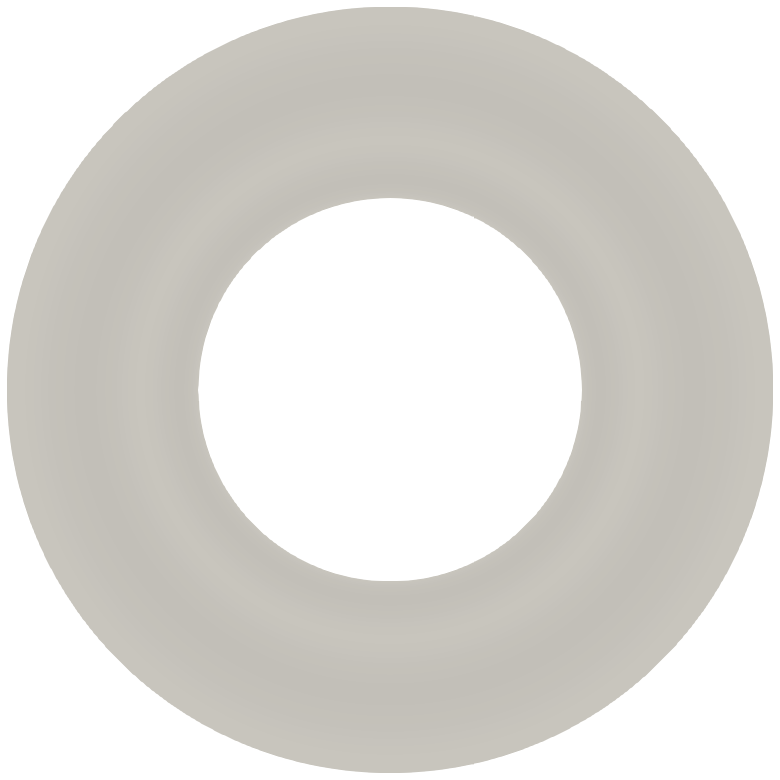}}
  \qquad
  \subfloat[$t=10$]{\includegraphics[width=0.16\textwidth]{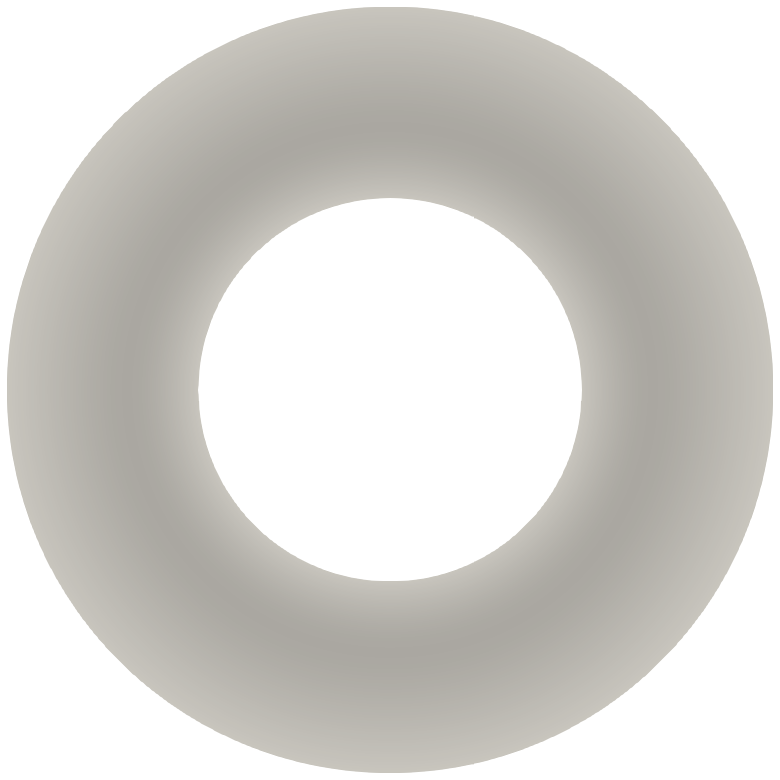}}
  \\
  \subfloat{\includegraphics[width=0.16\textwidth]{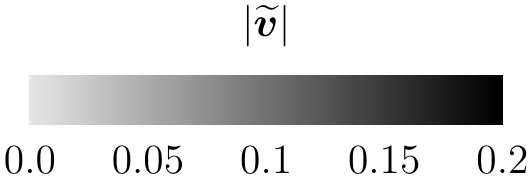}}
  \caption{Scenario B, snapshots of $|\widetilde{\vecv}|$ at different time instants.}% (lightest gray 0, black 2)
  \label{fig:snapshots2b}
\end{figure}

In the second scenario we start with a nonzero velocity perturbation $\widetilde{\vecv}$, and the $\lcgnc$ field is kept unchanged,
\begin{equation}
  \label{eq:137}
  \left. \widetilde{\lcgnc} \right|_{t=0} = \tensorq{O}.
\end{equation}
The initial velocity $\vecv$ is prescribed as\footnote{%
  Formally, we apply the same procedure as discussed in Footnote~\ref{fn:2}. The initial condition is $\vec{v} = \Omega r \cobvec{\hatp}$ inside the domain $\Omega$, and~\eqref{eq:bc-no-slip-steady} on the boundary of~$\Omega$. The actual computation starts after the first (formal) time step, when the discrete velocity field is divergence-free and it fulfills the boundary condition.}
\begin{equation}
  \label{eq:138}
  \left. \vecv \right|_{t=0} = \Omega r \cobvec{\hatp},
\end{equation}
where the angular velocity is the arithmetic mean of the two angular velocities $\Omega = \frac{\Omega_1+\Omega_2}{2}$. Again, as in the previous case, the non-zero perturbation in one unknown field ($\widetilde{\vecv}$) triggers for $t>0$ a nontrivial evolution of the other unknown field ($\widetilde{\lcgnc}$), see Figure~\ref{fig:snapshots2a} and Figure~\ref{fig:snapshots2b}.

\begin{figure}[h]
  \centering
  \subfloat[Net ``elastic energy'' of the perturbation, $\int_{\Omega} \frac{\Xi}{4} \absnorm{\widetilde{\lcgnc}}^2\,\cvolumee$.]{\includegraphics[width=0.45\textwidth]{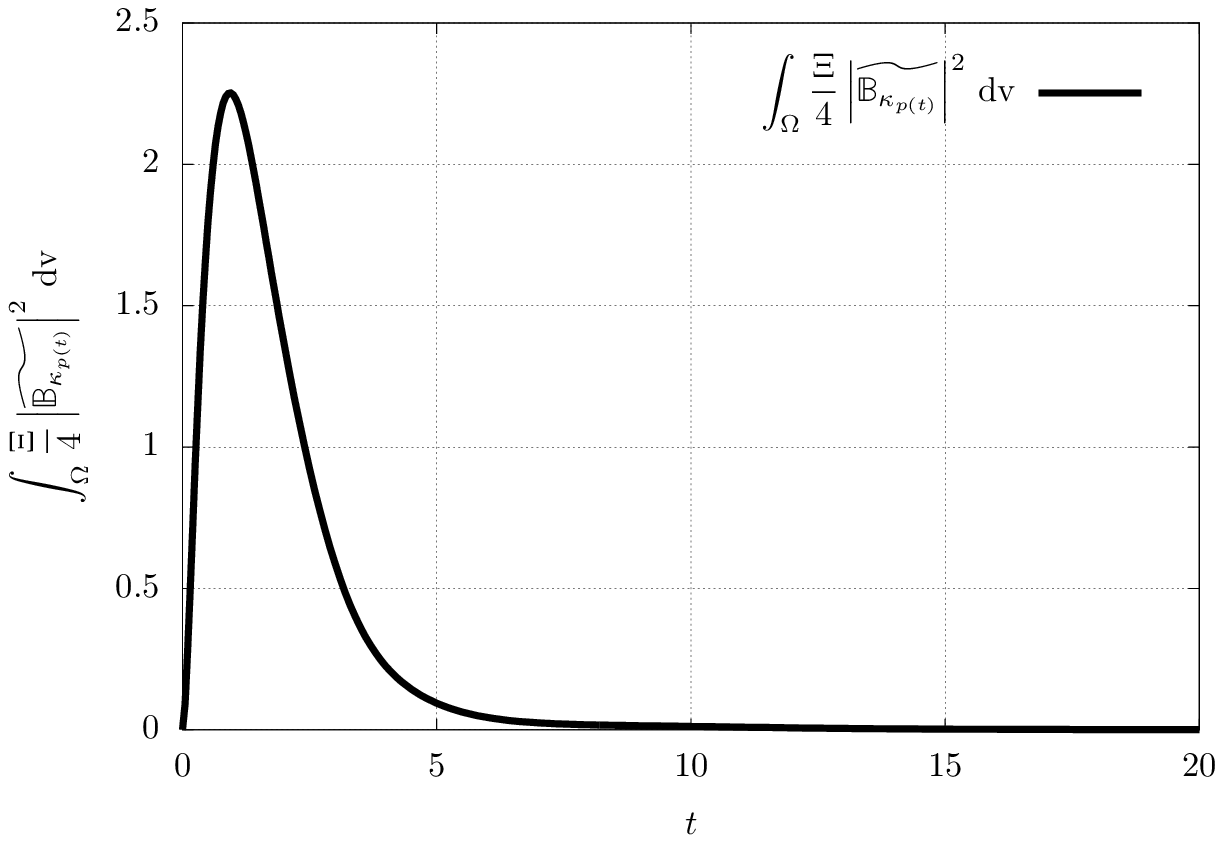}}
  \qquad
  \subfloat[\label{fig:graphs_perturb_velocity-b}Net ``kinetic energy'' of the perturbation, $\int_{\Omega}\frac{1}{2}\absnorm{\widetilde{\vec{v}}}^2\,\cvolumee$.]{\includegraphics[width=0.45\textwidth]{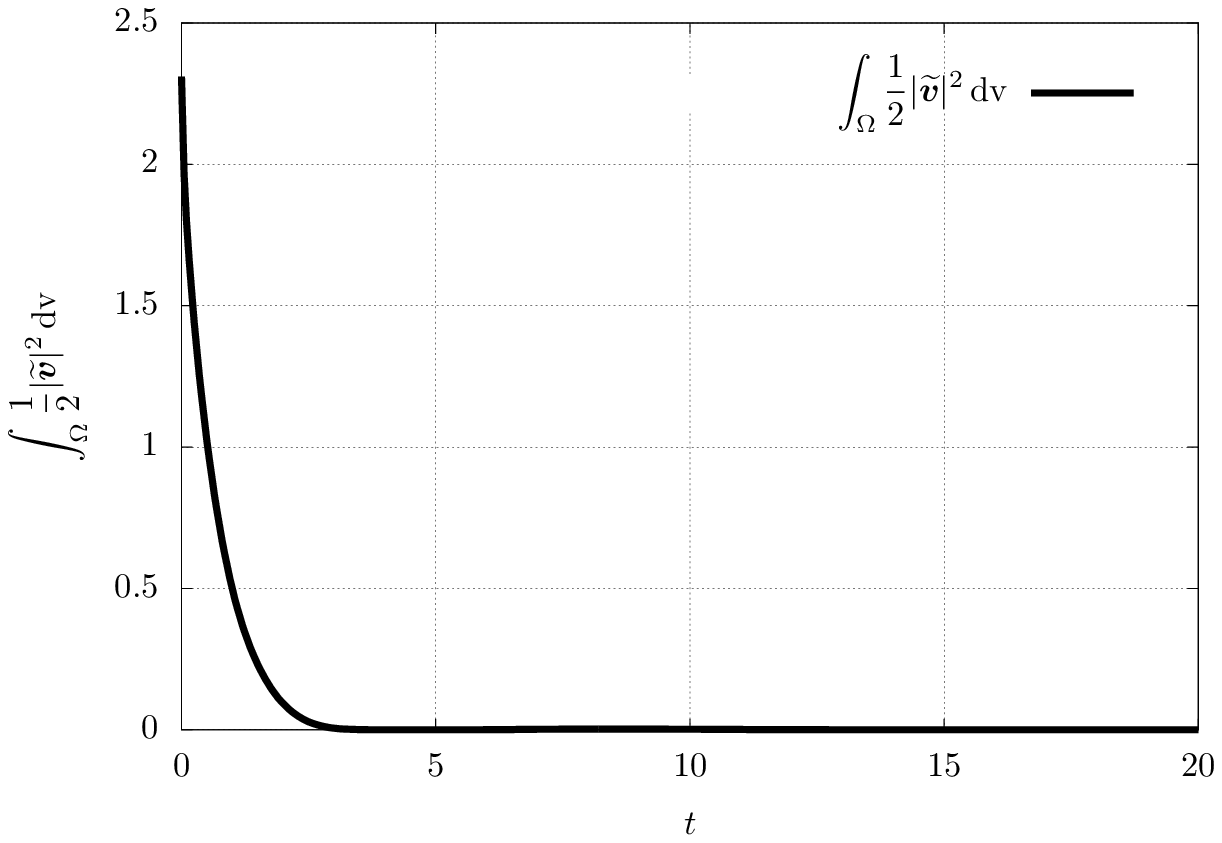}}
  \qquad
  \subfloat[\label{fig:graphs_perturb_velocity-c}Lyapunov type functional $\mathcal{V}_{\mathrm{neq}}$ and the functional $\mathcal{V}_{\mathrm{naive}}$.]{\includegraphics[width=0.45\textwidth]{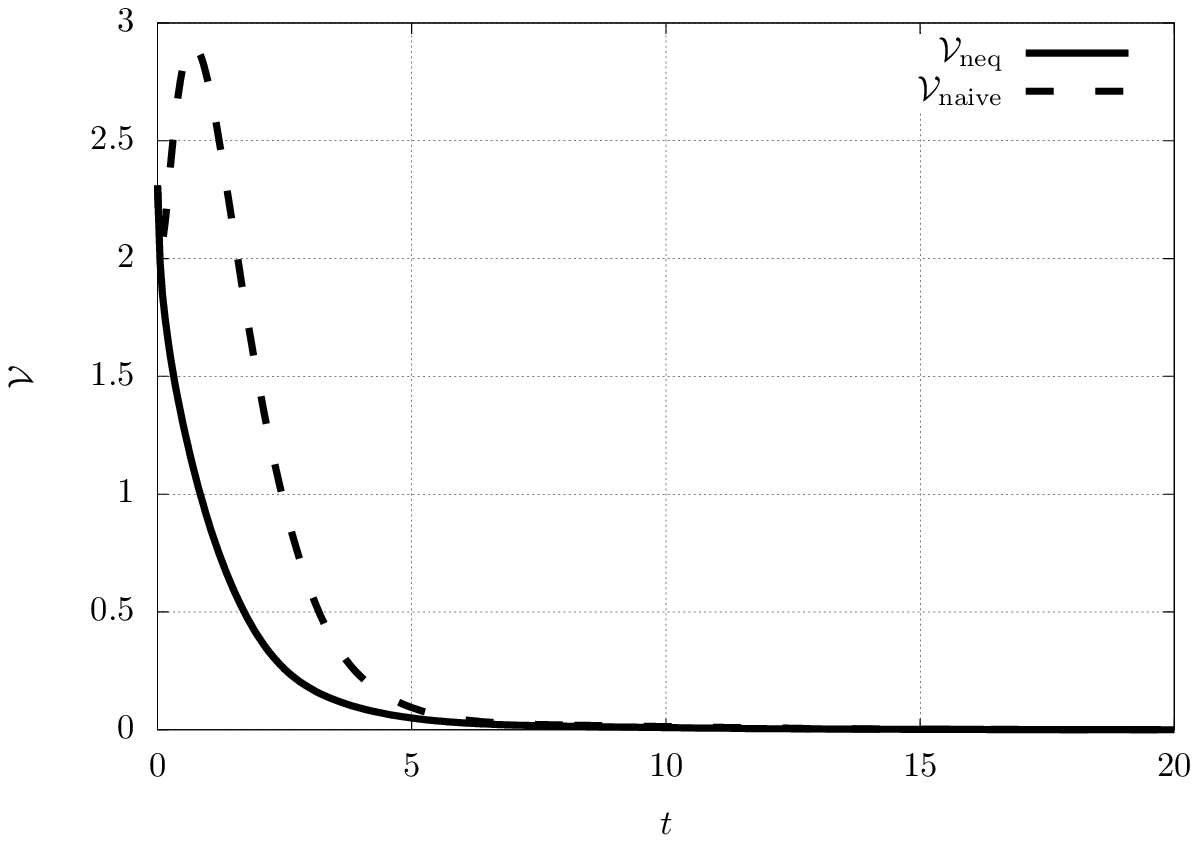}}
  \qquad
  \subfloat[\label{fig:graphs_perturb_velocity-d}Net mechanical energy flux through the boundary, \hbox{$\int_{\partial \Omega} \vectordot{\cstress \vec{v}}{\vec{n}} \, \csurfacees$}.]{\includegraphics[width=0.45\textwidth]{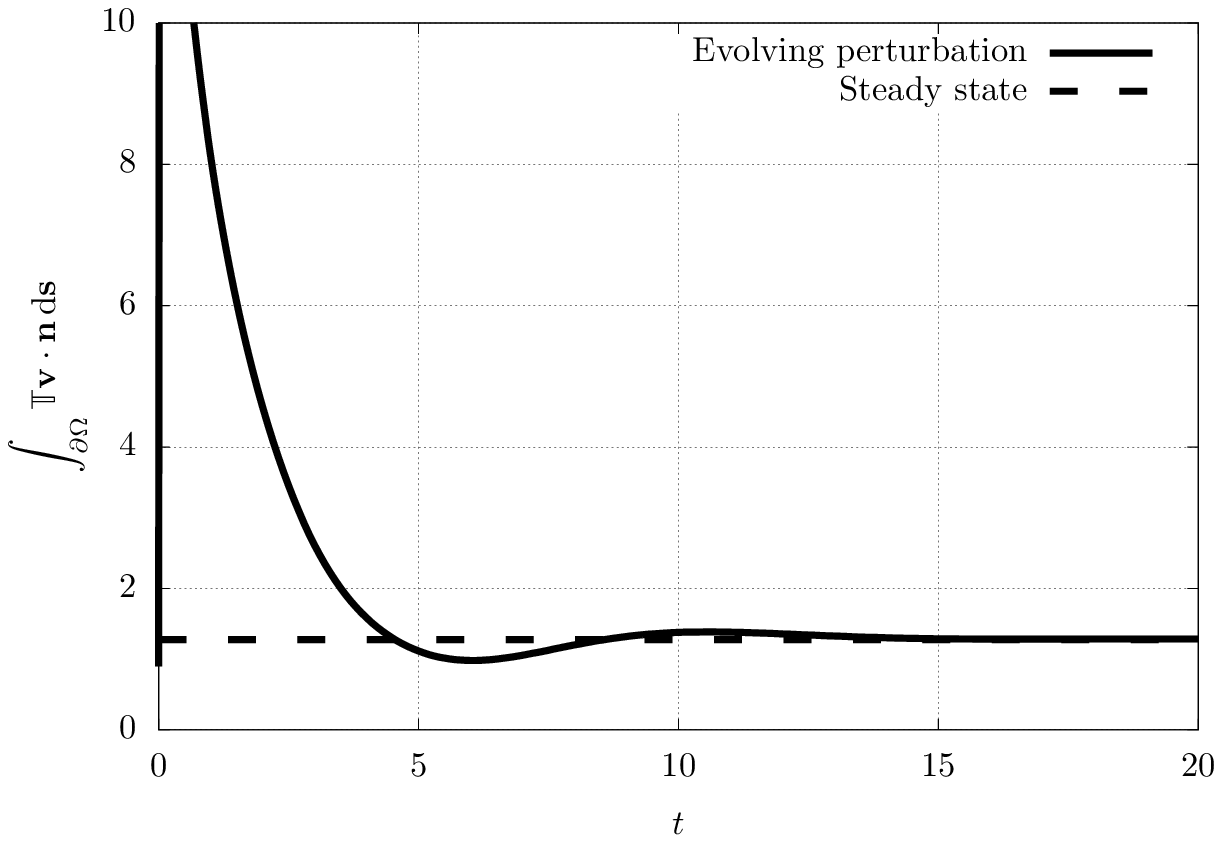}}
  \caption{Scenario B, time evolution of the net quantities.}
  \label{fig:graphs_perturb_velocity}
\end{figure}

Moreover, this numerical experiment is instructive from yet another reason. In Figure~\ref{fig:graphs_perturb_velocity-c} we plot the time evolution of the values of the functionals $\mathcal{V}_{\mathrm{neq}}$ and $\mathcal{V}_{\mathrm{naive}}$. Clearly, the functional $\mathcal{V}_{\mathrm{naive}}$, see~\eqref{eq:121}, which is a naive candidate for the Lyapunov type functional, experiences a transitional growth. Interestingly, the proposed complex Lyapunov type functional~$\mathcal{V}_{\mathrm{neq}}$ is still a decreasing function, although the Reynolds number/Weissenberg number values are outside the region, where we have actually proven the decay of the functional. This further indicates that the functional~$\mathcal{V}_{\mathrm{naive}}$ is indeed not a good candidate for a Lyapunov type functional, see also Section~\ref{sec:main-result} for further discussion.

\section{Conclusion}
\label{sec:conclusion}
We have investigated stability of spatially inhomogeneous non-equilibrium steady states (flows) of viscoelastic fluids described by the Giesekus model. We have derived bounds on the values of the Reynolds number and the Weissenberg number that guarantee the flow stability subject to \emph{any finite perturbation}. The stability has been investigated using the Lyapunov type functional given by the formula
\begin{equation}
  \label{eq:77}
      \mathcal{V}_{\mathrm{neq}}
      \left(
        \left.
          \widetilde{\vec{W}}
        \right\|
        \widehat{\vec{W}}
      \right)
      =
      \int_{\Omega}
      \frac{1}{2}
      \absnorm{\widetilde{\vec{v}}}^2
      \,
      \cvolumee
      +
      \int_{\Omega}
      \frac{\Xi}{2}
      \left[
        -
        \ln \det \left( \identity + \inverse{\widehat{\lcgnc}}\widetilde{\lcgnc} \right)
        +
        \Tr 
        \left( 
          \inverse{\widehat{\lcgnc}} 
          \widetilde{\lcgnc}
        \right)
      \right]
      \,
      \cvolumee
      .
\end{equation}
A few observations concerning the proposed Lyapunov type functional are at hand.

\emph{First}, the proposed Lyapunov type functional has a relatively complicated form. In particular it is \emph{not} quadratic in the perturbation~$\widetilde{\lcgnc}$, and it \emph{depends} on the spatially inhomogeneous non-equilibrium state~$\widehat{\lcgnc}$. This makes it remarkably different from a naive Lyapunov type functional of the form
\begin{equation}
  \label{eq:101}
  \mathcal{V}_{\mathrm{naive}}
      \left(
        \left.
          \widetilde{\vec{W}}
        \right\|
        \widehat{\vec{W}}
      \right)
      =
      \int_{\Omega}
      \frac{1}{2}
      \absnorm{\widetilde{\vec{v}}}^2
      \,
      \cvolumee
      +
      \int_{\Omega}
      \frac{\Xi}{4}
      \absnorm{
        \widetilde{\lcgnc}
      }^2
      \,
      \cvolumee
      ,
\end{equation}
which might be a first try if the stability problem was analysed using the popular ``energy method''. However, as we have shown, the complicated structure of the proposed functional~$\mathcal{V}_{\mathrm{neq}}$ leads to a relatively simple and well structured formula for its time derivative, which in turn allows one to formulate conditions that guarantee the negativity of the time derivative. Furthermore, the complicated structure of the proposed functional~$\mathcal{V}_{\mathrm{neq}}$ also leads to a simple relation between the functional and the metric
\begin{equation}
  \label{eq:48}
  \distance{\widehat{\vec{W}}}{\vec{W}}
  =_{\bydefinition}
  \left(
    \norm[\sleb{2}{\Omega}]{\widehat{\vec{v}} - \vec{v}}^2
    +
    \left[
      \distance[\tensorq{P}_{\Omega}(d), \, \mathrm{BW}]{\widehat{\lcgnc}}{\lcgnc}
    \right]^2
  \right)^{\frac{1}{2}}
\end{equation}
on the set of spatially distributed symmetric positive definite matrices.

\emph{Second}, the Lyapunov type functional has been used in the investigation of stability of solution to the \emph{complete system of nonlinear governing equations}. In particular, the evolution equations for the perturbation have been investigated without any simplification. This makes the present approach different from the ``energy budget'' analysis, see for example~\cite{joo.yl.shaqfeh.esg:effects}, \cite{byars.ja.oztekin.a.ea:spiral}, \cite{ganpule.hk.khomami.b:investigation}, \cite{smith.md.joo.yl.ea:linear}, \cite{karapetsas.g.tsamopoulos.j:on}, \cite{pettas.d.karapetsas.g.ea:on}, and especially \cite{grillet.am.bogaerds.acb.ea:stability} who have investigated the Giesekus model. The ``energy budget'' analysis, although valuable in the discussion of the nature of the instability mechanisms, is based on the \emph{linearised} momentum equation for the perturbation and \emph{linearised} constitutive equation for the ``polymeric stress''. Consequently, the standard ``energy budget'' analysis is, unlike the present approach, a tool that cannot be used in the finite amplitude stability analysis of the \emph{complete system of nonlinear governing equations}. One might also note that despite the complexity of the proposed Lyapunov type functional, the formula for its time derivative is in fact quite simple compared to the formulae in the ``energy budget'' analysis. This happens even though the ``energy budget'' formulae paradoxically stem from various simplifications of the original system of governing equations.

% \emph{Third}, some methods developed for the detailed nonlinear stability analysis of flows of the standard incompressible Navier--Stokes fluid rely on an optimisation technique that allows one to find the perturbation of least amplitude necessary for transition from the base steady state to another state, see~\cite{kerswell.rr:nonlinear}. In the Navier--Stokes case the objective functional used in the optimisation procedure is tantamount to the ``kinetic energy'' of the perturbation, $\int_{\Omega} \frac{1}{2}\absnorm{\widetilde{\vecv}}^2 \, \cvolumee$. In the case of Giesekus fluid, the counterpart of the ``kinetic energy functional'' is the functional~\eqref{eq:77}. Consequently, if the optimisation technique such as that presented in~\cite{kerswell.rr:nonlinear} is to be generalised to the case of Giesekus fluid, then the suitable objective functional might be the functional~\eqref{eq:77}.

\emph{Third}, the Lyapunov type functional has been designed using thermodynamical arguments. In fact the proposed Lyapunov type functional has been constructed from the net mechanical energy functional $\netmenergy$, see~\eqref{eq:netmenergy-dimless}, via the formula~\eqref{eq:lyapunov-functional-definition}. This makes the construction quite general, and one might speculate that a similar approach is very likely applicable to other popular viscoelastic rate-type models such as the PTT model, see~\cite{phan-thien.n.tanner.ri:new}, or the FENE-P model, see \cite{bird.rb.dotson.pj.ea:polymer}, as well as complex viscoelastic rate-type models with, for example, stress diffusion terms, see~\cite{malek.j.prusa.v.ea:thermodynamics} or~\cite{dostalk.m.prusa.v.ea:on}. %In principle, one needs to identify the specific Helmholtz free energy and the entropy production, and then show that the entropy production mechanisms are in a certain parameter range strong enough to make the time derivative of the Lyapunov type functional negative.
Further, the construction of the Lyapunov type functional has been based on the method proposed by~\cite{bulcek.m.malek.j.ea:thermodynamics}, and this method is speculated to work even for complex coupled thermo-mechanical systems. This naturally calls for the investigation of the applicability of the method in more complex settings such as flows of viscoelastic rate-type fluids with temperature dependent material parameters.

\emph{Fourth}, thermodynamical type considerations such as the identification of the energy storage mechanisms and entropy producing mechanisms are known to play an important role in the rigorous mathematical theory of nonlinear partial differential equations governing the motion of viscoelastic fluids, see for example~\cite{hu.d.lelievre.t:new}, \cite{boyaval.s.lelievre.t.ea:free-energy-dissipative}, \cite{barrett.jw.boyaval.s:existence}, \cite{barrett.j.boyaval.s:finite}, \cite{barrett.jw.suli.e:existence*5} or \cite{bulvcek.m.malek.j.ea:pde-analysis}. On the other hand, rigorous mathematical analysis of \emph{long-time behaviour} of viscoelastic fluids is usually done without a direct appeal to thermodynamics, and the available results are quite limited especially if one considers thermodynamically open systems, see for example~\cite{guillope.c.saut.jc:global}, \cite{nohel.j.pego.r:nonlinear}, \cite{jourdain.b.le-bris.c.ea:long-time} or~\cite{renardy.m:some*1}. (Usually, only stability of unidirectional steady flows in simple geometries is considered.) Consequently, the approach proposed in the current contribution might be of interest from the rigorous mathematical perspective as well. Meaning that one should deal with the weak solution to the governing equations, and that one should investigate the applicability of the presented arguments for a solution/perturbation that has only the smoothness that can be actually proven.% In particular, the proposed Lyapunov type functional could provide a handy tool for characterisation of the proximity of the perturbation to the steady state.

\appendix

\section{Distance between positive definite matrices and its generalization to spatially distributed tensor fields}
\label{sec:bures-wass-dist}

The function $\distance[\tensorq{P}(d),\, \mathrm{BW}]{\tensorq{A}}{\tensorq{B}}$ defined as
\begin{equation}
  \label{eq:11}
  \distance[\tensorq{P}(d),\, \mathrm{BW}]{\tensorq{A}}{\tensorq{B}}
  =_{\bydefinition}
  \left\{
    \Tr \tensorq{A}
    +
    \Tr \tensorq{B}
    -
    2 \Tr
    \left[
      \left(
        \tensorq{A}^{\frac{1}{2}}
        \tensorq{B}
        \tensorq{A}^{\frac{1}{2}}
      \right)^{\frac{1}{2}}
    \right]
  \right\}^{\frac{1}{2}},
\end{equation}
and referred to as the \emph{Bures--Wasserstein distance}, is known to be a metric on the manifold of symmetric (Hermitian) positive semidefinite matrices $\tensorq{P}(d) \subset \R^{d \times d}$ of arbitrary dimension~$d$, see~\cite{bhatia.r.jain.t.ea:on}. (See also~\cite{bhatia.r:positive} for a thorough discussion of properties of symmetric positive semidefinite matrices.) Another possibility regarding the definition of a metric on the manifold of symmetric positive definite matrices is
\begin{equation}
  \label{eq:37}
  \distance[\tensorq{P}(d),\, \delta_2]{\tensorq{A}}{\tensorq{B}}
  =_{\bydefinition}
  \absnorm{\ln \left(\tensorq{A}^{-\frac{1}{2}} \tensorq{B} \tensorq{A}^{-\frac{1}{2}}\right)}
  ,
\end{equation}
see~\cite{bhatia.r:positive}, where $\absnorm{\cdot}$ denotes the Frobenius norm, $\absnorm{\tensorq{M}} =_{\bydefinition} \left( \Tr \left( \tensorq{M} \conjugate{\tensorq{M}}\right) \right)^{\frac{1}{2}}$. (See~\cite{graham.md:polymer} for comments on the use of~\eqref{eq:37} in the context of fluid mechanics. For further discussion regarding the metric on the set $\tensorq{P}(d)$ see also~\cite{hiai.f.petz.d:riemannian*1} and \cite{hiai.f.petz.d:riemannian}.) Using the metric~\eqref{eq:11} or~\eqref{eq:37}, one can introduce a metric on the set of spatially distributed symmetric positive semidefinite matrices.

\begin{Definition}
  \label{dfn:1}
  Let $\tensorq{P}_{\Omega}(d) =_{\bydefinition} \left\{ \tensorq{X}: \vec{x} \in \Omega \subset \R^d \mapsto \tensorq{X}(\vec{x}) \in \tensorq{P}(d) \subset \R^{d \times d} \right\}$ denote the set of mappings from the spatial domain $\Omega \subset \R^d$ to the set of symmetric positive semidefinite matrices $\tensorq{P}(d)$, then for $\tensorq{A}, \tensorq{B} \in \tensorq{P}_{\Omega}(d)$ the function
\begin{equation}
  \label{eq:14}
  \distance[\tensorq{P}_{\Omega}(d), \, \cdot]{\tensorq{A}}{\tensorq{B}}
  =
  _{\bydefinition}
  \left[
    \int_{\Omega \subset \R^{d}} \left(\distance[\tensorq{P}(d), \, \cdot]{\tensorq{A}}{\tensorq{B}}\right)^2 \cvolumee
  \right]^{\frac{1}{2}},
\end{equation}
defines a metric on the set $\tensorq{P}_{\Omega}(d)$.
\end{Definition}
 The three fundamental properties of the metric---nonnegativity, symmetry and identity of indiscernibles---are clearly satisfied in virtue of the properties of the pointwise  Bures--Wasserstein/$\delta_2$ metric $\distance[\tensorq{P}(d), \, \cdot]{\tensorq{A}}{\tensorq{B}}$. It remains to show that the newly introduced metric satisfies the triangle inequality, which is shown in Lemma~\ref{lm:1}.
\begin{Lemma}[Triangle inequality]
  \label{lm:1}
  Let $\distance[\tensorq{P}_{\Omega}(d),\, \cdot]{\tensorq{A}}{\tensorq{B}}$ be the function defined via~\eqref{eq:14}, where we use either the Bures--Wasserstein metric~\eqref{eq:11} or the $\delta_2$ metric introduced in~\eqref{eq:37}. The function satisfies the triangle inequality
  \begin{equation}
    \label{eq:15}
    \forall \tensorq{A}, \tensorq{B}, \tensorq{C} \in \tensorq{P}_{\Omega}(d):
    \distance[\tensorq{P}_{\Omega}(d),\, \cdot]{\tensorq{A}}{\tensorq{B}}
    \leq
    \distance[\tensorq{P}_{\Omega}(d),\, \cdot]{\tensorq{A}}{\tensorq{C}}
    +
    \distance[\tensorq{P}_{\Omega}(d),\, \cdot]{\tensorq{C}}{\tensorq{B}}
    .
\end{equation}
\end{Lemma}

\begin{proof}
  We see that
  \begin{multline}
    \label{eq:16}
    \int_{\Omega} \left(\distance[\tensorq{P}(d),\, \cdot]{\tensorq{A}}{\tensorq{B}}\right)^2 \cvolumee
    =
    \int_{\Omega} \distance[\tensorq{P}(d),\, \cdot]{\tensorq{A}}{\tensorq{B}}\distance[\tensorq{P}(d),\, \cdot]{\tensorq{A}}{\tensorq{B}} \cvolumee
    \\
    \leq
    \int_{\Omega} \distance[\tensorq{P}(d),\, \cdot]{\tensorq{A}}{\tensorq{C}}\distance[\tensorq{P}(d),\, \cdot]{\tensorq{A}}{\tensorq{B}} \cvolumee
    +
    \int_{\Omega} \distance[\tensorq{P}(d),\, \cdot]{\tensorq{C}}{\tensorq{B}}\distance[\tensorq{P}(d),\, \cdot]{\tensorq{A}}{\tensorq{B}} \cvolumee
    ,
  \end{multline}
  where we have used the triangle inequality for the Bures--Wasserstein/$\delta_2$ metric $\distance[\tensorq{P}(d),\, \cdot]{\tensorq{A}}{\tensorq{B}}$. Using the H\"older inequality in both integrals on the right-hand side, we arrive at
  \begin{multline}
    \label{eq:17}
    \int_{\Omega} \left(\distance[\tensorq{P}(d),\, \cdot]{\tensorq{A}}{\tensorq{B}}\right)^2 \cvolumee
    \leq
    \left\{
      \left[
        \int_{\Omega} \left( \distance[\tensorq{P}(d),\, \cdot]{\tensorq{A}}{\tensorq{C}} \right)^2 \cvolumee
      \right]^{\frac{1}{2}}
      +
      \left[
        \int_{\Omega} \left( \distance[\tensorq{P}(d),\, \cdot]{\tensorq{C}}{\tensorq{B}} \right)^2 \cvolumee
      \right]^{\frac{1}{2}}
    \right\}
    \\
    \times
    \left[
      \int_{\Omega}
      \left(
        \distance[\tensorq{P}(d),\, \cdot]{\tensorq{A}}{\tensorq{B}}
      \right)^2
      \cvolumee
    \right]^{\frac{1}{2}}
  ,
\end{multline}
which upon dividing both sides by~$\distance[\tensorq{P}_{\Omega}(d),\, \cdot]{\tensorq{A}}{\tensorq{B}}$ yields the desired result.
\end{proof}

Now we are in position to define a metric that will allow us to characterize the distance between the states $\overline{\vec{Y}} = [ \overline{\vec{v}}, \overline{\lcgnc}]$ and $\vec{Y} = [\vec{v}, \lcgnc ]$ of the dynamical system given by equations~\eqref{eq:viscoelastic-model-governing-equations-intro}. Note that in Definition~\ref{dfn:2} we \emph{do not} assume that some of the states $\overline{\vec{Y}}$ or $\vec{Y}$ is a steady state.

\begin{Definition}
  \label{dfn:2}
  Let $\overline{\vec{Y}}$ and $\vec{Y}$ be the states of the dynamical system governed by equations~\eqref{eq:governing-equations-dimless}. The function
  \begin{equation}
    \label{eq:19}
    \distance{\overline{\vec{Y}}}{\vec{Y}}
    =_{\bydefinition}
    \left(
      \norm[\sleb{2}{\Omega}]{\overline{\vec{v}} - \vec{v}}^2
      +
      \left[
        \distance[\tensorq{P}_{\Omega}(d), \, \mathrm{BW}]{\overline{\lcgnc}}{\lcgnc}
      \right]^2
    \right)^{\frac{1}{2}}
    ,
  \end{equation}
  defines a metric on the set of states. Here $\distance[\tensorq{P}_{\Omega}(d), \, \mathrm{BW}]{\cdot}{\cdot}$ denotes the metric on the set of spatially distributed symmetric positive semidefinite matrices introduced in Definition~\ref{dfn:1}, and $\norm[\sleb{2}{\Omega}]{\cdot}$ denotes the standard norm on the Lebesgue space $\sleb{2}{\Omega}$.
  
\end{Definition}

\begin{proof}It is straightforward to verify that~\eqref{eq:19} indeed defines a metric. The three fundamental properties of the metric---nonnegativity, symmetry and identity of indiscernibles---are clearly satisfied in virtue of the properties of the $\sleb{2}{\Omega}$ norm and the metric $\distance[\tensorq{P}_{\Omega}(d), \, \mathrm{BW}]{\cdot}{\cdot}$. The triangle inequality follows from the discrete Minkowski inequality
$
  (\sum_{i=1}^n (a_i + b_i)^p )^{\frac{1}{p}}
  \leq
  (\sum_{i=1}^n a_i^p)^{\frac{1}{p}}
  +
  (\sum_{i=1}^n b_i^p)^{\frac{1}{p}}
$,
see for example~\cite{evans.lc:partial}, and the triangle inequalities for the metric induced by the $\sleb{2}{\Omega}$ norm and the metric $\distance[\tensorq{P}_{\Omega}(d), \, \mathrm{BW}]{\cdot}{\cdot}$.
\end{proof}

Besides the metric~\eqref{eq:19} we also make use of the conventional norm/metric introduced via formula~\eqref{eq:55} below. (The nomenclature ``shifted state space'' is explained in Section~\ref{sec:constr-lyap-funct}.) The fact that~\eqref{eq:55} defines a norm is straightforward to show using the same argument as in the proof of correctness of Definition~\ref{dfn:2}.

\begin{Definition}[Norm/metric on the shifted state space]
  \label{dfn:3}
  Let $\overline{\vec{Z}} =_{\bydefinition} \left[ \overline{\vec{v}}, \overline{\lcgncs} \right] $ and $\vec{Z} =_{\bydefinition} \left[ \vec{v}, \lcgncs \right] $ be the states of the dynamical system governed by equations~\eqref{eq:viscoelastic-model-governing-equations-intro} with the transformation introduced in~\eqref{eq:shifted-state}. The function
  \begin{equation}
    \label{eq:55}
    \norm[\mathrm{st}]{\vec{Z}}
    =_{\bydefinition}
    \left(
      \norm[\sleb{2}{\Omega}]{\vec{v}}^2
      +
      \int_{\Omega}
      \absnorm{
        \lcgncs
      }^2
      \,
      \cvolumee
    \right)^{\frac{1}{2}}
    ,
  \end{equation}
  defines a norm on the corresponding shifted state space, and the function
  \begin{equation}
    \label{eq:53}
    \norm[\mathrm{st}]{\overline{\vec{Z}}- \vec{Z}}
    =_{\bydefinition}
    \left(
      \norm[\sleb{2}{\Omega}]{\overline{\vec{v}} - \vec{v}}^2
      +
      \int_{\Omega}
      \absnorm{
        \overline{\lcgncs}
        -
        \lcgncs
      }^2
      \,
      \cvolumee
    \right)^{\frac{1}{2}}
    ,
  \end{equation}
  is the metric induced by the norm~\eqref{eq:55}.
\end{Definition}

\begin{Lemma}[Pointwise inequality for Bures--Wasserstein metric I]
  \label{lm:4}
  Let  $\tensorq{A}, \tensorq{B} \in \tensorq{P}(d)$, and let $\tensorq{A}$ be an invertible matrix. If $\absnorm{\cdot}$ denotes the standard Frobenius norm, then
  \begin{equation}
    \label{eq:38}
    \distance[\tensorq{P}(d),\, \mathrm{BW}]{\tensorq{A}}{\tensorq{B}}
    \leq
    \absnorm{\tensorq{A}^{\frac{1}{2}}}
    \distance[\tensorq{P}(d),\, \mathrm{BW}]{\identity}{\tensorq{A}^{-\frac{1}{2}} \tensorq{B} \tensorq{A}^{-\frac{1}{2}}}.
  \end{equation}
  
\end{Lemma}

\begin{proof}
  \cite{bhatia.r.jain.t.ea:on}, Theorem 1, have shown that the Bures--Wasserstein metric can be equivalently defined as
  \begin{equation}
    \label{eq:39}
    \distance[\tensorq{P}(d),\, \mathrm{BW}]{\tensorq{A}}{\tensorq{B}}
    =
    \min_{\tensorq{U} \in \tensorq{U}(d)}
    \absnorm{\tensorq{A}^{\frac{1}{2}} - \tensorq{B}^{\frac{1}{2}} \tensorq{U}},
  \end{equation}
  where $\tensorq{U}$ is a matrix that belongs to the group of unitary matrices $\tensorq{U}(d)$ of dimension $d \times d$. Making use of the submultiplicativity of the Frobenius norm, we see that
  \begin{multline}
    \label{eq:40}
    \distance[\tensorq{P}(d),\, \mathrm{BW}]{\tensorq{A}}{\tensorq{B}}
    =
    \min_{\tensorq{U} \in \tensorq{U}(d)}
    \absnorm{\tensorq{A}^{\frac{1}{2}} - \tensorq{B}^{\frac{1}{2}} \tensorq{U}}
    \leq
    \absnorm{
      \tensorq{A}^{\frac{1}{2}}
    }
    \min_{\tensorq{U} \in \tensorq{U}(d)}
    \absnorm{\identity - \tensorq{A}^{-\frac{1}{2}} \tensorq{B}^{\frac{1}{2}} \tensorq{U}}
    \\
    =
    \absnorm{
      \tensorq{A}^{\frac{1}{2}}
    }
    \min_{\tensorq{U} \in \tensorq{U}(d)}
    \left\{
    \Tr
      \left(\identity - \tensorq{A}^{-\frac{1}{2}} \tensorq{B}^{\frac{1}{2}} \tensorq{U} \right)
      \left(\identity - \conjugate{\tensorq{U}} \tensorq{B}^{\frac{1}{2}} \tensorq{A}^{-\frac{1}{2}} \right)
    \right\}^{\frac{1}{2}}
    % \\
    % =
    % \absnorm{
    %   \tensorq{A}^{\frac{1}{2}}
    % }
    % \left\{
    %   \Tr \identity
    %   +
    %   \Tr
    %   \left[
    %     \tensorq{A}^{-\frac{1}{2}} \tensorq{B} \tensorq{A}^{-\frac{1}{2}}
    %   \right]
    %   -
    %   \max_{\tensorq{U} \in \tensorq{U}(d)}
    %   \Tr
    %   \left[
    %     \conjugate{\tensorq{U}}
    %     \tensorq{B}^{\frac{1}{2}} 
    %     \tensorq{A}^{-\frac{1}{2}}
    %     +
    %     \tensorq{A}^{-\frac{1}{2}}
    %     \tensorq{B}^{\frac{1}{2}}
    %     \tensorq{U}
    %   \right]
    % \right\}^{\frac{1}{2}}
    \\
    =
    \absnorm{
      \tensorq{A}^{\frac{1}{2}}
    }
    \left\{
      \Tr \identity
      +
      \Tr
      \left[
        \tensorq{A}^{-\frac{1}{2}} \tensorq{B} \tensorq{A}^{-\frac{1}{2}}
      \right]
      -
      \max_{\tensorq{U} \in \tensorq{U}(d)}
      \Tr
      \left[
        \conjugate{\tensorq{U}}
        \tensorq{X}
        +
        \conjugate{\tensorq{X}}
        \tensorq{U}
      \right]
    \right\}^{\frac{1}{2}}
    ,
  \end{multline}
  where $\tensorq{X}=_{\bydefinition} \tensorq{B}^{\frac{1}{2}} \tensorq{A}^{-\frac{1}{2}}$. Following the argument given in \cite{bhatia.r.jain.t.ea:on}, one can show that the maximum value of the term
  $
  \Tr
    \left[
      \conjugate{\tensorq{U}}
      \tensorq{X}
      +
      \conjugate{\tensorq{X}}
      \tensorq{U}
    \right]
    $
    is attained for the unitary matrix $\tensorq{U}$ that corresponds to the unitary matrix $\tensorq{V}$ in the polar decomposition of $\tensorq{X}$, $\tensorq{X} = \tensorq{V} \tensorq{P}$. The matrix $\tensorq{V}$ in the polar decomposition of $\tensorq{X}$ can be found explicitly as $\tensorq{V} = \tensorq{X} \left(\conjugate{\tensorq{X}} \tensorq{X} \right)^{-\frac{1}{2}}$ which yields
    \begin{equation}
      \label{eq:42}
      \tensorq{V}
      =
      \tensorq{B}^{\frac{1}{2}} \tensorq{A}^{-\frac{1}{2}}
      \left( \tensorq{A}^{-\frac{1}{2}} \tensorq{B} \tensorq{A}^{-\frac{1}{2}} \right)^{-\frac{1}{2}}
      .
    \end{equation}
    If we set $\tensorq{U} =_{\bydefinition} \tensorq{V}$ in the last term on the right-hand side of~\eqref{eq:40}, then we get
    \begin{equation}
      \label{eq:43}
      \max_{\tensorq{U} \in \tensorq{U}(d)}
      \Tr
      \left[
        \conjugate{\tensorq{U}}
        \tensorq{X}
        +
        \conjugate{\tensorq{X}}
        \tensorq{U}
      \right]
      =
      2
      \Tr
      \left[
        \left(
          \tensorq{A}^{-\frac{1}{2}} \tensorq{B} \tensorq{A}^{-\frac{1}{2}}
        \right)^{\frac{1}{2}}
      \right]
      ,
    \end{equation}
    which implies~\eqref{eq:38} since according to the definition of the Bures--Wasserstein metric, see~\eqref{eq:11}, we have
    \begin{equation}
      \label{eq:50}
      \distance[\tensorq{P}(d),\, \mathrm{BW}]{\identity}{\tensorq{A}^{-\frac{1}{2}} \tensorq{B} \tensorq{A}^{-\frac{1}{2}}}
      =
      \Tr \identity
      +
      \Tr \left( \tensorq{A}^{-\frac{1}{2}} \tensorq{B} \tensorq{A}^{-\frac{1}{2}} \right)
      -
      2
      \Tr
      \left[
        \left(
          \tensorq{A}^{-\frac{1}{2}} \tensorq{B} \tensorq{A}^{-\frac{1}{2}}
        \right)^{\frac{1}{2}}
      \right]
      .
    \end{equation}
\end{proof}

\begin{Lemma}[Pointwise inequality for Bures--Wasserstein metric II]
  \label{lm:2}
  Let  $\tensorq{A}, \tensorq{B} \in \tensorq{P}(d)$, and let $\tensorq{A}$ and $\tensorq{B}$ be invertible matrices, then
  \begin{equation}
    \label{eq:18}
    \left[
      \distance[\tensorq{P}(d),\, \mathrm{BW}]{\identity}{\tensorq{A}^{-\frac{1}{2}} \tensorq{B} \tensorq{A}^{-\frac{1}{2}}}
    \right]^2
    \leq
    \Tr
    \left(
      \tensorq{A}^{-\frac{1}{2}} \tensorq{B} \tensorq{A}^{-\frac{1}{2}}
    \right)
    -
    d
    -
    \ln
    \det
    \left(
      \tensorq{A}^{-\frac{1}{2}}
      \tensorq{B}
      \tensorq{A}^{-\frac{1}{2}}
    \right)
    .
  \end{equation}
  Moreover, if $\absnorm{\cdot}$ denotes the Frobenius norm, then 
  \begin{equation}
    \label{eq:49}
    \distance[\tensorq{P}(d),\, \mathrm{BW}]{\identity}{\tensorq{A}^{-\frac{1}{2}} \tensorq{B} \tensorq{A}^{-\frac{1}{2}}}
    =
    \absnorm{
      \identity
      -
      \left(
        \tensorq{A}^{-\frac{1}{2}} \tensorq{B} \tensorq{A}^{-\frac{1}{2}}
      \right)^{\frac{1}{2}}
    }
    .
  \end{equation}
\end{Lemma}

\begin{proof}
  The proof follows from the spectral decomposition. Matrix
  $
  \tensorq{A}^{-\frac{1}{2}}
  \tensorq{B}
  \tensorq{A}^{-\frac{1}{2}}
  $
  is a symmetric positive definite matrix, hence it is unitarily similar to a diagonal matrix with the eigenvalues of
  $
  \tensorq{A}^{-\frac{1}{2}}
  \tensorq{B}
  \tensorq{A}^{-\frac{1}{2}}
  $
  at the diagonal. Consequently, if $\left\{ \lambda_i \right\}_{i=1}^d$ are eigenvalues of
  $
  \tensorq{A}^{-\frac{1}{2}}
  \tensorq{B}
  \tensorq{A}^{-\frac{1}{2}}
  $,
  then the right-hand side of~\eqref{eq:18} reads
  \begin{equation}
    \label{eq:20}
        \Tr
    \left(
      \tensorq{A}^{-\frac{1}{2}} \tensorq{B} \tensorq{A}^{-\frac{1}{2}}
    \right)
    -
    d
    -
    \ln
    \det
    \left(
      \tensorq{A}^{-\frac{1}{2}}
      \tensorq{B}
      \tensorq{A}^{-\frac{1}{2}}
    \right)
    =
    \sum_{i=1}^d \lambda_i - d  - \sum_{i=1}^d \ln \lambda_i
    ,
  \end{equation}
  while on the left-hand side of~\eqref{eq:18} we use the definition of the metric, see~\eqref{eq:11}, and we get
  \begin{equation}
    \label{eq:22}
    \left[
      \distance[\tensorq{P}(d),\, \mathrm{BW}]{\identity}{\tensorq{A}^{-\frac{1}{2}} \tensorq{B} \tensorq{A}^{-\frac{1}{2}}}
    \right]^2
    =
    \Tr
    \identity
    +
    \Tr
    \tensorq{A}^{-\frac{1}{2}}
    \tensorq{B}
    \tensorq{A}^{-\frac{1}{2}}
    -
    2
    \Tr
    \left[
      \left(
        \tensorq{A}^{-\frac{1}{2}}
        \tensorq{B}
        \tensorq{A}^{-\frac{1}{2}}
      \right)^{\frac{1}{2}}
    \right]
    =
    d +  \sum_{i=1}^d \lambda_i - 2 \sum_{i=1}^d \sqrt{\lambda_i}
    .
  \end{equation}
  Instead of~\eqref{eq:18} we therefore need to prove the following inequality for the eigenvalues
  \begin{equation}
    \label{eq:23}
    d +  \sum_{i=1}^d \lambda_i - 2 \sum_{i=1}^d \sqrt{\lambda_i}
    \leq
    \sum_{i=1}^d \lambda_i - d  - \sum_{i=1}^d \ln \lambda_i
    ,
  \end{equation}
  which reduces to
  \begin{equation}
    \label{eq:25}
    2 \left(d -  \sum_{i=1}^d \sqrt{\lambda_i}\right) + \sum_{i=1}^d \ln \lambda_i \leq 0. 
  \end{equation}
  However, if $\lambda_i > 0$, which is our case since we are dealing with positive definite matrices, then it is straightforward to check that
  \begin{equation}
    \label{eq:24}
    2 \left(1 -  \sqrt{\lambda_i}\right) + \ln \lambda_i \leq 0.
  \end{equation}
  Indeed, the function on the left-hand side of~\eqref{eq:24} is a function that vanishes for $\lambda_i= 1$, while its derivative reads $-\frac{1}{\sqrt{\lambda_i}} + \frac{1}{\lambda_i}$, which implies that the function is increasing for $0<\lambda_i <1$, and it is decreasing for $\lambda_i>1$. The point $\lambda_i=1$ is therefore the point where $ 2 (1 -  \sqrt{\lambda_i}) + \ln \lambda_i$ reaches its global maximum, and~\eqref{eq:24} holds. Formula~\eqref{eq:25} and consequently also~\eqref{eq:23} then follows via summation of~\eqref{eq:24} over the individual eigenvalues. Equality~\eqref{eq:49}  is a straightforward consequence of the definition of the Frobenius norm,
  \begin{equation}
    \label{eq:51}
    \absnorm{
      \identity
      -
      \left(
        \tensorq{A}^{-\frac{1}{2}} \tensorq{B} \tensorq{A}^{-\frac{1}{2}}
      \right)^{\frac{1}{2}}
    }^2
    =
    \Tr
    \left[
      \identity
      -
      2
      \left(
        \tensorq{A}^{-\frac{1}{2}} \tensorq{B} \tensorq{A}^{-\frac{1}{2}}
      \right)^{\frac{1}{2}}
      +
      \tensorq{A}^{-\frac{1}{2}} \tensorq{B} \tensorq{A}^{-\frac{1}{2}}
    \right]
    ,
  \end{equation}
  and the definition of the Bures--Wasserstein metric, see~\eqref{eq:50}.
\end{proof}

\begin{Lemma}[Inequality for the Frobenius norm]
  \label{lm:8}
  Let $\tensorq{A}, \tensorq{B} \in \tensorq{P}(d)$, and let $\tensorq{A}$ and $\tensorq{B}$ be invertible matrices, and let $\absnorm{\cdot}$ denote the Frobenius norm, then
  \begin{equation}
    \label{eq:61}
    \absnorm{\tensorq{A} - \tensorq{B}}
    \geq
    \absnorm{\tensorq{A}^{-\frac{1}{2}}}^{-2}
    \absnorm{\identity - \left(\tensorq{A}^{-\frac{1}{2}}\tensorq{B}\tensorq{A}^{-\frac{1}{2}}\right)^{\frac{1}{2}}}.
  \end{equation}
\end{Lemma}

\begin{proof} We observe that
  $
  \absnorm{\tensorq{A}^{-\frac{1}{2}}} \absnorm{\tensorq{A} - \tensorq{B}} \absnorm{\tensorq{A}^{-\frac{1}{2}}}
  \geq
  \absnorm{\identity - \tensorq{A}^{-\frac{1}{2}}\tensorq{B}\tensorq{A}^{-\frac{1}{2}}}
  $,
  where we have used the submultiplicativity of the Frobenius norm, $\absnorm{\tensorq{X}\tensorq{Y}} \leq \absnorm{\tensorq{X}}\absnorm{\tensorq{Y}}$. Next we use the spectral decomposition of the symmetric positive definite matrix $\tensorq{A}^{-\frac{1}{2}}\tensorq{B}\tensorq{A}^{-\frac{1}{2}}$, and we find that
  \begin{equation}
    \label{eq:29}
    \absnorm{\identity - \tensorq{A}^{-\frac{1}{2}}\tensorq{B}\tensorq{A}^{-\frac{1}{2}}}^2
    =
    \sum_{i=1}^d \left(1 - \lambda_i\right)^2
    =
    \sum_{i=1}^d \left(1 - \lambda_i^{\frac{1}{2}}\right)^2 \left(1 + \lambda_i^{\frac{1}{2}}\right)^2
    \geq
    \sum_{i=1}^d \left(1 - \lambda_i^{\frac{1}{2}}\right)^2
    =
    \absnorm{\identity - \left(\tensorq{A}^{-\frac{1}{2}}\tensorq{B}\tensorq{A}^{-\frac{1}{2}}\right)^{\frac{1}{2}}}^2
    ,
  \end{equation}
  where $\left\{ \lambda_i \right\}_{i=1}^d$ denote the positive eigenvalues of $\tensorq{A}^{-\frac{1}{2}}\tensorq{B}\tensorq{A}^{-\frac{1}{2}}$.
\end{proof}

% \begin{multline}
%   \label{eq:41}
%   \mathcal{V}_{\mathrm{neq}}
%   \left(
%     \left.
%       \widetilde{\vec{W}}
%     \right\|
%     \widehat{\vec{W}}
%   \right)
%   \geq
%   \frac{1}{2}
%   \norm[\sleb{2}{\Omega}]{\widehat{\vec{v}} - \left(\widehat{\vec{v}} + \widetilde{\vec{v}}\right)}^2
%   +
%   \frac{\Xi}{2}
%   \int_{\Omega}
%   \frac{\left( \distance[\tensorq{P}(d),\, \mathrm{BW}]{\widehat{\lcgnc}}{\widehat{\lcgnc} + \widetilde{\lcgnc}} \right)^2}{\absnorm{\widehat{\lcgnc}^{\frac{1}{2}}}^2}
%   \,
%   \cvolumee

% \\
% =
% \min
% \left\{
%   \frac{1}{2}
%   ,
%   \frac{\Xi}{2}
%   \min_{\vec{x} \in \Omega}
%   \absnorm{\widehat{\lcgnc}^{\frac{1}{2}}}^{-2}
% \right\}
% \left(\distance{\widehat{\vec{W}}}{\vec{W}}\right)^2
% .
% \end{multline}

\section{Formula for the time derivative}
\label{sec:form-time-deriv}

Let us now derive formula~\eqref{eq:9} for the time derivative of the proposed Lyapunov type functional. Straightforward differentiation of~\eqref{eq:lyapunov-functional-explicit-formula} under the integral sign yields
\begin{equation}
  \label{eq:lyapunov-time-derivative}
  \dd{\mathcal{V}_{\mathrm{neq}}}{t}
  \left(
    \left.
      \widetilde{\vec{W}}
    \right\|
    \widehat{\vec{W}}
  \right)
  =
  \int_{\Omega}
  \vectordot{\widetilde{\vec{v}}}{\pd{\widetilde{\vec{v}}}{t}}
  \,
  \cvolumee
  -
  \int_{\Omega}
  \frac{\Xi}{2} \Tr
  \left[ 
    \inverse{\left( \widehat{\lcgnc} + \widetilde{\lcgnc} \right)} \pd{\widetilde{\lcgnc}}{t} 
  \right]
  \,
  \cvolumee
  +
  \int_{\Omega}
  \frac{\Xi}{2} \Tr
  \left( 
    \inverse{\widehat{\lcgnc}}\pd{\widetilde{\lcgnc}}{t}
  \right)
  \,
  \cvolumee
  .
\end{equation}
In the following we shall treat the individual terms of \eqref{eq:lyapunov-time-derivative} separately. In order to proceed further, we need to formulate the evolution equations for the perturbation $\widetilde{\vec{W}}$, which will give us formulae for the partial time derivatives of $\widetilde{\vec{v}}$ and $\widetilde{\lcgnc}$.

\subsection{Evolution equations for perturbation}
\label{sec:evolution-equations-perturbation}
The perturbed field $\vec{W} = \widehat{\vec{W}} + \widetilde{\vec{W}}$ must satisfy the governing equations~\eqref{eq:governing-equations-dimless}, which means that
\begin{subequations}
  \label{eq:0}
  \begin{align}
    \label{eq:70}
    \divergence \left( \widehat{\vec{v}} + \widetilde{\vec{v}} \right) &= 0, \\
    \label{eq:1}
    \pd{}{t} (\widehat{\vec{v}} + \widetilde{\vec{v}})
    +
    \left[ \vectordot{(\widehat{\vec{v}} + \widetilde{\vec{v}})}{\nabla} \right]
    (\widehat{\vec{v}} + \widetilde{\vec{v}})
    &=
    \divergence \cstress(\widehat{\vec{W}} + \widetilde{\vec{W}})
    ,
  \end{align}
  and
  \begin{multline}
    \label{eq:2}
    \pd{}{t} \left( \widehat{\lcgnc} + \widetilde{\lcgnc} \right)
    +
    \left[ \vectordot{(\widehat{\vec{v}} + \widetilde{\vec{v}})}{\nabla} \right]
    \left( \widehat{\lcgnc} + \widetilde{\lcgnc} \right)
    -
    \left( \widehat{\gradvl} + \widetilde{\gradvl}  \right) \left( \widehat{\lcgnc} + \widetilde{\lcgnc} \right)
    - 
    \left( \widehat{\lcgnc} + \widetilde{\lcgnc} \right) \transpose{\left( \widehat{\gradvl} + \widetilde{\gradvl}  \right)}
    \\
    =
    -
    \frac{1}{\Weissenberg}
    \left[ 
      \alpha \left( \widehat{\lcgnc} + \widetilde{\lcgnc} \right)^2 
      + 
      (1 - 2 \alpha) \left( \widehat{\lcgnc} + \widetilde{\lcgnc} \right) 
      - 
      (1 - \alpha) \identity
    \right]
    ,
  \end{multline}
where
\begin{equation}
  \label{eq:74}
  \cstress(\widehat{\vec{W}} + \widetilde{\vec{W}})
  =
  \frac{1}{\Reynolds} \left(\widehat{\mns} + \widetilde{\mns} \right) \identity 
  +
  \frac{2}{\Reynolds} \left(\widehat{\gradsym} + \widetilde{\gradsym} \right)
  + 
  \Xi \, \traceless{\left( \widehat{\lcgnc} + \widetilde{\lcgnc} \right)}
  ,
\end{equation}
\end{subequations}
and where we have used the notation $\widehat{\gradvl} = \nabla \widehat{\vec{v}}$, $\widetilde{\gradvl} = \nabla \widetilde{\vec{v}}$ and similarly for the symmetric part of the velocity gradient $\widehat{\gradsym}$ and $\widetilde{\gradsym}$. Now we are in position to exploit the fact that the non-equilibrium steady state $\widehat{\vec{W}}$ solves~\eqref{eq:steady-state-equations}. Using~\eqref{eq:steady-state-equations} in~\eqref{eq:0} yields
\begin{subequations}
  \label{eq:evolution-equations-perturbation}
  \begin{align}
    \label{eq:evolution-equation-velocity-pertubation-incompressibility}
    \divergence \widetilde{\vec{v}} &= 0, \\
    \label{eq:evolution-equation-velocity-pertubation}
    \pd{\widetilde{\vec{v}}}{t}
    &=
    \divergence \cstress(\widetilde{\vec{W}})
    -
    \left(
      \vectordot{\widetilde{\vec{v}}}{\nabla}
    \right) \widehat{\vec{v}}
    -
    \left[
      \vectordot{(\widehat{\vec{v}} + \widetilde{\vec{v}})}{\nabla}          
    \right]
    \widetilde{\vec{v}}
    ,
  \end{align}
  where
  \begin{equation}
    \label{eq:78}
    \cstress(\widetilde{\vec{W}})
    =
    \frac{1}{\Reynolds} \widetilde{\mns}  \identity 
    +
    \frac{2}{\Reynolds} \widetilde{\gradsym}
    + 
    \Xi \, \traceless{\left( \widetilde{\lcgnc} \right)}
    ,
  \end{equation}
  and
  \begin{multline}
    \label{eq:evolution-equation-lcgnc-perturbation}
    \pd{\widetilde{\lcgnc}}{t}
    =
    - 
    \left[ 
      \vectordot{(\widehat{\vec{v}} + \widetilde{\vec{v}})}{\nabla} 
    \right]
    \left( 
      \widehat{\lcgnc} + \widetilde{\lcgnc} 
    \right)
    +
    \left(
      \widehat{\gradvl} + \widetilde{\gradvl}
    \right)
    \left(
      \widehat{\lcgnc} + \widetilde{\lcgnc}
    \right)
    + 
    \left(
      \widehat{\lcgnc} + \widetilde{\lcgnc}
    \right)
    \transpose{\left(
        \widehat{\gradvl} + \widetilde{\gradvl}
      \right)}
    \\
    - 
    \frac{1}{\Weissenberg} 
    \left[ 
      \alpha \left( \widehat{\lcgnc} + \widetilde{\lcgnc} \right)^2 
      + 
      (1 - 2 \alpha) \left( \widehat{\lcgnc} + \widetilde{\lcgnc} \right) 
      - 
      (1 - \alpha) \identity
    \right]
    ,
  \end{multline}
\end{subequations}
  which can be in virtue of~\eqref{eq:steady-state-equation-2} further simplified to
  \begin{multline}
    \label{eq:123}
    \pd{\widetilde{\lcgnc}}{t}
    =
    -
    \left(
      \vectordot{\widetilde{\vec{v}}}{\nabla}
    \right)
    \widetilde{\lcgnc}
    -
    \left(
      \vectordot{\widehat{\vec{v}}}{\nabla}
    \right)
    \widetilde{\lcgnc} 
    - 
    \left(
      \vectordot{\widetilde{\vec{v}}}{\nabla}
    \right)
    \widehat{\lcgnc} 
    +
    \widehat{\gradvl}
    \widetilde{\lcgnc}
    + 
    \widetilde{\gradvl}
    \widehat{\lcgnc}
    +
    \widetilde{\gradvl}
    \widetilde{\lcgnc}
    +
    \widetilde{\lcgnc}
    \transpose{\widehat{\gradvl}}
    \\
    + 
    \widehat{\lcgnc}
    \transpose{\widetilde{\gradvl}}
    +
    \widetilde{\lcgnc}
    \transpose{\widetilde{\gradvl}}
    - 
    \frac{1}{\Weissenberg} 
    \left[
      \alpha \widetilde{\lcgnc}^2
      +
      \alpha \left( \widehat{\lcgnc}\widetilde{\lcgnc} +  \widetilde{\lcgnc} \widehat{\lcgnc} \right) 
      + 
      (1 - 2 \alpha) \widetilde{\lcgnc}
    \right]
    .
  \end{multline}
  In the subsequent analysis it will be however more convenient to work with~\eqref{eq:evolution-equation-lcgnc-perturbation} instead of~\eqref{eq:123}. Equation~\eqref{eq:123} is exploited only in Section~\ref{sec:main-result}. System \eqref{eq:evolution-equations-perturbation} of evolution equations for the perturbation~$\widetilde{\vec{W}}$ must be solved subject to boundary condition~\eqref{eq:bc-perturbation-gamma2} and periodic boundary condition on $\Gamma_1$.

Having identified the formulae for the time derivatives of~$\widetilde{\vec{v}}$ and $\widetilde{\lcgnc}$, we can go back to~\eqref{eq:lyapunov-time-derivative}, and we can start to evaluate the individual terms on the right-hand side of the formula~\eqref{eq:lyapunov-time-derivative} for the time derivative of the Lyapunov type functional.

\subsection{First term of \eqref{eq:lyapunov-time-derivative}}
\label{sec:lyapunov-time-derivative-first-term}

Using the evolution equation for the velocity perturbation \eqref{eq:evolution-equation-velocity-pertubation} we see that
\begin{equation}
  \label{eq:3}
  \int_{\Omega}
  \vectordot{\widetilde{\vec{v}}}{\pd{\widetilde{\vec{v}}}{t}}
  \,
  \cvolumee
  =
  \int_{\Omega}
  \vectordot{\widetilde{\vec{v}}}{\divergence \cstress(\widetilde{\vec{W}})}
  \,
  \cvolumee
  -
  \int_{\Omega}
  \vectordot{\widetilde{\vec{v}}}{\left[ \left( \vectordot{\widetilde{\vec{v}}}{\nabla} \right) \widehat{\vec{v}} \right]}
  \,
  \cvolumee
  -
  \int_{\Omega}
  \vectordot{\widetilde{\vec{v}}}{\left[ \left\{ \vectordot{\left( \widehat{\vec{v}} + \widetilde{\vec{v}} \right)}{\nabla} \right\} \widetilde{\vec{v}} \right]}
  \,
  \cvolumee
  ,
\end{equation}
The first term of the last equation can be manipulated as follows 
\begin{equation}
  \label{eq:3a}
  \int_{\Omega}
  \vectordot{\widetilde{\vec{v}}}{\divergence \cstress(\widetilde{\vec{W}})}
  \,
  \cvolumee
  =
  \int_{\Omega}
  \divergence \left( 
    \cstress(\widetilde{\vec{W}}) \widetilde{\vec{v}} 
  \right)
  \,
  \cvolumee
  -
  \int_{\Omega}
  \tensordot{\nabla \widetilde{\vec{v}}}{\cstress(\widetilde{\vec{W}})}
  \,
  \cvolumee
  =
  -
  \int_{\Omega}
  \tensordot{\nabla \widetilde{\vec{v}}}{\cstress(\widetilde{\vec{W}})}
  \,
  \cvolumee
  ,
\end{equation}
where we have used the Stokes theorem and the identity \eqref{eq:surface-integral-vanishes}. The second term on the right-hand side of~\eqref{eq:3} can be written as
\begin{equation}
  \label{eq:3aa}
  \int_{\Omega}
  \vectordot{\widetilde{\vec{v}}}{\left[ \left( \vectordot{\widetilde{\vec{v}}}{\nabla} \right) \widehat{\vec{v}} \right]}
  \,
  \cvolumee
  =
  \int_{\Omega}
  \vectordot{\widehat{\gradsym} \widetilde{\vec{v}}}{\widetilde{\vec{v}}}
  \,
  \cvolumee
  .
\end{equation}
The third term on the right-hand side of~\eqref{eq:3} vanishes in virtue of the standard manipulation
\begin{multline}
  \label{eq:3b}
  \int_{\Omega}
  \vectordot{\widetilde{\vec{v}}}{\left\{ \left[ \vectordot{\left( \widehat{\vec{v}} + \widetilde{\vec{v}} \right)}{\nabla} \right] \widetilde{\vec{v}} \right\} }
  \,
  \cvolumee
  =
  \frac{1}{2}
  \int_{\Omega}
  \left[ \vectordot{\left( \widehat{\vec{v}} + \widetilde{\vec{v}} \right)}{\nabla} \right] \absnorm{\widetilde{\vec{v}}}^2
  \,
  \cvolumee
  \\
  =
  \frac{1}{2}
  \int_{\Omega}
  \divergence \left[ 
    \left( \widehat{\vec{v}} + \widetilde{\vec{v}} \right) \absnorm{\widetilde{\vec{v}}}^2 
  \right]
  \,
  \cvolumee
  -
  \frac{1}{2}
  \int_{\Omega}
  \absnorm{\widetilde{v}}^2 \divergence \left( 
    \widehat{\vec{v}} + \widetilde{\vec{v}}
  \right)
  \,
  \cvolumee
  =
  0
  ,
\end{multline}
where we have again used \eqref{eq:surface-integral-vanishes} as well as the incompressibility condition~\eqref{eq:70}. Substituting \eqref{eq:3a}, \eqref{eq:3aa} and \eqref{eq:3b} back into \eqref{eq:3} yields
%\begin{equation}
%  \label{eq:3c}
$
\int_{\Omega}
  \vectordot{\widetilde{\vec{v}}}{\pd{\widetilde{\vec{v}}}{t}}
  \,
  \cvolumee
  =
  -
  \int_{\Omega}
  \tensordot{\nabla \widetilde{\vec{v}}}{\cstress(\widetilde{\vec{W}})}
  \,
  \cvolumee
  -
  \int_{\Omega}
  \, \vectordot{\widehat{\gradsym} \widetilde{\vec{v}}}{\widetilde{\vec{v}}}
  \,
  \cvolumee
$.
%\end{equation}
Finally, using the explicit formula for the Cauchy stress tensor~\eqref{eq:78} we obtain
\begin{equation}
  \label{eq:lyapunov-time-derivative-first-term}
  \int_{\Omega}
  \vectordot{\widetilde{\vec{v}}}{\pd{\widetilde{\vec{v}}}{t}}
  \,
  \cvolumee
  =
  -
  \int_{\Omega}
  \frac{2}{\Reynolds} \tensordot{\widetilde{\gradsym}}{\widetilde{\gradsym}}
  \,
  \cvolumee
  -
  \int_{\Omega}
  \Xi \, \tensordot{\widetilde{\lcgnc}}{\widetilde{\gradsym}}
  \,
  \cvolumee
  -
  \int_{\Omega}
  \vectordot{\widehat{\gradsym} \widetilde{\vec{v}}}{\widetilde{\vec{v}}}
  \,
  \cvolumee
  .
\end{equation}

\subsection{Second term of \eqref{eq:lyapunov-time-derivative}}
\label{sec:lyapunov-time-derivative-second-term}
Using the evolution equation \eqref{eq:evolution-equation-lcgnc-perturbation} for $\widetilde{\lcgnc}$ yields 
\begin{multline}
  \label{eq:4}
  \int_{\Omega}
  \frac{\Xi}{2} \Tr
  \left[ 
    \inverse{\left( \widehat{\lcgnc} + \widetilde{\lcgnc} \right)} \pd{\widetilde{\lcgnc}}{t} 
  \right]
  \,
  \cvolumee
  \\
  =
  \int_{\Omega}
  \frac{\Xi}{2} \Tr
  \left[
    - \inverse{\left( \widehat{\lcgnc} + \widetilde{\lcgnc} \right)}
    \left\{
      \vectordot{\left( \widehat{\vec{v}} + \widetilde{\vec{v}} \right)}{\nabla}
    \right\}
    \left( \widehat{\lcgnc} + \widetilde{\lcgnc} \right)
  \right]
  \,
  \cvolumee
  \\
  +
  \int_{\Omega}
  \frac{\Xi}{2} \Tr
  \left[
    \inverse{\left( \widehat{\lcgnc} + \widetilde{\lcgnc} \right)} 
    \left( \widehat{\gradvl} + \widetilde{\gradvl} \right)
    \left( \widehat{\lcgnc} + \widetilde{\lcgnc} \right)
  \right]
  \,
  \cvolumee
  \\
  +
  \int_{\Omega}
  \frac{\Xi}{2} \Tr
  \left[
    \inverse{\left( \widehat{\lcgnc} + \widetilde{\lcgnc} \right)} 
    \left( \widehat{\lcgnc} + \widetilde{\lcgnc} \right)
    \transpose{\left( \widehat{\gradvl} + \widetilde{\gradvl} \right)}
  \right]
  \,
  \cvolumee
  \\
  -
  \int_{\Omega}
  \frac{\Xi}{2 \Weissenberg} \Tr
  \left[ 
    \inverse{\left( \widehat{\lcgnc} + \widetilde{\lcgnc} \right)} 
    \left(
      \alpha \left( \widehat{\lcgnc} + \widetilde{\lcgnc} \right)^2 
      + 
      (1 - 2 \alpha) \left( \widehat{\lcgnc} + \widetilde{\lcgnc} \right) 
      - 
      (1 - \alpha) \identity
    \right)
  \right]
  \,
  \cvolumee
  .
\end{multline}
We can immediately see that the second and the third terms vanish due to the incompressibility condition~\eqref{eq:70} and the invariance of the trace under cyclic permutations. The first term on the right-hand side of~\eqref{eq:4} can be shown to vanish as well via the standard manipulation
\begin{multline}
  \label{eq:4a}
  \int_{\Omega}
  \frac{\Xi}{2} \Tr
  \left[
    \inverse{\left( \widehat{\lcgnc} + \widetilde{\lcgnc} \right)}
    \left\{
      \vectordot{\left( \widehat{\vec{v}} + \widetilde{\vec{v}} \right)}{\nabla}
    \right\}
    \left( \widehat{\lcgnc} + \widetilde{\lcgnc} \right)
  \right]
  \,
  \cvolumee
  % \\
  % =
  % \int_{\Omega}
  % \frac{\Xi}{2}
  % \left[ 
  %   \vectordot{\left( \widetilde{\vec{v}} + \widehat{\vec{v}} \right)}{\nabla} 
  % \right]
  % \ln \det \left( \widehat{\lcgnc} + \widetilde{\lcgnc} \right)
  % \,
  % \cvolumee
  \\
  =
  \int_{\Omega}
  \frac{\Xi}{2} \divergence
  \left[ 
    \left( \widetilde{\vec{v}} + \widehat{\vec{v}} \right)
    \ln \det \left( \widehat{\lcgnc} + \widetilde{\lcgnc} \right)
  \right]
  \,
  \cvolumee
  -
  \int_{\Omega}
  \frac{\Xi}{2} \divergence \left( \widetilde{\vec{v}} + \widehat{\vec{v}} \right)
  \ln \det \left( \widehat{\lcgnc} + \widetilde{\lcgnc} \right)
  \,
  \cvolumee
  =
  0.
\end{multline}
(The last equality again follows from the Stokes theorem, the identity~\eqref{eq:surface-integral-vanishes} and the incompressibility condition~\eqref{eq:70}. Moreover, we have also used the fact that $\vectordot{\vec{u}}{\left(\nabla \ln \det \generictensor\right)} = \Tr \left( \inverse{\generictensor} \left(\vectordot{\vec{u}}{\nabla} \right) \generictensor \right)$.) Finally, we see that
\begin{multline}
  \label{eq:lyapunov-time-derivative-second-term}
  \int_{\Omega}
  \frac{\Xi}{2} \Tr
  \left[ 
    \inverse{\left( \widehat{\lcgnc} + \widetilde{\lcgnc} \right)} \pd{\widetilde{\lcgnc}}{t} 
  \right]
  \,
  \cvolumee
  \\
  =
  -
  \int_{\Omega}
  \frac{\Xi}{2 \Weissenberg} \Tr
  \left[
    \alpha \left( \widehat{\lcgnc} + \widetilde{\lcgnc} \right)
    +
    (1 - 2 \alpha) \identity
    - 
    (1 - \alpha) \inverse{\left( \widehat{\lcgnc} + \widetilde{\lcgnc} \right)}
  \right]
  .
\end{multline}

\subsection{Third term of \eqref{eq:lyapunov-time-derivative}}
\label{sec:lyapunov-time-derivative-third-term}
Let us first make use of the equation for the steady flow \eqref{eq:steady-state-equation-2} to derive a useful identity. Multiplying~\eqref{eq:steady-state-equation-2} by $\inverse{\widehat{\lcgnc}} \widetilde{\lcgnc} \inverse{\widehat{\lcgnc}}$ from the left, taking the trace and integrating over the domain $\Omega$ yields
\begin{multline}
  \label{eq:555}
  \int_{\Omega}
  \Tr \left[
    \inverse{\widehat{\lcgnc}} \widetilde{\lcgnc} \inverse{\widehat{\lcgnc}}
    \left( \vectordot{\widehat{\vec{v}}}{\nabla} \right) {\widehat{\lcgnc}}
  \right]
  \,
  \cvolumee
  -
  \int_{\Omega}
  \Tr \left[
    \inverse{\widehat{\lcgnc}} \widetilde{\lcgnc} \inverse{\widehat{\lcgnc}}
    \left( 
      \widehat{\gradvl} \widehat{\lcgnc} + \widehat{\lcgnc} \transpose{\widehat{\gradvl}} 
    \right)
  \right]
  \,
  \cvolumee
  \\
  =
  -
  \int_{\Omega}
  \frac{1}{\Weissenberg} \Tr \left[
    \inverse{\widehat{\lcgnc}} \widetilde{\lcgnc} \inverse{\widehat{\lcgnc}}
    \left( 
      \alpha \widehat{\lcgnc}^2 + (1 - 2 \alpha) \widehat{\lcgnc} - (1 - \alpha) \identity
    \right)
  \right]
  \,
  \cvolumee
  .
\end{multline}
Consequently, using the invariance of the trace under cyclic permutations and rearranging the terms we obtain the identity
\begin{multline}
  \label{eq:auxiliary-identity}
  \int_{\Omega}
  \Tr \left[
    \inverse{\widehat{\lcgnc}} \widetilde{\lcgnc} \inverse{\widehat{\lcgnc}}
    \left( \vectordot{\widehat{\vec{v}}}{\nabla} \right) {\widehat{\lcgnc}}
  \right]
  \,
  \cvolumee
  =
  \int_{\Omega}
  \Tr \left[
    \inverse{\widehat{\lcgnc}} 
    \left(
      \widehat{\gradvl} \widetilde{\lcgnc} 
      + 
      \widetilde{\lcgnc} \transpose{\widehat{\gradvl}}
    \right)
  \right]
  \,
  \cvolumee
  \\	
  -
  \int_{\Omega}
  \frac{1}{\Weissenberg} \Tr \left[
    \alpha \widetilde{\lcgnc}	    
    +
    (1 - 2 \alpha) \inverse{\widehat{\lcgnc}} \widetilde{\lcgnc}
    -
    (1 - \alpha) \inverse{\widehat{\lcgnc}} \widetilde{\lcgnc} \inverse{\widehat{\lcgnc}}
  \right]
  \,
  \cvolumee
  .
\end{multline}

Having the identity, let us now go back to~\eqref{eq:lyapunov-time-derivative}, and let us manipulate the third term on the right-hand side of~\eqref{eq:lyapunov-time-derivative}. Employing the evolution equation~\eqref{eq:evolution-equation-lcgnc-perturbation} for $\widetilde{\lcgnc}$ yields
\begin{multline}
  \label{eq:5}
  \int_{\Omega}
  \frac{\Xi}{2} \Tr
  \left( 
    \inverse{\widehat{\lcgnc}}\pd{\widetilde{\lcgnc}}{t}
  \right)
  \,
  \cvolumee
  =
  \int_{\Omega}
  \frac{\Xi}{2} \Tr
  \left[
    - \inverse{\widehat{\lcgnc}}
    \left\{
      \vectordot{\left( \widehat{\vec{v}} + \widetilde{\vec{v}} \right)}{\nabla}
    \right\}
    \left( \widehat{\lcgnc} + \widetilde{\lcgnc} \right)
  \right]
  \,
  \cvolumee
  \\
  +
  \int_{\Omega}
  \frac{\Xi}{2} \Tr
  \left[
    \inverse{\widehat{\lcgnc}} 
    \left( \widehat{\gradvl} + \widetilde{\gradvl} \right)
    \left( \widehat{\lcgnc} + \widetilde{\lcgnc} \right)
  \right]
  \,
  \cvolumee
  +
  \int_{\Omega}
  \frac{\Xi}{2} \Tr
  \left[
    \inverse{\widehat{\lcgnc}} 
    \left( \widehat{\lcgnc} + \widetilde{\lcgnc} \right)
    \transpose{\left( \widehat{\gradvl} + \widetilde{\gradvl} \right)}
  \right]
  \,
  \cvolumee
  \\
  -
  \int_{\Omega}
  \frac{\Xi}{2 \Weissenberg} \Tr
  \left[
    \inverse{\widehat{\lcgnc}}
    \left\{
      \alpha \left( \widehat{\lcgnc} + \widetilde{\lcgnc} \right)^2 
      + 
      (1 - 2 \alpha) \left( \widehat{\lcgnc} + \widetilde{\lcgnc} \right) 
      - 
      (1 - \alpha) \identity
    \right\}
  \right]
  \,
  \cvolumee
  .
\end{multline}
The first term on the right-hand side of~\eqref{eq:5} reduces to
\begin{equation}
  \label{eq:434}
  \int_{\Omega}
  \frac{\Xi}{2} \Tr
  \left[
    - \inverse{\widehat{\lcgnc}}
    \left\{
      \vectordot{\left( \widehat{\vec{v}} + \widetilde{\vec{v}} \right)}{\nabla}
    \right\}
    \left( \widehat{\lcgnc} + \widetilde{\lcgnc} \right)
  \right]
  \,
  \cvolumee
  =
  -
  \int_{\Omega}
  \frac{\Xi}{2} \Tr
  \left[
    \inverse{\widehat{\lcgnc}}
    \left\{
      \vectordot{\left( \widehat{\vec{v}} + \widetilde{\vec{v}} \right)}{\nabla}
    \right\}
    \widetilde{\lcgnc}
  \right]
  \,
  \cvolumee
  ,
\end{equation}
where we have used a similar manipulation as in \eqref{eq:4a}. Moreover, the expression can be further transformed as follows
\begin{multline}
  \label{eq:823}
  \int_{\Omega}
  \frac{\Xi}{2} \Tr
  \left[
    \inverse{\widehat{\lcgnc}}
    \left\{
      \vectordot{\left( \widehat{\vec{v}} + \widetilde{\vec{v}} \right)}{\nabla}
    \right\}
    \widetilde{\lcgnc}
  \right]
  \,
  \cvolumee
  \\
  =
  \int_{\Omega}
  \frac{\Xi}{2}
  \left\{
    \vectordot{\left( \widehat{\vec{v}} + \widetilde{\vec{v}} \right)}{\nabla}
  \right\}
  \Tr \left( \inverse{\widehat{\lcgnc}} \widetilde{\lcgnc} \right)
  \,
  \cvolumee
  -
  \int_{\Omega}
  \frac{\Xi}{2} \Tr
  \left[
    \widetilde{\lcgnc}
    \left\{
      \vectordot{\left( \widehat{\vec{v}} + \widetilde{\vec{v}} \right)}{\nabla}
    \right\}
    \inverse{\widehat{\lcgnc}}
  \right]
  \,
  \cvolumee
  \\
  =
  \int_{\Omega}
  \frac{\Xi}{2}
  \divergence \left[
    \left( \widehat{\vec{v}} + \widetilde{\vec{v}} \right)
    \Tr \left( \inverse{\widehat{\lcgnc}} \widetilde{\lcgnc} \right)
  \right]
  \,
  \cvolumee
  -
  \int_{\Omega}
  \frac{\Xi}{2}
  \divergence \left( \widehat{\vec{v}} + \widetilde{\vec{v}} \right)
  \Tr \left( \inverse{\widehat{\lcgnc}} \widetilde{\lcgnc} \right)
  \,
  \cvolumee
  \\
  -
  \int_{\Omega}
  \frac{\Xi}{2} \Tr
  \left[
    \widetilde{\lcgnc}
    \left\{
      \vectordot{\left( \widehat{\vec{v}} + \widetilde{\vec{v}} \right)}{\nabla}
    \right\}
    \inverse{\widehat{\lcgnc}}
  \right]
  \,
  \cvolumee
  =
  -
  \int_{\Omega}
  \frac{\Xi}{2} \Tr
  \left[
    \widetilde{\lcgnc}
    \left\{
      \vectordot{\left( \widehat{\vec{v}} + \widetilde{\vec{v}} \right)}{\nabla}
    \right\}
    \inverse{\widehat{\lcgnc}}
  \right]
  \,
  \cvolumee
  \\
  =
  \int_{\Omega}
  \frac{\Xi}{2} \Tr
  \left[
    \inverse{\widehat{\lcgnc}} \widetilde{\lcgnc} \inverse{\widehat{\lcgnc}}
    \left\{
      \vectordot{\left( \widehat{\vec{v}} + \widetilde{\vec{v}} \right)}{\nabla}
    \right\}
    \widehat{\lcgnc}
  \right]
  \,
  \cvolumee
  ,
\end{multline} 
where we have used the Stokes theorem, the identity~\eqref{eq:surface-integral-vanishes}, the incompressibility condition~\eqref{eq:70}, and the identity
$
\Tr
\left[
  \generictensor
  \left( \vectordot{\vec{u}}{\nabla} \right)
  \inverse{\tensorq{B}}
\right]
=
-
\Tr
\left[
  \generictensor
  \inverse{\tensorq{B}}
  \left\{
    \left(
      \vectordot{\vec{u}}{\nabla}
    \right)
    \tensorq{B}
  \right\}
  \inverse{\tensorq{B}}
\right]
$
that follows from the fact that $\nabla \left( \inverse{\tensorq{B}} \right)= -\inverse{\tensorq{B}} \left( \nabla \tensorq{B} \right) \inverse{\tensorq{B}}$. Note that a part of the expression on the right-hand side of~\eqref{eq:823} is the same as the left-hand side of~\eqref{eq:auxiliary-identity}.

So far we have found that
\begin{multline}
  \label{eq:80}
    \int_{\Omega}
  \frac{\Xi}{2} \Tr
  \left( 
    \inverse{\widehat{\lcgnc}}\pd{\widetilde{\lcgnc}}{t}
  \right)
  \,
  \cvolumee
  =
  -
  \int_{\Omega}
  \frac{\Xi}{2} \Tr
  \left[
    \inverse{\widehat{\lcgnc}} \widetilde{\lcgnc} \inverse{\widehat{\lcgnc}}
    \left\{
      \vectordot{\left( \widehat{\vec{v}} + \widetilde{\vec{v}} \right)}{\nabla}
    \right\}
    \widehat{\lcgnc}
  \right]
  \,
  \cvolumee
  \\
  +
  \int_{\Omega}
  \frac{\Xi}{2} \Tr
  \left[
    \inverse{\widehat{\lcgnc}} 
    \left( \widehat{\gradvl} + \widetilde{\gradvl} \right)
    \left( \widehat{\lcgnc} + \widetilde{\lcgnc} \right)
  \right]
  \,
  \cvolumee
  +
  \int_{\Omega}
  \frac{\Xi}{2} \Tr
  \left[
    \inverse{\widehat{\lcgnc}} 
    \left( \widehat{\lcgnc} + \widetilde{\lcgnc} \right)
    \transpose{\left( \widehat{\gradvl} + \widetilde{\gradvl} \right)}
  \right]
  \,
  \cvolumee
  \\
  -
  \int_{\Omega}
  \frac{\Xi}{2 \Weissenberg} \Tr
  \left[
    \inverse{\widehat{\lcgnc}}
    \left\{
      \alpha \left( \widehat{\lcgnc} + \widetilde{\lcgnc} \right)^2 
      + 
      (1 - 2 \alpha) \left( \widehat{\lcgnc} + \widetilde{\lcgnc} \right) 
      - 
      (1 - \alpha) \identity
    \right\}
  \right]
  \,
  \cvolumee
  ,
\end{multline}
which---upon exploiting the identity \eqref{eq:auxiliary-identity} in the first term---reduces to
\begin{multline}
  \label{eq:lyapunov-time-derivative-third-term}
  \int_{\Omega}
  \frac{\Xi}{2} \Tr
  \left( 
    \inverse{\widehat{\lcgnc}}\pd{\widetilde{\lcgnc}}{t}
  \right)
  \,
  \cvolumee
  \\
  =
  -
  \int_{\Omega}
  \frac{\Xi}{2} \Tr
  \left[
    \inverse{\widehat{\lcgnc}} \widetilde{\lcgnc} \inverse{\widehat{\lcgnc}}
    \left(
      \vectordot{\widetilde{\vec{v}}}{\nabla}
    \right)
    \widehat{\lcgnc}
  \right]
  \,
  \cvolumee
  +
  \int_{\Omega}
  \frac{\Xi}{2}
  \tensordot{\inverse{\widehat{\lcgnc}}}{
    \left(
      \widetilde{\gradvl} \widetilde{\lcgnc} 
      + \widetilde{\lcgnc} \transpose{\widetilde{\gradvl}}
    \right)}
  \,
  \cvolumee
  \\
  -
  \int_{\Omega}
  \frac{\Xi}{2 \Weissenberg} \Tr
  \left[
    \alpha 
    \left( 
      \widehat{\lcgnc} + \widetilde{\lcgnc} 
      + \inverse{\widehat{\lcgnc}} \widetilde{\lcgnc}^2 
    \right)
    +
    (1 - \alpha) 
    \left( 
      \inverse{\widehat{\lcgnc}} \widetilde{\lcgnc} \inverse{\widehat{\lcgnc}} 
      - 
      \inverse{\widehat{\lcgnc}} 
    \right)
    +
    (1 - 2 \alpha) \identity
  \right]
  \,
  \cvolumee
  ,
\end{multline}
where we have again used the incompressibility condition.

\subsection{Explicit formula for the time derivative of Lyapunov type functional}
\label{sec:time-derivative-lyapunov-functional-explicit}
Using~\eqref{eq:lyapunov-time-derivative-first-term}, \eqref{eq:lyapunov-time-derivative-second-term} and \eqref{eq:lyapunov-time-derivative-third-term} in \eqref{eq:lyapunov-time-derivative} we obtain
\begin{multline}
  \label{eq:6}
  \dd{\mathcal{V}_{\mathrm{neq}}}{t}
  \left(
    \left.
      \widetilde{\vec{W}}
    \right\|
    \widehat{\vec{W}}
  \right)
  =
  -
  \int_{\Omega}
  \frac{2}{\Reynolds} \tensordot{\widetilde{\gradsym}}{\widetilde{\gradsym}}
  \,
  \cvolumee
  -
  \int_{\Omega}
  \Xi \, \tensordot{\widetilde{\lcgnc}}{\widetilde{\gradsym}}
  \,
  \cvolumee
  -
  \int_{\Omega}
  \vectordot{\widehat{\gradsym} \widetilde{\vec{v}}}{\widetilde{\vec{v}}}
  \,
  \cvolumee
  \\
  -
  \int_{\Omega}
  \frac{\Xi}{2} \Tr
  \left[
    \inverse{\widehat{\lcgnc}} \widetilde{\lcgnc} \inverse{\widehat{\lcgnc}}
    \left(
      \vectordot{\widetilde{\vec{v}}}{\nabla}
    \right)
    \widehat{\lcgnc}
  \right]
  \,
  \cvolumee
  +
  \int_{\Omega}
  \frac{\Xi}{2}
  \tensordot{\inverse{\widehat{\lcgnc}}}{
    \left(
      \widetilde{\gradvl} \widetilde{\lcgnc} 
      + \widetilde{\lcgnc} \transpose{\widetilde{\gradvl}}
    \right)}
  \,
  \cvolumee
  \\
  -
  \int_{\Omega}
  (1 - \alpha) \frac{\Xi}{2 \Weissenberg} \Tr
  \left[
    \inverse{\left( \widehat{\lcgnc} + \widetilde{\lcgnc} \right)}
    - \inverse{\widehat{\lcgnc}}
    + \inverse{\widehat{\lcgnc}}\widetilde{\lcgnc}\inverse{\widehat{\lcgnc}}
  \right]
  \,
  \cvolumee
  \\
  -
  \int_{\Omega}
  \alpha \frac{\Xi}{2 \Weissenberg} \Tr
  \left[
    \inverse{\widehat{\lcgnc}} \widetilde{\lcgnc}^2
  \right]
  \,
  \cvolumee
  ,
\end{multline}
and using the resolvent identity 
$
\inverse{\generictensor_1} - \inverse{\generictensor_2}
=
\inverse{\generictensor_1}
\left(
  \generictensor_2
  -
  \generictensor_1
\right)
\inverse{\generictensor_2}
$
we can rearrange the next to the last term in~\eqref{eq:6}, and we see that the time derivative of the proposed Lyapunov type functional is indeed~\eqref{eq:9}.

\section{Estimate on the time derivative}
\label{sec:estim-time-deriv-1}

\begin{proof}[Proof of Lemma~\ref{lm:6}]
  For the sake of completeness, let us recall that the Korn equality reads
\begin{equation}
  \label{eq:102}
  2
  \intcvolume
  {
    \tensordot{\gradsym_{\vec{u}}}{\gradsym_{\vec{u}}}
  }
  =
  \intcvolume
  {
    \tensordot{\nabla \vec{u}}{\nabla \vec{u}}
  }
  +
  \intcvolume
  {
    \left(
      \divergence \vec{u}
    \right)^2
  }
  ,
\end{equation}
where $\vec{u}$ is a (smooth) vector field that vanishes on $\Gamma_2$ and satisfies the periodic boundary condition on $\Gamma_1$, and~$\gradsym_{\vec{u}}$ denotes the symmetric part of the corresponding gradient $\nabla \vec{u}$. The Friedrichs--Poincar\'e inequality reads
$
{\norm{\vec{u}}}_{\sleb{2}{\Omega}}^2
  \leq
  C_P
  {\norm{\nabla \vec{u}}}_{\sleb{2}{\Omega}}^2,
$
where $C_P$ is the domain dependent constant, $\vec{u}$ is a (smooth) vector field that vanishes on $\Gamma_2$ and satisfies the periodic boundary condition on $\Gamma_1$, and ${\norm{\vec{w}}}_{\sleb{2}{\Omega}}^2 =_{\bydefinition} \int_{\Omega} \absnorm{\vec{w}}^2 \, \cvolumee$ denotes the standard Lebesgue space norm, and $\absnorm{\vec{w}}$ denotes the standard Euclidean norm. 

Now we are in position to find an estimate on the right-hand side of~\eqref{eq:9}. In estimating the time derivative~\eqref{eq:9} we can completely ignore the next-to-the last nonpositive term on the right-hand side of~\eqref{eq:9}. We can thus write
\begin{multline}
  \label{eq:7}
  \dd{\mathcal{V}_{\mathrm{neq}}}{t}
  \left(
    \left.
      \widetilde{\vec{W}}
    \right\|
    \widehat{\vec{W}}
  \right)
  \leq
  -
  \int_{\Omega}
  \frac{2}{\Reynolds} \tensordot{\widetilde{\gradsym}}{\widetilde{\gradsym}}
  \,
  \cvolumee
  -
  \int_{\Omega}
  \Xi \, \tensordot{\widetilde{\lcgnc}}{\widetilde{\gradsym}}
  \,
  \cvolumee
  -
  \int_{\Omega}
  \vectordot{\widehat{\gradsym} \widetilde{\vec{v}}}{\widetilde{\vec{v}}}
  \,
  \cvolumee
  \\
  -
  \int_{\Omega}
  \frac{\Xi}{2} \Tr
  \left[
    \inverse{\widehat{\lcgnc}} \widetilde{\lcgnc} \inverse{\widehat{\lcgnc}}
    \left(
      \vectordot{\widetilde{\vec{v}}}{\nabla}
    \right)
    \widehat{\lcgnc}
  \right]
  \,
  \cvolumee
  +
  \int_{\Omega}
  \frac{\Xi}{2}
  \tensordot{\inverse{\widehat{\lcgnc}}}{
    \left(
      \widetilde{\gradvl} \widetilde{\lcgnc} 
      + \widetilde{\lcgnc} \transpose{\widetilde{\gradvl}}
    \right)}
  \,
  \cvolumee
  \\
  -
  \int_{\Omega}
  \alpha \frac{\Xi}{2 \Weissenberg} \Tr
  \left(
    \inverse{\widehat{\lcgnc}} \widetilde{\lcgnc}^2
  \right)
  \,
  \cvolumee
  .
\end{multline}
This means that we lose a term that has the negative sign, and that the estimate on the stability range will be more demanding than it would have to be. We also need to restrict ourselves to $\alpha \not =0$. Let us now bound the individual terms involved in~\eqref{eq:7}. The first term can be in virtue of Korn equality rewritten as
\begin{equation}
  \label{eq:first-bound}
  -
  \int_{\Omega}
  \frac{2}{\Reynolds} \tensordot{\widetilde{\gradsym}}{\widetilde{\gradsym}}
  \,
  \cvolumee
  =
  -
  \frac{1}{\Reynolds} {\norm{\nabla \widetilde{\vec{v}}}}_{\sleb{2}{\Omega}}^2
  .
\end{equation}
The third term on the right-hand side of~\eqref{eq:7} can be estimated using the spectral estimate
$
  \lambda_{\min}( \widehat{\gradsym}) \absnorm{\widetilde{\vec{v}}}^2
  \leq
  \vectordot{\widehat{\gradsym} \widetilde{\vec{v}}}{\widetilde{\vec{v}}}
  \leq
  \lambda_{\max} ( \widehat{\gradsym}) \absnorm{\widetilde{\vec{v}}}^2
$
for the symmetric matrix $\widehat{\gradsym}$, where $\lambda_{\min}(\widehat{\gradsym})$ denotes the smallest eigenvalue of~$\widehat{\gradsym}$ and~$\lambda_{\max}(\widehat{\gradsym})$ denotes the largest eigenvalue of~$\widehat{\gradsym}$ at the given spatial point $\vec{x}$. The spectral estimate yields
\begin{equation}
  \label{eq:110}
  -
  \int_{\Omega}
  \vectordot{\widehat{\gradsym} \widetilde{\vec{v}}}{\widetilde{\vec{v}}}
  \,
  \cvolumee
  \leq
  -
  \int_{\Omega}
  \lambda_{\min}(\widehat{\gradsym})
  \absnorm{\widetilde{\vec{v}}}^2
  \,
  \cvolumee
  \leq
  \sup_{\vec{x} \in \Omega} \absnorm{\lambda_{\min}(\widehat{\gradsym})}
  {\norm{\widetilde{\vec{v}}}}_{\sleb{2}{\Omega}}^2
  .
\end{equation}
Further,  using the Poincar\'e inequality we get
\begin{equation}
  \label{eq:third-bound}
  -
  \int_{\Omega}
  \vectordot{\widehat{\gradsym} \widetilde{\vec{v}}}{\widetilde{\vec{v}}}
  \,
  \cvolumee
  \leq
  \sup_{\vec{x} \in \Omega} \absnorm{\lambda_{\min}(\widehat{\gradsym})}
  {\norm{\widetilde{\vec{v}}}}_{\sleb{2}{\Omega}}^2
  \leq
  C_P
  \sup_{\vec{x} \in \Omega} \absnorm{\lambda_{\min}(\widehat{\gradsym})}
  {\norm{\nabla \widetilde{\vec{v}}}}_{\sleb{2}{\Omega}}^2
  .
\end{equation}
The last term in~\eqref{eq:7} can be estimated as
\begin{equation}
  \label{eq:sixth-bound}
  -
  \int_{\Omega}
  \alpha \frac{\Xi}{2 \Weissenberg} \Tr
  \left(
    \inverse{\widehat{\lcgnc}} \widetilde{\lcgnc}^2
  \right)
  \,
  \cvolumee
  \leq
  -
  \alpha \frac{\Xi}{2 \Weissenberg}  
  \inf_{\vec{x} \in \Omega} \lambda_{\min}(\inverse{\widehat{\lcgnc}}) 
  {\norm{\widetilde{\lcgnc}}}_{\sleb{2}{\Omega}}^2
  ,
\end{equation}
where $\lambda_{\min}(\widehat{\lcgnc})$ denotes the smallest eigenvalue of the given symmetric positive definite matrix~$\widehat{\lcgnc}$ at the given spatial point $\vec{x}$. (Note that $\lambda_{\min}(\widehat{\lcgnc})$ is in virtue of the positivity of $\widehat{\lcgnc}$ a positive number.)

Estimates on the rest of the terms are obtained easily by the application of Cauchy--Schwarz and Young inequalities and the submultiplicative property of the matrix norm. First, we group the second and the fifth term in~\eqref{eq:7}, and we get
\begin{subequations}
  \label{eq:positive-bounds}
  \begin{multline}
    \label{eq:second-bound}
    -
    \int_{\Omega}
    \Xi \, \tensordot{\widetilde{\lcgnc}}{\widetilde{\gradsym}}
    \,
    \cvolumee
    +
    \int_{\Omega}
    \frac{\Xi}{2} 
    \tensordot{\inverse{\widehat{\lcgnc}}}{
      \left(
        \widetilde{\gradvl} \widetilde{\lcgnc} 
        + \widetilde{\lcgnc} \transpose{\widetilde{\gradvl}}                    
      \right)}
    \,
    \cvolumee
    \\
    =
    -
    \int_{\Omega}
    \frac{\Xi}{2}
    \tensordot{\identity}
    {
      \left(
        \widetilde{\gradvl} \widetilde{\lcgnc} 
        +
        \widetilde{\lcgnc} \transpose{\widetilde{\gradvl}}                    
      \right)      
    }
    \,
    \cvolumee
    +
    \int_{\Omega}
    \frac{\Xi}{2} 
    \tensordot{\inverse{\widehat{\lcgnc}}}
    {
      \left(
        \widetilde{\gradvl} \widetilde{\lcgnc} 
        +
        \widetilde{\lcgnc} \transpose{\widetilde{\gradvl}}                    
      \right)
    }
    \,
    \cvolumee
    \\
    =
    \int_{\Omega}
    \frac{\Xi}{2} 
    \tensordot
    {
      \left(
        \inverse{\widehat{\lcgnc}} - \identity
      \right)
    }
    {
      \left(
        \widetilde{\gradvl} \widetilde{\lcgnc} 
        +
        \widetilde{\lcgnc} \transpose{\widetilde{\gradvl}}                    
      \right)
    }
    \,
    \cvolumee
    \leq
    \frac{\Xi}{2} \sup_{\vec{x} \in \Omega} \absnorm{\inverse{\widehat{\lcgnc}} - \identity}
    \left(
      {\norm{\widetilde{\lcgnc}}}_{\sleb{2}{\Omega}}^2 
      +
      {\norm{\nabla \widetilde{\vec{v}}}}_{\sleb{2}{\Omega}}^2
    \right)
    .
  \end{multline}
  The fourth term in~\eqref{eq:7} is in virtue of the Poincar\'e inequality estimated as
  \begin{multline}
    \label{eq:fourth-bound}
    -
    \int_{\Omega}
    \frac{\Xi}{2} \Tr
    \left[
      \inverse{\widehat{\lcgnc}} \widetilde{\lcgnc} \inverse{\widehat{\lcgnc}}
      \left(
        \vectordot{\widetilde{\vec{v}}}{\nabla}
      \right)
      \widehat{\lcgnc}
    \right]
    \,
    \cvolumee
    \\
    \leq
    \frac{\Xi}{4}
    \sup_{\vec{x} \in \Omega}
    \absnorm{\inverse{\widehat{\lcgnc}}}^2
    \sup_{\vec{x} \in \Omega}
    \absnorm{\nabla \widehat{\lcgnc}}
    \left(
      {\norm{\widetilde{\lcgnc}}}_{\sleb{2}{\Omega}}^2 
      +
      {\norm{\widetilde{\vec{v}}}}_{\sleb{2}{\Omega}}^2
    \right)
    \\
    \leq
    \frac{\Xi}{4}
    \sup_{\vec{x} \in \Omega}
    \absnorm{\inverse{\widehat{\lcgnc}}}^2
    \sup_{\vec{x} \in \Omega}
    \absnorm{\nabla \widehat{\lcgnc}}
    \left(
      {\norm{\widetilde{\lcgnc}}}_{\sleb{2}{\Omega}}^2 
      +
      C_P
      {\norm{\nabla \widetilde{\vec{v}}}}_{\sleb{2}{\Omega}}^2
    \right)
    ,
  \end{multline}
\end{subequations}
where we introduce the notation $\absnorm{\nabla \widehat{\lcgnc}}^2 =_{\bydefinition} \sum_{k,l,m = 1}^3 \left( \pd{\tensor{{\widehat{\lcgncc_{\nplacer}}}}{^k^l}}{x_m} \right)^2$ for the norm of the corresponding third order tensor. Altogether, the estimates \eqref{eq:first-bound}, \eqref{eq:sixth-bound} and \eqref{eq:positive-bounds} give us~\eqref{eq:lyapunov-derivative-estimate}.  
\end{proof}

%%% Local Variables:
%%% mode: latex
%%% TeX-master: "viscoelastic-stability-mdpi"
%%% End: